\documentclass[]{aa}
\usepackage{txfonts}
\usepackage{graphicx} 
\numberwithin{equation}{section}
\usepackage{hyperref}
\usepackage{natbib}
\usepackage{subcaption}
\usepackage{color}
\usepackage[normalem]{ulem}
\usepackage{gensymb}

\usepackage{extarrows}

\newcommand{\Nat}{\mathbb N} 

\newcommand{\Ha}{\mathcal H} 
\newcommand{\GravC}{\mathcal G} 
\newcommand{\D}[1]{\ensuremath{\operatorname{d}\!{#1}}} 

\newcommand{\M}{\ell} 
\newcommand{\VARPI}{\varpi} 
\newcommand{\LAMBDA}{\lambda} 
\newcommand{\GAMMA}{\gamma}

\newcommand{\LAMBDONA}{\Lambda}
\newcommand{\GAMMONA}{\Gamma}

\newcommand{\KAPPONA}{\mathcal K}

\newcommand{\THETA}{\theta}

\newcommand{\PSI}{\psi}
\newcommand{\PSIONA}{\Psi}

\newcommand{\ANGMOM}{\mathcal L}

\newcommand{\Z}{\mathbb Z} 

\newcommand{\bfr}{\mathbf r} 

\newcommand{\bfp}{\mathbf p} 

\newcommand{\bfu}{\mathbf u} 

\newcommand{\orb}{\mathrm{orb}} 
\newcommand{\res}{\mathrm{res}} 
\newcommand{\syn}{\mathrm{syn}} 
 
\newcommand{\const}{\mathrm{const}}
\newcommand{\mpl}{m_{\mathrm{pl}}}

\newcommand{\mEarth}{M_\oplus}

\newcommand{\kepl}{\mathrm{kepl}}
\newcommand{\Hakepl}{{\Ha_{\mathrm{kepl}}}}
\newcommand{\Hapert}{{\Ha_{\mathrm{pert}}}}
\newcommand{\Hares}{{\Ha_{\mathrm{res}}}}
\newcommand{\Hasyn}{{\Ha_{\mathrm{syn}}}}
\newcommand{\eq}{\mathrm{eq}}
\newcommand{\crit}{\mathrm{crit}}
\newcommand{\mig}{\mathrm{mig}} 
\newcommand{\damp}{\mathrm{damp}}
\newcommand{\wave}{\mathrm{wave}} 
\newcommand{\alphaSigma}{\alpha_\Sigma}

\newcommand{\betah}{{\beta_\mathrm{f}}}
\newcommand{\secres}{\mathrm{scnd.res}}
\newcommand{\pert}{\mathrm{pert}} 

\newcommand{\bfalpha}{{\boldsymbol\alpha}}

\newcommand{\sincos}{\left\{\begin{matrix}\sin\\ \cos\end{matrix}\right\}}
\newcommand{\phase}{\mathrm{phase}}

\newcommand{\bfm}{\mathbf m}

\newcommand{\bfh}{\mathbf h}

\newcommand{\chisyn}{{\chi_\mathrm{syn}}}
\newcommand{\barchisyn}{{\bar{\chi}_\mathrm{syn}}}

\newcommand{\grad}{\boldsymbol\nabla} 

\usepackage{mathtools}

\DeclarePairedDelimiter{\floor}{\lfloor}{\rfloor}

\begin{document}
\font\courier=pcrr at 10pt
\font\codefont=pcrr

\title{The onset of instability in resonant chains}
\author{
Gabriele Pichierri\inst{1},
\and
Alessandro Morbidelli\inst{2}
          }

\institute{
(1) Max-Planck-Institut f{\"u}r Astronomie, K{\"o}nigstuhl 17, 69117 Heidelberg\\
\email{pichierri@mpia.de}\\
(2) Laboratoire Lagrange, UMR7293, Universit{\'e} C{\^o}te d'Azur, CNRS, Observatoire de la C{\^o}te d’Azur, Boulevard de l’Observatoire, 06304 Nice Cedex 4, France
             }
\authorrunning{Pichierri, Morbidelli}
\date{\today}
\abstract{
There is evidence that most chains of mean motion resonances of type $k$:$k-1$ among exoplanets become unstable once the dissipative action from the gas is removed from the system, particularly for large $N$ (the number of planets) and $k$ (indicating how compact the chain is). We present a novel dynamical mechanism that can explain the origin of these instabilities and thus the dearth of resonant systems in the exoplanet sample. It relies on the emergence of secondary resonances between a fraction of the synodic frequency $2 \pi (1/P_1-1/P_2)$ and the libration frequencies in the mean motion resonance. These secondary resonances excite the amplitudes of libration of the mean motion resonances thus leading to an instability. We detail the emergence of these secondary resonances by carrying out an explicit perturbative scheme to second order in the planetary masses and isolating the harmonic terms that are associated with them. Focusing on the case of three planets in the 3:2 -- 3:2 mean motion resonance as an example, a simple but general analytical model of one of these resonances is obtained which describes the initial phase of the activation of one such secondary resonance. The dynamics of the excited system is also briefly described. This scheme shows how one can obtain analytical insight into the emergence of these resonances, and into the dynamics that they trigger. Finally, a generalisation of this dynamical mechanism is obtained for arbitrary $N$ and $k$. This leads to an explanation of previous numerical experiments on the stability of resonant chains, showing why the critical planetary mass allowed for stability decreases with increasing $N$ and $k$.
}


\maketitle

\section{Introduction}\label{sec:Introduction}
The formation of planetary systems is one of the key questions of planetary science, however it remains to this date observationally poorly constrained. One can nonetheless contemplate fully formed planetary systems, of which we have examples galore thanks to exoplanet-hunting missions such as HARPS and Nasa's Kepler surveys, and consider what physical and dynamical mechanisms can produce them. Despite our limited knowledge on the real nature of exoplanetary systems, there are some clear trends in the exoplanet sample, which thus impose constraints on formation scenarios.

A very common type of exoplanet, which was unknown in our Solar System, is what we now call Super-Earths or Mini-Neptunes. These are planets having a mass of about $1$ to $20 \mEarth$ (Earth's masses) and are found on relatively short orbital periods, of less than about 200 days. They are estimated to orbit a third to a half of all Sun-like stars (\citealt{2011arXiv1109.2497M,2012ApJS..201...15H,2013ApJ...766...81F,2013PNAS..11019273P}), and multi-planetary systems are not rare (of the order of a few hundred).
Given that several of them host H/He gaseous atmospheres that cannot be explained by production of volatiles after the formation of the planet (e.g.\ \citealt{2015ApJ...801...41R,2019PNAS..116.9723Z}), these planets are believed to form within the lifetime of their protoplanetary disc, and therefore interact dynamically with it. This type of interaction is called type-I migration: on the one hand the eccentricities of the orbits are damped by the disc, on the other hand (and on longer timescales) the orbit's size changes over time, usually shrinking, so that the planet is seen to migrate inward towards the star. At the inner edge of the disc, carved by the magnetic activity of the host star itself, another torque is activated which halts inward migration (a so-called planetary trap, e.g.\ \citealt{2006ApJ...642..478M}). In systems with multiple planets, when the inner planet has reached the inner edge of the disc, the second planet is still migrating inward, so the two planets are approaching each other. A preferred outcome of this convergent migration is the formation of compact chains of mean motion resonances, where the period ratio of neighbouring planets is close to a ratio of simple integers (\citealt{2007ApJ...654.1110T,2008A&A...482..677C,2008A&A...478..929M,2015A&A...578A..36O,2017MNRAS.470.1750I,2019arXiv190208772I,2018CeMDA.130...54P}).

Since these are transiting planets, their orbital period is known with extremely good precision; the period ratio distribution is therefore one of the best constrained distribution for exoplanets. Within the observed Super-Earth population, we do observe relatively long, coplanar resonant chains of planets, such as Trappist-1 (\citealt{2016Natur.533..221G,2017Natur.542..456G,2017NatAs...1E.129L}) and Kepler-223 \citep{2016Natur.533..509M}. However, an initially puzzling realisation is that the overall distribution of the period ratios is marked by systems that show little preference for near-integer period ratios, hosting planets with much wider orbital separations than those characterising resonant chains (e.g.\ \citealt{2015ARA&A..53..409W}). This appears at first in striking contradiction with the type-I migration scenario for Super-Earths and Mini-Neptunes, which naturally produces resonant chains.
This paradox is however only apparent, as pointed out by \cite{2017MNRAS.470.1750I}. Their analysis showed that many orbital properties of observed Kepler systems (including orbital spacing and multiplicity distribution) are very well reproduced if a large fraction of resonant systems eventually become unstable in the Gy evolution following the dissipation of the disc, with instability rates of $\sim95\%$. The remaining stable systems naturally represent the observed resonant systems, such as Trappist-1, Kepler-223, etc. 
In the original \cite{2017MNRAS.470.1750I} paper, only a limited fraction of resonant systems constructed via type-I migration went unstable within reasonable systems' lifetimes after the removal of the disc. However, in \cite{2019arXiv190208772I} these high rates of post-disc phase instabilities needed to explain the Kepler data are actually recovered, especially in the simulations where the formed systems are more massive and more compact. They therefore conclude that the final number of planets in the chain, the compactness of the system and the planets' masses are crucial parameters that differentiate between systems that remain stable after disc removal (for total integration times of 50 -- 300 My) and system that suffer dynamical instabilities (collisions or ejections).

The results of \cite{2017MNRAS.470.1750I,2019arXiv190208772I} motivate a careful dynamical analysis on the threshold of stability in mean motion resonant chains, and in this paper we focus on the dynamical mechanisms leading to the instability even in absence of external perturbations\footnote{
External perturbations have also been invoked to increase the fraction of unstable systems, such as the turbulence in the disc (which prevents capture in deep resonance, \cite{2017AJ....153..120B}) or the scattering of left-over planetesimals from the planetary region \cite{2016IAUFM..29A..30C}.
}. On this subject, an important numerical study was performed by \cite{2012Icar..221..624M}. There, the authors studied numerically the stability of resonant multi-planetary systems for high-integer first-order mean motion resonances. They built the desired resonant configuration by simulating the convergent type-I migration phase in a protoplanetary disc of gas; then they slowly depleted the disc. They observed that there is a critical number of planets $N_\crit$ above which the resonant systems go naturally unstable, with a crossing time comparable to that of non-resonant systems, and studied how this number changes with the planetary masses ($\mpl/M_*$, where $M_*$ is the stellar mass) and compactness of the chain (index $k$ of the $k$:$k-1$ resonance). More specifically, they demonstrated numerically that the critical number $N_\crit$ which guarantees stability decreases with increasing compactness of the chain (increasing $k$) and increasing planetary mass $\mpl$. The dynamical reason of the instability, however, was not discussed, nor the exact scaling law that links $N_\crit$, $\mpl$ and $k$.\\

The main goal of this paper is to investigate both analytically and numerically the dynamical mechanisms at the origin of the onset of instability in resonant chains, in order to explain the result of \cite{2012Icar..221..624M} and the large instability fraction of resonant chains observed in the \citep{2017MNRAS.470.1750I,2019arXiv190208772I} simulations. More precisely we focus on the stability of resonant configurations with small amplitude of libration around a resonant equilibrium point. These configurations are the resonant states less susceptible to instabilities \citep{2018CeMDA.130...54P}, and therefore represent a natural testing ground to assess the limits of stability of resonant chains. Because we intend to work analytically, and since the planetary Hamiltonian is not a continuous function of the number of planets $N$, it is convenient to rephrase the findings of \cite{2012Icar..221..624M} with the following equivalent statement: given the number $N$ of planets and the compactness of the system (the resonant index $k$), there is a limit mass $(\mpl/M_*)_{\crit}$ for stability, which decreases with increasing $N$ and $k$.
Thus, in this paper we address the question of why resonant chains at an initial state of low amplitude of libration become unstable if the planets are too massive, for different values of $N$ and $k$. This work is the continuation of our previous paper \cite{2018CeMDA.130...54P}, in which we considered the stability of two deeply resonant planets as a function of the planetary mass.

In order to fix ideas, as in the case of two resonant planets, we consider systems of planets of the same mass, $m_i\equiv\mpl$, $\forall i=1,\dots,N$. This is a useful simplification which allows one to grasp the main points having to work with only one parameter. We note also that individual Kepler systems seem to show a homogeneity in planetary masses (\citealt{2018AJ....155...48W,2017ApJ...849L..33M}), so this simplification does not constitute a major inconvenience. 
We will also consider coplanar orbits for simplicity. Indeed, if the chains that we intend to study are the result of capture in mean motion resonances during the disc phase, any significant mutual inclinations would have been damped out by the disc. Moreover, the few confirmed truly resonant systems (such as Trappist-1 or Kepler-223) show very small mutual inclinations. This suggests that resonant chains form in a relatively planar orbital configuration.

The remainder of this paper is organised as follows. In Section \ref{sec:NumericalMaps} we detail the setup for our numerical investigations, similar to the one used in \cite{2018CeMDA.130...54P}. 
In the $(N+1)$-body simulations with $N=3$ resonant planets, a new dynamical phenomenon is observed which was not present in the case $N=2$, that triggers the instability of the resonant chains. 
In Section \ref{sec:OriginOfInstability} we give a phenomenological description of this dynamical feature, and how it can explain the dependence of the limit mass for stability with the number $N$ of the planets and the index $k$ of the resonance, thus elucidating the numerical findings of \cite{2012Icar..221..624M,2019arXiv190208772I}.
In Section \ref{sec:HamiltonianModel}, we give a detailed analytical description of this dynamical phenomenon in the exemplifying case $N=3$, $k=3$, and in Section \ref{sec:StabilityN>3} we generalise the analytical scheme to arbitrary $N$ and $k$. Our conclusions are presented in Section \ref{sec:Conclusions}. Finally, in Appendix \ref{app:NBodyCapture} we summarise the main aspects of the numerical setup which allows to capture planets into mean motion resonance at different desired eccentricities.

\section{Numerical maps of stability of resonant planets}\label{sec:NumericalMaps}
In this section we describe the setup of our numerical investigation of resonant chains.
Motivated by the results of \cite{2012Icar..221..624M}, we investigate the stability of planets in chains of first order mean motion resonances in terms of the critical planetary mass $m_\crit$ allowed for stability. Specifically, we want to understand why $m_\crit$ decreases with the number of the planets $N$ and the index of the resonance $k$ along the chain.

The setup of our numerical experiments is the same as in our previous paper \cite{2018CeMDA.130...54P} on two resonant planets, and we review it here briefly for ease of reading but refer to the first paper for the details. The underlying idea is similar to that of \cite{2012Icar..221..624M} (see also for example \citealt{2017A&A...602A.101R,2015ApJ...810..119D,2018MNRAS.481.1538X}): planets are captured into mean motion resonance by running $(N+1)$-body simulations with added dissipative forces that mimic disc-planet interactions of the type-I migration regime (relevant for Super-Earths and Mini-Neptunes). However, unlike \cite{2012Icar..221..624M}, we do not attempt resonant capture experiments with different masses. The reason is that for relatively large planetary masses, close to the instability limit, the capture itself can become quite chaotic which may lead to large amplitudes of libration. Then, it becomes difficult to compare the long-term stability of these systems with large amplitude of libration with those with smaller masses that settle near the resonant equilibrium point. Instead, for a theoretical understanding of stability of a resonant chain as a function of planetary mass only, it is preferable to capture all the planets in resonance at low libration amplitudes at small masses and then, after gas removal, slowly increase the planetary masses until an instability is achieved. We stress that this growth in mass should not be interpreted as a physical process. It is just a numerical artifice to explore resonant dynamics as a function of the planetary mass and achieve an analytic understanding of the instability process. 

Our numerical experiments to probe the stability of resonant planets thus consist of two phases. First, the desired number of planets is captured deeply in the desired resonant chain at low planetary mass, and we consider planets of the same mass for simplicity. We implement a planetary trap at the inner edge in order to ensure convergent migration which is needed for the planets to capture (e.g.\ \citealt{2006ApJ...642..478M}). Then the disc is slowly dissipated away, leaving the system in a state of small libration around a resonant equilibrium point (\citealt{2018CeMDA.130...54P}, see also Appendix \ref{app:NBodyCapture}) and only the pure conservative dynamics remains. In the second phase, the value of $\mpl$ is slowly increased at each time-step, maintaining the small amplitude of oscillation around the resonance, until un instability is reached (usually, the instability results in planetary collisions); again, this mass increase is purely fictitious and serves the only purpose to study for each value of $\mpl$ the stability of resonant configurations with the same level of excitation of the resonant degrees of freedom.

Built around the same numerical setup, we review below a few important aspects discussed in \cite{2018CeMDA.130...54P} on the case of two resonant planets, as they will turn out to be relevant for the case of three and more planets as well. We then discuss the application of the numerical simulations on the stability of three resonant planets in Subsection \ref{subsec:StabilityMapsForN=3}.

\subsection{Review on the case of two resonant planets}\label{sec:ReviewCaseN=2}
There are some points that should be revisited from our previous paper \cite{2018CeMDA.130...54P} on the stability of mean motion resonances in two-planets systems. We summarise them below, and refer to \cite{2018CeMDA.130...54P} for a full discussion.

The first point is that, when two planets are in a first order mean motion resonance $k$:$k-1$, there are two frequencies associated with the libration of the system around the resonant equilibrium point, which we call resonant frequencies and indicate with $\omega_{\res,i}$, $i=1,2$. These frequencies dictate the evolution of the system over long timescales, and are essentially associated to the evolution of the two resonant angles $\psi_i=k\LAMBDA_2-(k-1)\LAMBDA_1-\VARPI_i$, $i=1,2$, which indeed have slow variation under the assumption that the system is in the $k$:$k-1$ mean motion resonance. Instead, on shorter timescales, the evolution is dominated by the non-resonant combination $\delta\LAMBDA_{1,2}=\LAMBDA_1-\LAMBDA_2$ of the mean longitudes $\LAMBDA_i$, which is a fast-evolving angle; this angle is called synodic angle, and its frequency is called the synodic frequency $\omega_\syn$. Since $\dot\LAMBDA_i=n_i$, $n_1/n_2\simeq k/(k-1)$ by the resonance condition, and $n_i$ is linked to the semi-major axis by Kepler's third law $n_i=\sqrt{\GravC M_*/a^3}$, we have that the synodic frequency $\omega_\syn=\dot\delta\LAMBDA_{1,2}=n_1/k=\sqrt{\GravC M_*/a^3}/k$. Thus, the synodic frequency is independent of the planetary mass and only depends on the nominal separation of the planets to the star. Instead, the resonant frequencies grow with the planetary mass $\mpl$: for example, in a simple pendulum approximation of the mean motion resonant dynamics, the resonant frequencies are expected to grow as $\sqrt{\mpl}$ (see Subsect.\ \ref{sec:StabilityN=3.subsec:PurelyResonantDynamics}). Thus, for small enough planetary masses, the synodic frequency is much higher than the resonant frequencies, so that the two contributions happen on totally different timescales and are perfectly decoupled: then, the fast synodic degree of freedom can be averaged out and only the purely resonant evolution (the combination of both resonant frequencies) matters over a long time. However, at large enough planetary masses, the resonant frequencies might become comparable with the synodic frequency. When the ratio between the synodic frequency and resonant frequencies is close to an integer ratio, a secondary resonance is encountered: this means that the purely resonant degrees of freedom can now exchange energy with the synodic degree of freedom. Therefore, these secondary resonances between the synodic and resonant frequencies could in principle destabilise a resonant pair of planets.
In \cite{2018CeMDA.130...54P} we found that, in the case of two planets in first order mean motion resonance, these secondary resonances are active at such high planetary masses that the system actually becomes unstable at smaller values of $\mpl$ because of close encounters between the planets. Therefore, we concluded that these secondary resonances are not responsible for instability in a system of two resonant planets.

The second point thus concerns the instability caused by close encounters in the case of resonant planets. This type of planetary instability is a well understood phenomenon, so that we can discriminate the orbital configurations that are stable with respect to close encounters (also called Hill-stable) and those that are not (e.g.\ \citealt{1993Icar..106..247G,1982CeMec..26..311M,2018A&A...617A..93P}). Following for example the approximation for initially circular and coplanar planets made in \cite{1993Icar..106..247G}, one has that (for a general, non-resonant system) if the orbital distance $d=a_2-a_1$ satisfies 
\begin{equation}\label{eq:GladmanStability}
d\geq d_\crit = 2\sqrt{3} r_{\mathrm{H},1,2} \simeq 3.46 r_{\mathrm{H},1,2},
\end{equation}
then the system is Hill-stable (see also \citealt{2017Icar..293...52O}).\footnote{One should note that the resonant condition is a condition on the angles which prevents the closest approach along the two planets' orbits to happen. This means that resonant systems are expected to be more protected than non-resonant ones with respect to close encounters. However, in \cite{2018CeMDA.130...54P}, we used the actual minimal approach distance $d$ rather than the orbital distance $d_\orb$ in \eqref{eq:GladmanStability} to measure the limits of stability against close encounters, and to compare the result with the general case of non-resonant systems.} Here $r_{\mathrm{H},1,2}$ is the mutual Hill radius of the two planets, defined as 
\begin{equation}
r_{\mathrm{H},1,2}=\frac{a_1+a_2}{2}\left(\frac{m_1+m_2}{3M_*}\right)^{1/3}.
\end{equation} 
We found in \cite{2018CeMDA.130...54P} that resonant planets are more stable with respect to close encounters than non-resonant ones, in the sense that, to suffer mutual scattering, the planets need to approach to each other significantly closer than $d_\crit$; however we did find that close encounters destabilise the systems at lower planetary masses than the aforementioned secondary resonances would. Moreover, we found that the larger the amplitude of oscillation associated with the resonant motion around the resonant equilibrium point, the closer to $d_\crit$ is the minimal physical distance for instability (the same remains true with respect to the more general criterion found e.g.\ in \citealt{1982CeMec..26..311M,2018A&A...617A..93P}).

It will be important to keep these two points in mind even in the case of three and more resonant planets, as they will be relevant for understanding their stability. We investigate the case $N\geq3$ below.

\subsection{Numerical stability maps for three resonant planets}\label{subsec:StabilityMapsForN=3}
The first step is to perform numerical experiments as explained at the beginning of Section \ref{sec:NumericalMaps}.
We refer to \cite{2018CeMDA.130...54P} for a more in-depth discussion on the setup for capture into mean motion resonance (including an analytical understanding of this process which is consistent with the Hamiltonian formalism and adiabatic theory), the subsequent phase of fictitious mass growth and how it can be understood analytically. 
There is only one small difference to be pointed out in the capture phase of our simulations. In \cite{2018CeMDA.130...54P} we could obtain any desired value of $e_2$ (equivalently, $e_1$) by changing the value of the eccentricity damping timescale $\tau_e$. By setting a large value for $\tau_e$, large planetary eccentricities could be obtained (cfr.\ Equation \eqref{eq:e2EquilibriumWithTrap}). Here, because the planets capture in resonance in sequence (first planet 1 and 2, then planet 3) if $\tau_e$ is large, $e_1$ and $e_2$ can grow significantly before planet 3 enters in resonance. This can force large secular eccentricity oscillations of planet 3, which may preclude its resonant capture (see e.g.\ \citealt{2015MNRAS.451.2589B} on criteria for resonant capture). We give the details of the setup for capture in Appendix \ref{app:NBodyCapture} and describe a numerical recipe to overcome this difficulty, which poses no problem at all in the context of the second phase where we actually investigate the stability of the chains as a function of planetary mass.

\subsubsection{Numerical stability maps for $N=3$ and $k=3$}\label{subsec:StabilityN=3.subsec:NumericalStabilityMapsk=3}
\begin{figure*}[!t]
\centering
\begin{subfigure}[b]{0.49 \textwidth}
\centering
\includegraphics[scale=0.55]{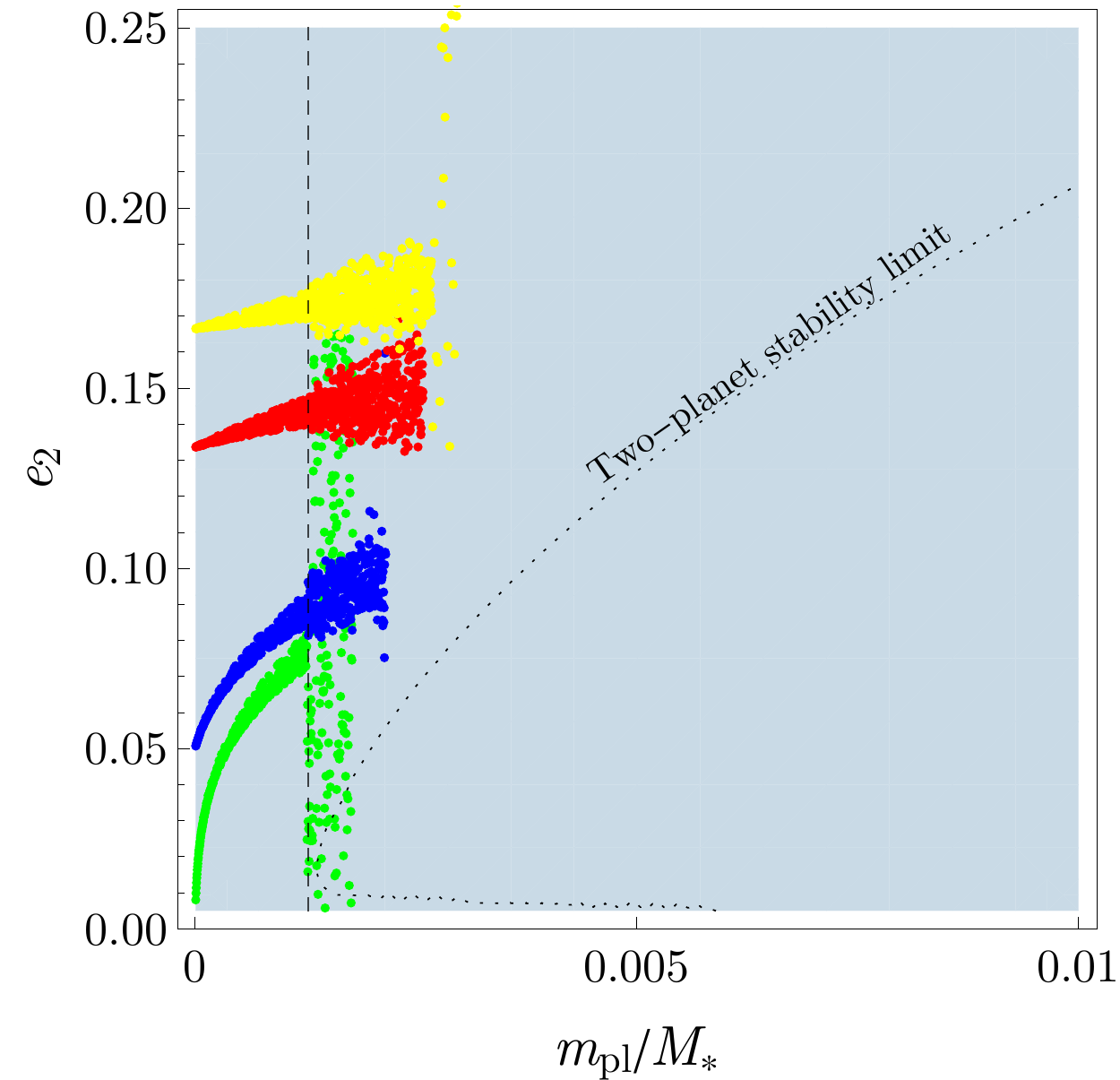}
\caption{}
\label{fig:3-2_3-2Stability.subfig:3-2_3-2GlobalStabilityMap} 
\end{subfigure}
\hfill
\begin{subfigure}[b]{0.49 \textwidth}
\centering
\includegraphics[scale=0.5]{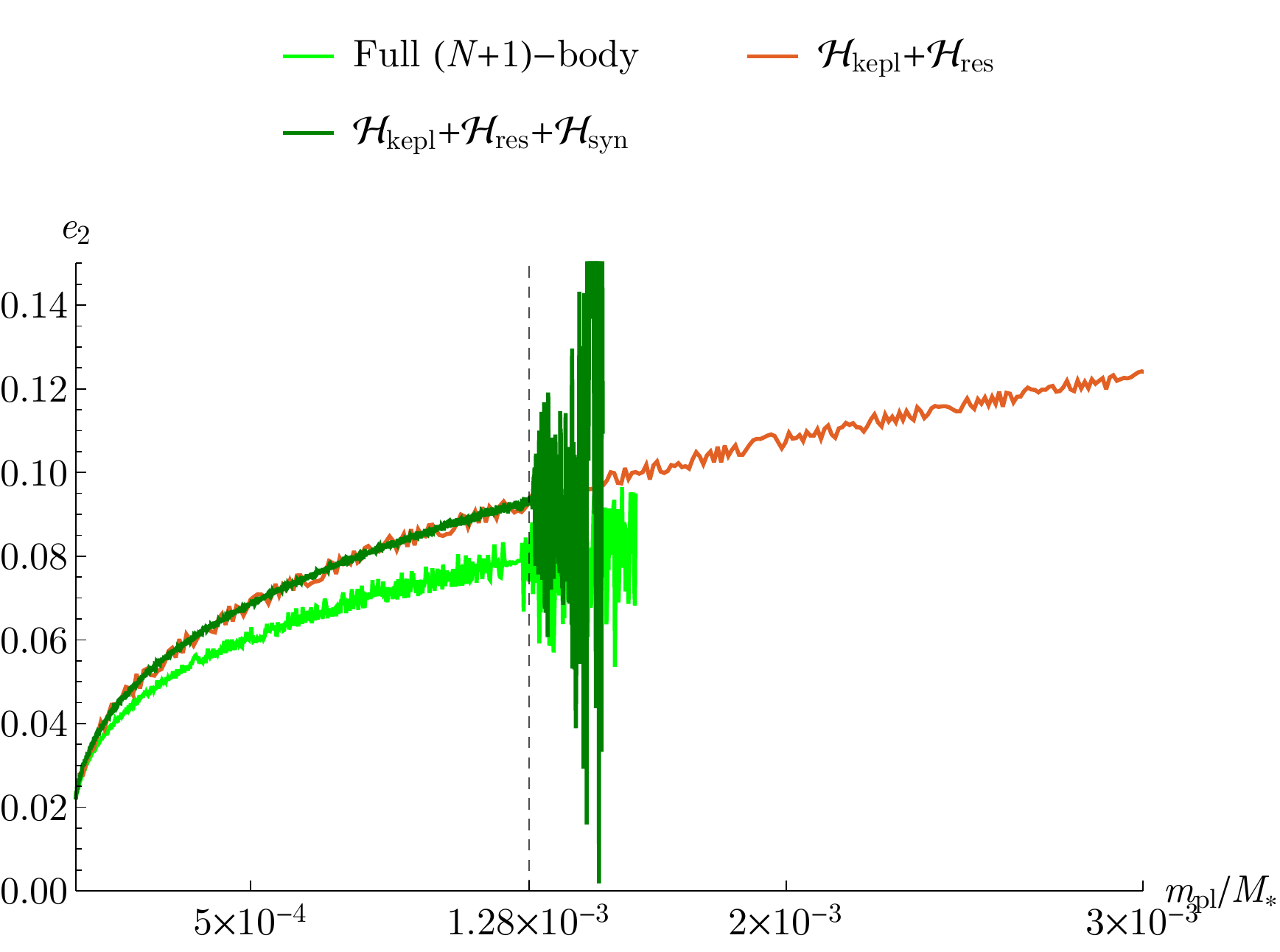}
\vspace{.5cm}
\caption{}
\label{fig:3-2_3-2Stability.subfig:3-2_3-2_InstabOnsetOne2Plot} 
\end{subfigure}

\caption%
{Numerical investigation of the stability of three planets deep in the 3:2 -- 3:2 mean motion resonance chain, as a function of the planetary mass $\mpl$, equal for all planets. In panel (a), four numerical simulations (the coloured markers) are performed starting from low-mass planets ($\mpl=10^{-5}M_*$) and slowly increasing the planetary mass until an instability occurs (a collision in all cases). The dotted curve indicates the limit of stability for a system of two planets deep in the 3:2 mean motion resonance \citep{2018CeMDA.130...54P}: this shows that three resonant planets go unstable at lower masses than two resonant planets, in accord with \cite{2012Icar..221..624M}. As explained in the main text, the anticipated instability is unlikely caused by close encounters, which were causing the instability in the the two-planet case. Indeed, in the case of three resonant planets a new dynamical phenomenon appears which is not observed in simulations of two planets: the system experiences an excitation in amplitude of oscillation before going unstable. This excitation, starting at $\mpl/M_*\simeq\protect\input{MassAtPhenom3-2_3-2.dat}$ (vertical dashed line) is more clearly visible in panel (b), where the result of one such numerical simulation is shown in light green. 
In panel (b) this simulation is also compared with the integration of two simplified models (dark green and orange lines), with the same initial conditions as the numerical simulation of the complete equations of motion. In both simplified models, only terms up to first order in the eccentricities are considered (cfr.\ Subsect.\ \ref{subsec:RescaledHamiltonianAndNewSetOfCanonicalVariables}). The orange line represents the evolution of the averaged equations of motion where all non-resonant terms have been dropped: the evolution is initially qualitatively similar to the complete simulation, however no excitation is observed (cfr.\ Subsect.\ \ref{sec:StabilityN=3.subsec:PurelyResonantDynamics}). 
The dark green line represents the evolution of a model with both resonant and synodic interaction terms for each planet pair: although only terms up to order one in the eccentricities have been considered, we see that the excitation at $\mpl/M_*\simeq\protect\input{MassAtPhenom3-2_3-2.dat}$ is well reproduced in this simplified system (cfr.\ Subsect.\ \ref{sec:StabilityN=3.subsec:SynodicContribution}).}
\label{fig:3-2_3-2Stability} 
\end{figure*}

We show in Figure \ref{fig:3-2_3-2Stability.subfig:3-2_3-2GlobalStabilityMap} the result of four simulations of the second phase of our numerical experiments for the case $N=3$ and the 3:2 -- 3:2 resonant chain, starting from different initial eccentricities. On the horizontal axis we report the (increasing) planetary mass, while on the vertical axis we show the evolution of the eccentricity. The simulations are stopped when an instability occurs (a collision in all cases). This plot is to be compared with the similar Figures 9 and 10 in \cite{2018CeMDA.130...54P} for the case of $N=2$ and the same resonance index $k=3$, and uses the same scale on both axes to allow for an easier comparison. The approximate location of the observed instability for two planets in the same resonance is represented in Figure \ref{fig:3-2_3-2Stability.subfig:3-2_3-2GlobalStabilityMap} by a dotted line.
Comparing the cases $N=2$ and $N=3$, there are two important observations to make.
The first is that the instabilities occur at lower masses in the case $N=3$ than in the case $N=2$. This is in agreement with the results of \cite{2012Icar..221..624M}. 
This anticipated instability, in terms of planetary mass, is unlikely to be due to too-close encounters between pairs of planets as it was the case $N=2$. This is because a resonant chain repeats the same orbital geometry between adjacent planets of a two-planet resonance of the same order. Thus, if the critical mass $m_\crit$ corresponding to the instability in the case $N=3$ is smaller, the minimal approach distance between each pair of neighbouring planets is necessarily larger in terms of mutual Hill radii than that causing an instability for $N=2$. There is no apparent reason for which the threshold distance for destabilising two-body encounters should significantly change with the number $N$ of planets in the system. So, the instability is likely to have a different cause.  
Upon close examination of the $(N+1)$-body integrations shown in Figure \ref{fig:3-2_3-2Stability.subfig:3-2_3-2GlobalStabilityMap}, one notices that an interesting phenomenon is evident. For $\mpl/M_*<1.28\times 10^{-3}$, the amplitude of oscillation of the eccentricity grows linearly with the planets' mass. This is due to the increasing amplitude of the fast-frequency term associated to the synodic terms (the same effect was present in the case of two planets); instead, the amplitude of libration associated to the purely resonant dynamics is conserved adiabatically. 
Then, at $\mpl/M_*\simeq 1.28\times 10^{-3}$ (the dashed vertical line in the figure) there is a sudden excitation of the amplitude of eccentricity oscillations. Upon close inspection of the numerical output with high temporal resolution, we realise that this excitation is now due to an actual increase of the amplitude of libration inside the resonance, as will be clear below.
After the excitation at $\mpl/M_*\simeq 1.28\times 10^{-3}$, the systems temporarily remain in resonance, albeit with an increased libration amplitude of the resonant angles; soon after, while the planetary mass is still increasing, the systems finally become unstable as the planets experience close encounters, eventually leading to collisions. This is observed in all simulations.

We have seen in \cite{2018CeMDA.130...54P} (see also Sect.\ \ref{sec:ReviewCaseN=2}) that, with increasing amplitude of libration, the planets need to farther away from each other (in terms of mutual Hill radius) to be stable. On the other hand, the larger is the libration amplitude in the resonance, the closer the planets approach each other during their evolution. Thus, in order to remain stable, the planetary masses have to be smaller, so that the mutual Hill radius $r_{\mathrm{H},1,2}$ shrinks and their minimal physical distance in terms of $r_{\mathrm{H},1,2}$ remains large. In other words, we concluded that more excited resonant states become unstable at smaller planetary masses.
So, our interpretation for the anticipated instability in the $N=3$ case is the following: first some dynamical process excites the libration amplitude; then the planets become encounter-unstable because the threshold distance for instability exceeds the actual minimal distance of approach between planet pairs. 
Thus, below we will look for the dynamical mechanism increasing the libration amplitude.   
It should be noticed that if such mechanism exists, it would also preclude capture in the resonance at small libration amplitude for the corresponding planetary mass, which is what was observed by \cite{2012Icar..221..624M}.

\subsubsection{Numerical and analytical investigation of the phenomenon}\label{subsec:InvestigationOfThePhenomenon}
In the previous subsection we have underlined the importance of the observed increase in the amplitude of libration around the equilibrium point in the $(N+1)$-body simulations, and its relevance for triggering the instability of resonant chains. In the following we aim at better understanding the dynamical origin of this growth of libration amplitude.

Our approach is to find a simplified $N$-planets Hamiltonian model which captures the main features of the dynamics that are observed in the complete $(N+1)$-body integrations. This is because the complete model contains a virtually infinite number of harmonics, making it extremely hard to proceed analytically or to obtain any insights from the observed evolution. 
If we are able to observe the same phenomenon in a simplified problem it will be easier to isolate its origin. Thus, in the following we start from a Hamiltonian planetary model that has only a minimal number of terms (harmonics) and we progressively add more terms until we observe in the integration of the considered Hamiltonian the same phenomenon that we have seen in the full numerical integration. 
The Hamiltonian models are integrated numerically, while slowly increasing the mass of the planets at each integration time step in accordance with the $(N+1)$-body simulations in Subsection \ref{subsec:StabilityN=3.subsec:NumericalStabilityMapsk=3}. Only when the numerical integrations show very good agreement with the full $(N+1)$-body integrations, will we consider the corresponding Hamiltonian as a good approximation to the full one and work directly with the former. Before we get into the technicalities of our investigation, we plan out our methodology below.\\

The first reasonable choice for the numerical integrations is to consider the averaged equations of motion, expanded to some order in the eccentricity. This corresponds to dropping all non-resonant harmonics from the planetary Hamiltonian (cfr.\ Subsect.\ \ref{subsec:PlanetaryHamiltonian}) and only keeping resonant harmonics up to some order in $e$. This results in a system governed by a Hamiltonian $\bar\Ha:=\Hakepl+\Hares$; this approach is presented in Subsection \ref{sec:StabilityN=3.subsec:PurelyResonantDynamics}. By doing so, one realises that these terms cannot alone be responsible for the increase in amplitude of libration observed in the $(N+1)$-body integrations. This fact is anticipated in Figure \ref{fig:3-2_3-2Stability.subfig:3-2_3-2_InstabOnsetOne2Plot}, where we plot with a dark orange line the evolution of the system governed by $\bar\Ha$ over one of the full $(N+1)$-body integration with the same initial conditions; we see that at first the two simulations are qualitatively equivalent (the slight differences emerge solely from the expansion up to first order in the eccentricities made in the truncated model $\bar\Ha$), but the averaged model does not reproduce the excitation observed in the $(N+1)$-body simulation at the location of the dashed vertical line. Actually, we will show that such excitation in the purely averaged model is not possible at any value of the planetary mass $\mpl$. This is the first main result of this section: \emph{the purely resonant system $\bar\Ha=\Hakepl+\Hares$ with initial conditions at vanishing amplitude around a resonant equilibrium point is (Lyapunov) stable for all planetary masses}. 

The next step is therefore to include additional non-resonant terms, which were naturally present in the full Hamiltonian that governs the evolution of the $(N+1)$-body integrations. Maintaining for simplicity the expansion to first order in the eccentricity (which should be valid at least when all eccentricities are small enough), we then add synodic terms. In the case of three planets, these include the harmonics $\LAMBDA_1-\LAMBDA_2$ and $\LAMBDA_2-\LAMBDA_3$, which we add in an additional interaction Hamiltonian $\Hasyn$. As we show in Subsection \ref{sec:StabilityN=3.subsec:SynodicContribution}, the introduction of these terms is responsible for the same phenomenon observed in Figure \ref{fig:3-2_3-2Stability.subfig:3-2_3-2GlobalStabilityMap}. This fact is anticipated in Figure \ref{fig:3-2_3-2Stability.subfig:3-2_3-2_InstabOnsetOne2Plot}, where we plot with a darker colour the evolution of the system governed by $\Ha^*:=\Hakepl+\Hares+\Hasyn$ over one of the full $(N+1)$-body integration with the same initial conditions, and we see that there is good qualitative agreement between the two evolutions. We also investigate the possibility of adding only one of the two synodic terms, but show that both are needed to reproduce the phenomenon at similar planetary masses, which is a result that we will also explain analytically (cfr.\ Subsect.\ \ref{subsubsec:TheOmpl^2contribution}). In the light of this, we will use the evolution yielded by the simplified model $\Ha^*=\Hakepl+\Hares+\Hasyn$ as a guide to understand the relevant dynamics contained in the full $(N+1)$-body integrations. At the same time, working with a controlled number of interaction terms allows us to proceed analytically (see Subsect.\ \ref{subsubsec:EliminatingOmplSynodicTerms}) and to understand what is the dynamical mechanism that gives rise to the increase in amplitude of libration around the resonant equilibrium point. Carrying out the calculation explicitly in the specific case of $N=3$ planets and for the 3:2 -- 3:2 chain, we show in Subsect.\ \ref{subsubsec:TheOmpl^2contribution} that this is due to a set of secondary resonances between a fraction of the synodic frequency (which remains relatively constant with increasing $\mpl$) and specific combinations of the libration frequencies around the equilibrium point (which increase with $\mpl$, as we will show). Considering relevant canonical action-angle variables centred at the equilibrium, such secondary resonances have the effect of exciting the action to values farther and farther away from the origin. This is the second main result of this section: \emph{the synodic contribution introduces terms of order $\mathcal O(\mpl^2)$ which include secondary resonances between a fraction the synodic frequency and the resonant libration frequencies, which are responsible for the excitation of the system and eventually for its instability}. In Subsect.\ \ref{subsubsec:ModelForSecondaryResonancedeltalambda12+2phi2}
we build a model for the secondary resonance that is encountered in the specific case $N=3$ and $k=3$, but the method can be easily generalised to the other secondary resonances that can in principle be encountered.
Finally, we proceed to generalise this result to more populated and/or more compact resonant chains in Section \ref{sec:StabilityN>3}.

\section{The origin of instability in resonant chains}\label{sec:OriginOfInstability}
In the next section, we will begin a careful analysis of the dynamics for three planets in a chain of mean motion resonances based on the insights elucidated above, which were lead by numerical integrations such as those of Figure \ref{fig:3-2_3-2Stability}. In particular, we aim at gaining a deep understanding of the process which causes the sudden excitation in the systems shown in Figure \ref{fig:3-2_3-2Stability}. As anticipated at the end of the last subsection, this process involves secondary resonances between some fraction of the synodic frequency $\omega_\syn=\frac{\D {}}{\D t}(\LAMBDA_1-\LAMBDA_2)$ and the resonant frequencies $\omega_{\res,l}$ associated with the libration of the system around the resonant equilibrium point.
Before we delve into the dynamical details of these secondary resonances, let us delineate in a more general and practical sense why they are relevant for the problem of the stability of resonant chains of $N$ planets.

The idea is that, normally, the synodic evolution (with characteristic frequency $\omega_\syn$) and the purely resonant evolution (with characteristic frequency $\omega_{\res,l}\ll\omega_\syn$) happen on such different timescales that there can be no interaction between them (as we already recalled in Subsection \ref{sec:ReviewCaseN=2}).  However, a secondary resonance between them effectively allows energy to be transferred between the synodic and resonant degrees of freedom, and can ultimately cause an excitation of the latter which in turn makes the chain unstable to close encounters between the planets.

Now, in the case of two planets, the resonant frequencies were too small compared to $\omega_\syn$ and grew too slowly with $\mpl$, so that secondary resonances were active at such high planetary masses that the system was already unstable to close encounters (cfr.\ Subsect.\ \ref{sec:ReviewCaseN=2}). Note that for the same planetary mass $\mpl$ and for the same $k$, the libration frequencies for two and three resonant planets are roughly similar for similar eccentricities. However, the key point is that in the case $N\geq3$ there is a \emph{fraction} of the synodic frequency which appears in the Hamiltonian (in terms at second order in the planetary masses). In the case of three planets, this fraction is $\omega_\syn/k$ where $k$ as usual is the index of the resonance\footnote{
A simple explanation for why this fraction of the synodic frequency naturally pops up in the equations of motion (that is in the Hamiltonian) at second order in $\mpl$ is the following. The Hamiltonian of three planets contains both the $\delta\LAMBDA_{1,2}=\LAMBDA_1-\LAMBDA_2$ and $\delta\LAMBDA_{2,3}=\LAMBDA_2-\LAMBDA_3$ harmonics. If both planet pairs are in the $k$:$k-1$ mean motion resonance, one can write $\delta\LAMBDA_{2,3}$ as $(k-1)\delta\LAMBDA_{1,2}/k$ plus some correction harmonic terms that only depend on the resonant angles (cfr.\ \eqref{eq:ThreePlanetsSynodicHarmonicsInx} with constant index $k$ along the chain); this can be easily understood by noting that $\dot{\delta\LAMBDA_{2,3}}$ should be comparable to $(k-1)\dot{\delta\LAMBDA_{1,2}}/k$ in a $k$:$k-1$ chain. Then, the two angles $\delta\LAMBDA_{1,2}$ and $(k-1)\delta\LAMBDA_{1,2}/k$ get combined at second order in $\mpl$ which yields a harmonic containing $\delta\LAMBDA_{1,2}/k$ plus purely resonant harmonics (cfr.\ \eqref{eq:NeededSecondaryResonanceHarmonicsInqk1=k2=k}).}. Thus, in the case of three planets, in order to reach a secondary resonance involving synodic and resonant degrees of freedom, the resonant frequencies do not have to be as large, that is, the planetary masses do not have to be as large as in the two-planets case. This is why for $N\geq3$ these secondary resonances can be relevant while they were not in the case $N=2$.

To extend this principle to the general case $N\geq3$, one can easily calculate that the smallest fraction of the synodic frequency that appears in the case of $N$ planets in a $k$:$k-1$ resonant chain is $\frac{1}{k} \left(\frac{k-1}{k}\right)^{N-3}\omega_\syn$ (cfr.\ Eq.\ \eqref{eq:SmallestSynodicFrequencyAtOrder2InmplN>=3}). Again, this frequency can resonate with the resonant frequencies $\omega_\res$, and, just as before, $\frac{1}{k} \left(\frac{k-1}{k}\right)^{N-3}\omega_\syn\simeq \frac{1}{k^2} \left(\frac{k-1}{k}\right)^{N-3}n_1$ is independent of the planetary masses, and for fixed orbital separation (fixed $n_1$) decreases with $k$ and $N$. Finally the resonant frequencies $\omega_\res$ still increase with $\mpl$ (and with $k$), more or less independently on the number of planets. Thus there will be a critical mass after which a regime of secondary resonances is encountered, which can excite the system and cause its subsequent instability by close encounters. Since the factor $\frac{1}{k} \left(\frac{k-1}{k}\right)^{N-3}$ multiplying $\omega_\syn$ decreases with increasing $N$ and with $k$, the conclusion is that {\it the regime of secondary resonances between synodic and resonant degrees of freedom is encountered at lower masses for increasing $k$ and/or increasing $N$, and therefore the critical mass $(\mpl/M_*)_\crit$ allowed for stability decreases with $N$ and with $k$}. This mechanism gives a dynamical explanation to the numerical findings of \cite{2012Icar..221..624M,2019arXiv190208772I}. In the rest of this paper, we give a detailed analytical description of the dynamical emergence of these secondary resonances.

\section{Hamiltonian model}\label{sec:HamiltonianModel}
In this section we describe the analytical tools used to investigate the emergence of secondary resonances between synodic and resonant degrees of freedom. We begin introducing the general planetary Hamiltonian and the customary notation in Subsection \ref{subsec:PlanetaryHamiltonian}, and we then consider the relevant harmonic terms in the Hamiltonian that interest us in Subsection \ref{subsec:RescaledHamiltonianAndNewSetOfCanonicalVariables}. Then, in Subsections \ref{sec:StabilityN=3.subsec:PurelyResonantDynamics} and \ref{sec:StabilityN=3.subsec:SynodicContribution} respectively we consider the averaged model $\bar\Ha$ and the model $\Ha^*$ which includes synodic terms. There, we give an analytical descriptions of the main dynamical features of the simulations shown in Figure \ref{fig:3-2_3-2Stability}.

\subsection{Planetary Hamiltonian}\label{subsec:PlanetaryHamiltonian}
We start with the Hamiltonian $\Ha$ of $N$ planets of masses $m_i$, $i=1,\dots,N$ orbiting a star of mass $M_*$. We let $\bfu_i$ be the inertial barycentric cartesian coordinate of each planet, and $\tilde\bfu_i=m_i\dot\bfu_i$ the conjugated momentum.  We write $\Ha$ in canonical heliocentric variables $(\bfp_i, \bfr_i)$, $i=1,\dots,N$, defined from the inertial barycentric canonical variables $(\bfu,\tilde\bfu_i)$ as 
\begin{alignat}{2}\label{eq:CanonicalAstrocentricCoordinatesPP}
\bfp_0	&=\sum_{i=0}^N\tilde\bfu_i,	&&\quad \bfr_0=\bfu_0,\nonumber\\
\bfp_i	&=\tilde\bfu_i,				&&\quad \bfr_i=\bfu_i-\bfu_0,~i=1,\dots,N.
\end{alignat}
(e.g.\ \citealt{1892mnm..book.....P,1990mmmc.conf...63L}). Doing so, the Hamiltonian can be split as
\begin{equation}\label{eq:HPPInAstrocentricCoordinatesWithoutBarycentre}
\begin{split}
\Ha(\bfp,\bfr)	&=\Hakepl+\Hapert,\\
\Hakepl		&=\sum_{i=1}^N\left(\frac{\|\bfp_i\|^2}{2\mu_i} - \frac{\GravC(M_*+m_i)\mu_i}{\|\bfr_i\|} \right)=\sum_{i=1}^N \Ha_{\kepl,i},\\
\Hapert		&=\sum_{1\leq i<j\leq N}\left(\frac{\bfp_i\cdotp\bfp_j}{M_*} - \frac{\GravC m_i m_j}{\|\bfr_i-\bfr_j\|}\right).
\end{split}
\end{equation} 
In other words, the Hamiltonian appears as a sum of two terms. 
One term is the sum of the Keplerian unperturbed Hamiltonians for each planet $\Ha_{\kepl,i}$, describing the planet-star interactions. The other is the perturbing Hamiltonian $\Hapert$ which describes all planet-planet interactions; $\Hapert$ itself is split into direct terms, $-\sum_{1\leq i<j\leq N}\GravC m_i m_j/\|\bfr_i-\bfr_j\|$, and indirect terms, $\sum_{1\leq i<j\leq N}\bfp_i\cdotp\bfp_j/M_*$, which come from having considered canonical heliocentric rather than barycentric variables. $\Hakepl=\sum_{i=1}^N \Ha_{\kepl,i}$ is integrable, while $\Hapert$ is of order $\mpl/M_*$ with respect to $\Hakepl$ (where $\mpl$ is the typical mass of the planets) so it can be seen as a small perturbation to the integrable Keplerian Hamiltonian. 
For each planet, the canonical modified Delaunay variables can be introduced, which are action-angle variables for the reference Keplerian problems $\Ha_{\kepl,i}$. We will consider only coplanar motion for the planets, so we only have two pairs of action-angle variables $(\LAMBDONA_i,\LAMBDA_i)$ and $(\GAMMONA_i,\GAMMA_i)$. Their definition in terms of the orbital elements is (e.g.\ \citealt{2002mcma.book.....M})
\begin{alignat}{2}\label{eq:ModifiedDelaunayVariablesPlanar}
\LAMBDONA_i	&=\mu_i\sqrt{\GravC(M_*+m_i)a_i},		&&\quad \LAMBDA_i=\M_i+\VARPI_i,\nonumber\\
\GAMMONA_i	&=\LAMBDONA_i(1-\sqrt{1-e_i^2})\sim\LAMBDONA_i e_i^2/2, 		&&\quad \GAMMA_i=-\VARPI_i.
\end{alignat}
As usual, for each planet $a_i$ is the semi-major axis, $e_i$ is the eccentricity, $\LAMBDA_i$ is the mean longitude, $\M_i$ is the mean anomaly $\VARPI_i$ is the longitude of the pericentre and $\mu_i=m_iM_*/(M_*+m_i)\simeq m_i$ is the reduced mass; the index $i=1,\dots,N$ refers to the $i$-th planet, with planets ordered with increasing semi-major axis. We note that, as in \cite{2018CeMDA.130...54P}, the orbital elements are defined starting from heliocentric positions and barycentric velocities \eqref{eq:CanonicalAstrocentricCoordinatesPP} (they are the so-called \emph{formal} osculating elements, \citealt{2002mcma.book.....M}).

In the modified Delaunay variables \eqref{eq:ModifiedDelaunayVariablesPlanar} the Keplerian part rewrites
\begin{equation}\label{eq:KeplerianPartInModifDelaunayVariablesNPlanets}
\Hakepl=-\GravC^2\sum_{i=1}^N \frac{\mu_i^3 (M_*+m_i)^2}{2\LAMBDONA_i^2},
\end{equation}
while no simple expression exists for $\Hapert$, which is usually expanded in Fourier series of the angles. In this expansion, there are only combinations of $\LAMBDA_i$ and $\GAMMA_i$ which satisfy the d'Alembert characteristics, and only harmonic terms combining angles from two planets. 
We won't go into the details of how this expansion is performed in general, which can be found in many works (e.g.\ \citealt{1995CeMDA..62..193L,1999ssd..book.....M}), and we will only concentrate on the specific terms that interest us below. 

\subsection{Rescaled Hamiltonian and new set of canonical variables}\label{subsec:RescaledHamiltonianAndNewSetOfCanonicalVariables}
In order to make the calculations and algebraic expressions less cumbersome, we start by performing the following simplifications. These are clearly general and are carried out here for any number $N$ of planets, but we will give specific examples to the case of 3 planets to fix ideas.\\

Firstly, since the instabilities for $N\geq3$ planets occur at much lower values of $\mpl/M_*$ than for 2 planets, we approximate the reduced mass $\mu=\frac{\mpl M_*}{M_*+\mpl}\sim\mpl$ and $M_*+\mpl\sim M_*$. Then, we recall that all the planets have the same mass $\mpl$, and we intend later on to make use of the tools of perturbation theory to study the dynamics of the resonant chains. It is therefore convenient to write the Hamiltonian in the form of a sum of an integrable part which does not depend on the small parameter $\mpl$, plus a small perturbation proportional to $\mpl$. 
The natural choice is to rescale all the actions $(\LAMBDONA,\GAMMONA)$ of the modified Delaunay variables by the planetary mass $\mpl$, which yields 
\begin{equation}
\begin{split}
\LAMBDONA	&= \sqrt{\GravC M_* a},\\
\GAMMONA	&= \LAMBDONA (1-\sqrt{1-e^2}),
\end{split}
\end{equation}
where for simplicity we have maintained the same notation as for the non-rescaled variables. In order to maintain the canonicity of the Hamiltonian, $\Ha$ itself must be rescaled by $\mpl$.
With this choice the reduced $N$-planets Hamiltonian takes the form (again, as for the canonical variables we do not change the notation for the rescaled Hamiltonian)
\begin{equation}\label{eq:RescaledHamiltonianNplanets}
\begin{split}
\Ha = \Hakepl + \Hapert,\\
\Hakepl = -\sum_{i=1}^N \frac{\GravC^2 M_*^2}{2 \LAMBDONA_i^2},
\end{split}
\end{equation}
where $\Hakepl$ is independent of $\mpl$, and the (rescaled) perturbation is of order $\mathcal O(\mpl)$:
\begin{equation}
\Hapert=\mpl \Hapert'.
\end{equation}

For a pair of neighbouring planets labelled by the indices $i$ and $i+1$ which are near a $k^{(i)}:(k^{(i)}-1)$ mean motion resonance, the perturbing resonant contribution to first order in the eccentricity takes the form
\begin{equation}\label{eq:RescaledResonantInteractionHamiltonian}
\begin{split}
\Hares^{(i)}	&= \mpl \left[\alpha_1^{(i)} e_i \cos\left(k^{(i)}\LAMBDA_{i+1}-(k^{(i)}-i)\LAMBDA_i+\GAMMA_i\right) \right. \\
			&\qquad~~ \left.+\alpha_2^{(i)} e_{i+1} \cos\left(k^{(i)}\LAMBDA_{i+1}-(k^{(i)}-i)\LAMBDA_i+\GAMMA_{i+1}\right)\right],
\end{split}
\end{equation}
where the (rescaled) coefficients $\alpha$ are
\begin{equation}\label{eq:RescaledResonantAlphaCoefficients}
\alpha_j^{(i)}= -\frac{\GravC^2 M_*}{\bar\LAMBDONA_{i+1}^2} f_\res^{(j,i)}(\alpha_\res^{(i)}),
\end{equation}
where $f_\res^{(j,i)}(\alpha_\res^{(i)})$ are functions of the Laplace coefficients $b_{s}^{(j)}$ \citep{1999ssd..book.....M}, themselves (weakly) depending on the semi-major axis ratios (they include both direct and indirect terms; indirect terms only appear in the 2:1 mean motion resonance).
Here as usual $\alpha_\res^{(i)}=\bar a_i/\bar a_{i+1}=\big((k^{(i)}-1)/k^{(i)}\big)^{2/3}$ is the nominal semi-major axis ratio corresponding to the resonance location in the Keplerian approximation, so the Laplace coefficients are the same for each pair of planets in a resonant chain repeating the $k$:$k-1$ commensurability. Moreover, we have evaluated the $\LAMBDONA_{i+1}^2$ at denominator at its nominal Keplerian value $\bar\LAMBDONA_{i+1}$ (because the terms in \eqref{eq:RescaledResonantAlphaCoefficients} are already of order $\mathcal O(e)$, e.g.\ \citealt{2013A&A...556A..28B}). Doing so, the coefficients $\alpha_j^{(i)}$ are effectively constants for a given chain and a given nominal orbital separation, and they represent the strengths of the resonances.

The other terms in the perturbing function $\Hapert$ that are of interest to us are the synodic terms for each neighbouring planet pair. At lowest order in the eccentricities and lowest harmonic order in $\LAMBDA_i-\LAMBDA_{i+1}$, they take the form 
\begin{equation}\label{eq:RescaledSynodicTermForPlanetPair}
\Hasyn^{(i)} = c_i \cos(\LAMBDA_i-\LAMBDA_{i+1}) = \mpl C_i \cos(\LAMBDA_i-\LAMBDA_{i+1}),
\end{equation} 
where the coefficients $C_i$ for the rescaled Hamiltonian are
\begin{equation}\label{eq:RescaledSynodicCCoefficients}
C_i =-\frac{\GravC^2 M_*}{\bar\LAMBDONA_{i+1}^2}\times \left[\frac{1}{2} b_{1/2}^{(1)}(\alpha_\res^{(i)})-\left(\alpha_\res^{(i)}\right)^{-1/2}\right],
\end{equation}
and have the same scaling in $\bar\LAMBDONA_{i+1}$ as the coefficients in \eqref{eq:RescaledResonantAlphaCoefficients} but a different dependence on the Laplace coefficients $b_{s}^{(j)}$ (e.g.\ \citealt{1999ssd..book.....M}; the term $-\left(\alpha_\res^{(i)}\right)^{-1/2}$ comes from the indirect term of the perturbing function).
Notice that \eqref{eq:RescaledSynodicTermForPlanetPair} is of order 0 in eccentricity. The term $\mathcal O(e)$ cannot exist, because it would not satisfy the d'Alembert rules. So, \eqref{eq:RescaledSynodicTermForPlanetPair} is all we have for the terms dependent on the difference of the mean longitudes of neighbouring planets $\LAMBDA_i-\LAMBDA_{i+1}$, but independent of the resonant angles, in an expansion up to $\mathcal O(e)$ of the original Hamiltonian. 
At order 1 in eccentricity, there are also terms coupling resonant and synodic angles (e.g.\ the terms $\big(k^{(i)}\LAMBDA_{i+1}-(k^{(i)}-i)\LAMBDA_i+\GAMMA_i\big)+j\big(\LAMBDA_i-\LAMBDA_{i+1}\big)$, for an arbitrary integer $j$). Because, in what follows, they would behave like those in $\Hasyn$ in \eqref{eq:RescaledSynodicTermForPlanetPair} but are $\mathcal O(e)$ smaller, we neglect them for simplicity. Notice also that in \eqref{eq:RescaledSynodicTermForPlanetPair} we can limit ourselves to the lowest multiples of $\LAMBDA_i-\LAMBDA_{i+1}$ because we are looking for the slowest possible synodic frequency, as explained in Section \ref{sec:OriginOfInstability}.

In the following we will want to consider the case of $N$ planets, each pair being near a $k^{(i)}:(k^{(i)}-1)$ mean motion resonance, and thus introduce the resonant angles as canonical coordinates. However, at the same time, we will want to make use of the the non-resonant synodic angles $\LAMBDA_i-\LAMBDA_{i+1}$, so it is preferable that one of them, say $\LAMBDA_1-\LAMBDA_2$, be also one of the canonical variables. 
The natural choice is to use as canonical positions the resonant angles $\PSI_1^{(i)} = \THETA^{(i)}+\GAMMA_i$ (where $\THETA^{(i)}=k^{(i)}\LAMBDA_{i+1}-(k^{(i)}-1)\LAMBDA_i$ is the longitude of conjunction for the $i$-th pair) and the apsidal differences $\delta\GAMMA_{i,i+1}=\GAMMA_i-\GAMMA_{i+1}$ for $i=1,\dots,N-1$, then define $\delta\LAMBDA_{1,2}=\LAMBDA_1-\LAMBDA_2$ and finally keep an angle which will not appear explicitly in the Hamiltonian, such as $\GAMMA_N$. These linear changes of variables for the positions are easily extended to a canonical transformation (the transformation on the actions is linear, with matrix equal to the transpose of the inverse of the matrix defining the transformation on the angles).
For example, in the case $N=3$ the new angles will be
\begin{alignat}{2}\label{eq:CanonicalVariablesqWithSynodicAngle}
\PSI_1^{(1)} 		&= k^{(1)}\LAMBDA_2-(k^{(1)}-1)\LAMBDA_1+\GAMMA_1, \nonumber\\
\PSI_1^{(2)} 		&= k^{(2)}\LAMBDA_3-(k^{(2)}-1)\LAMBDA_2+\GAMMA_2, \nonumber\\
\delta\GAMMA_{1,2}	&= \GAMMA_1-\GAMMA_2,\\
\delta\GAMMA_{2,3}	&= \GAMMA_2-\GAMMA_3, \nonumber\\
\delta\LAMBDA_{1,2}	&= \LAMBDA_1-\LAMBDA_2, \nonumber\\
\GAMMA_3'			&=-\GAMMA_3, \nonumber
\end{alignat}
while the new conjugated actions are\footnote{
A note on notation can help clarify the meaning of the names of these variables. Reading the definitions for the angles, the upper indices $(i)$ refer to which pair of planets is considered, so that $k^{(1)}$ is the index of the first order mean motion resonance for the inner pair and $\PSI^{(1)}$ refers to a resonant angle for that pair. Similarly $\THETA^{(i)}=k^{(i)}\LAMBDA_{i+1}-(k^{(i)}-1)\LAMBDA_i$ is the longitude of conjunction of the $i$-th pair. The subscript 1 in $\PSI_1^{(i)}$ signifies the fact that for each pair we choose to use the resonant angle which depends on the longitude of pericentre $\GAMMA$ of the innermost planet of the pair, so $\PSI_1^{(i)} = \THETA^{(i)} + \GAMMA_i$, while the other resonant angle would then be $\PSI_2^{(i)}= \THETA^{(i)} + \GAMMA_{i+1}$ (and it does not appear since we also use $\GAMMA_i-\GAMMA_{i+1}$ as canonical angles). The conventions for the other angles are evident. For the actions, we simply use an uppercase first letter to indicate to which angle each action is conjugated, except for the last action since it is just the orbital angular momentum, which we always indicate with $\ANGMOM$, and it is always a constant of motion.}
\begin{alignat}{2}\label{eq:CanonicalVariablespWithSynodicAngle}
\PSIONA_1^{(1)}			&=\LAMBDONA_1+\LAMBDONA_2+\frac{k^{(2)}-1}{k^{(2)}}\LAMBDONA_3, \nonumber\\
\PSIONA_1^{(2)}			&=\frac{1}{k^{(2)}}\LAMBDONA_3, \nonumber\\
\Delta\GAMMA_{1,2}		&=-(\LAMBDONA_1+\LAMBDONA_2+\frac{k^{(2)}-1}{k^{(2)}}\LAMBDONA_3)+\GAMMONA_1,\\
\Delta\GAMMA_{2,3}		&=-(\LAMBDONA_1+\LAMBDONA_2+\LAMBDONA_3)+\GAMMONA_1+\GAMMONA_2,\nonumber\\
\Delta\LAMBDA_{1,2}	&=k^{(1)}\LAMBDONA_1 + (k^{(1)}-1)\LAMBDONA_2 + \frac{(k^{(1)}-1)(k^{(2)}-1)}{k^{(2)}}\LAMBDONA_3, \nonumber\\
					&= k^{(1)}\KAPPONA, \nonumber\\
\ANGMOM			&=(\LAMBDONA_1+\LAMBDONA_2+\LAMBDONA_3)-(\GAMMONA_1+\GAMMONA_2+\GAMMONA_3);\nonumber
\end{alignat}
the canonicity of this transformation can easily be checked using the Poisson bracket criterion.
This canonical change of variable has the advantage of being easily generalisable to any number $N$ of planets and of having the specific angular momentum $\ANGMOM$ appearing as an explicit constant of motion, since its conjugated angle $\GAMMA_N'=-\GAMMA_N$ never appears explicitly in the transformed Hamiltonian (all the other angles satisfy the d'Alembert rules, while this one does not so it cannot appear in the Hamiltonian function, even the non-averaged one). We remark that $\ANGMOM$ is now the specific angular momentum because the actions have been rescaled by the planetary mass; this also entails that when integrating the system \eqref{eq:RescaledHamiltonianNplanets} with increasing $\mpl$, $\ANGMOM$ will always remain constant. Moreover, the action $\Delta\LAMBDA_{1,2}$ conjugated to the angle $\delta\LAMBDA_{1,2}$ is simply a factor away from the action $\KAPPONA$ used in \cite{2018CeMDA.130...54P} (see also e.g.\ \citealt{2013A&A...556A..28B}); this action has been called the ``spacing parameter'' \citep{2008MNRAS.387..747M} and is a constant in the averaged model where all non-resonant contributions to $\Hapert$ are dropped, yielding information on the nominal location $\bar\LAMBDONA$ of the resonance at hand.

We note that in these variables in the case of three resonant planets in a $k^{(1)}:(k^{(1)}-1)$ -- $k^{(2)}:(k^{(2)}-1)$ chain, the synodic harmonics for the two pairs of planets write
\begin{equation}\label{eq:ThreePlanetsSynodicHarmonicsInx}
\begin{split}
\LAMBDA_1-\LAMBDA_2	&=\delta\LAMBDA_{1,2},\\
\LAMBDA_2-\LAMBDA_3	&=\frac{1}{k^{(2)}}\big((k^{(1)}-1)\delta\LAMBDA_{1,2}+\PSI_1^{(1)}-\PSI_1^{(2)}-\delta\GAMMA_{1,2}\big).
\end{split}
\end{equation}
Then, in the new variables the Keplerian Hamiltonian \eqref{eq:RescaledHamiltonianNplanets}, the resonant contribution \eqref{eq:RescaledResonantInteractionHamiltonian} and the synodic contribution \eqref{eq:RescaledSynodicTermForPlanetPair} for pairs of neighbouring planets write
\begin{subequations}\label{eq:FullHamiltonianN=3Dissip3PlanetsInx}
\begin{alignat}{3}
\label{eq:FullHamiltonianN=3Dissip3PlanetsInx.subeq:Hakepl}
\Hakepl	&=-\frac{\GravC^2 M_*^2}{2 \left(-\Delta\LAMBDA_{1,2}+k^{(1)} \PSIONA_1^{(1)}-k^{(2)} \PSIONA_1^{(2)}+\PSIONA_1^{(2)}\right)^2}\\
		&\qquad-\frac{\GravC^2 M_*^2}{2 \left(\Delta\LAMBDA_{1,2}-k^{(1)} \PSIONA_1^{(1)}+\PSIONA_1^{(1)}\right)^2}-\frac{\GravC^2 M_*^2}{2 \left(k^{(2)} \PSIONA_1^{(2)}\right)^2},\nonumber\\
\label{eq:FullHamiltonianN=3Dissip3PlanetsInx.subeq:Hares}
\Hares	&=\mpl \Hares',\\
\label{eq:FullHamiltonianN=3Dissip3PlanetsInx.subeq:Hasyn}
\Hasyn	&=\mpl\Big[C_1 \cos(\delta\LAMBDA_{1,2})\nonumber\\
		&\qquad+C_2 \cos\left(\frac{1}{k^{(2)}}\big((k^{(1)}-1)\delta\LAMBDA_{1,2}+\PSI_1^{(1)}-\PSI_1^{(2)}-\delta\GAMMA_{1,2}\big)\right)\Big] \nonumber\\
		&=\mpl \Hasyn'.
\end{alignat}
\end{subequations}
$\Hakepl$ is independent of $\mpl$ and depends on the variables $\PSIONA_1^{(1)}$, $\PSIONA_1^{(2)}$ and $\Delta\LAMBDA_{1,2}$ only; one can introduce the frequencies (analogous to the mean motions $n$)
\begin{equation}\label{eq:EtaFrequenciesFromHkepl}
\begin{split}
\eta_{\PSIONA_1^{(1)}}	&:=\frac{\partial\Ha}{\partial\PSIONA_1^{(1)}},\\
\eta_{\PSIONA_1^{(2)}}	&:=\frac{\partial\Ha}{\partial\PSIONA_1^{(2)}},\\
\eta_{\Delta\LAMBDA_{1,2}}	&:=\frac{\partial\Ha}{\Delta\LAMBDA_{1,2}}. 
\end{split}
\end{equation}
$\Hares$ only depends on the angles through the harmonic terms $\cos\PSI_1^{(1)}$, $\cos(\PSI_1^{(1)}-\delta\GAMMA_{1,2})$, $\cos\PSI_1^{(2)}$ and $\cos(\PSI_1^{(2)}-\delta\GAMMA_{2,3})$, and each term has a coefficient depending on the actions \eqref{eq:CanonicalVariablespWithSynodicAngle} and the coefficients $\alpha_j^{(i)}$; the exact expression can easily be obtained by direct substitution. 
In \eqref{eq:FullHamiltonianN=3Dissip3PlanetsInx.subeq:Hares}, \eqref{eq:FullHamiltonianN=3Dissip3PlanetsInx.subeq:Hasyn} we use a prime ($'$) to indicate that the Hamiltonian term has been rescaled by $\mpl$ itself, so, it is $\mathcal O(0)$ in $\mpl$, and the dependence on $\mpl$ has been clearly expressed with a coefficient.
In the following we will also use the notation
\begin{equation}
\mathbf x=(\PSIONA_1^{(1)},\PSIONA_1^{(2)},\Delta\GAMMA_{1,2},\Delta\GAMMA_{2,3},\Delta\LAMBDA_{1,2},\PSI_1^{(1)},\PSI_1^{(2)},\delta\GAMMA_{1,2},\delta\GAMMA_{2,3},\delta\LAMBDA_{1,2})
\end{equation}
for the canonical variables that enter in $\Ha$ (except the pair $(\ANGMOM,\GAMMA_3')$, since $\GAMMA_3'$ does not appear in $\Ha$ and $\ANGMOM$ is a constant of motion); we write for the actions $\mathbf p=(\PSIONA_1^{(1)},\PSIONA_1^{(2)},\Delta\GAMMA_{1,2},\Delta\GAMMA_{2,3},\Delta\LAMBDA_{1,2})$ and for the angles $\mathbf q=(\PSI_1^{(1)},\PSI_1^{(2)},\delta\GAMMA_{1,2},\delta\GAMMA_{2,3},\delta\LAMBDA_{1,2})$.

\subsection{Purely resonant dynamics}\label{sec:StabilityN=3.subsec:PurelyResonantDynamics}
The purely resonant dynamics is the one governed by the Hamiltonian averaged over the fast angle $\delta\LAMBDA_{1,2}$, i.e.\ $\bar\Ha=\Hakepl+\Hares$. $\bar\Ha$ is now rewritten in terms of the new canonical variables (\ref{eq:CanonicalVariablesqWithSynodicAngle}, \ref{eq:CanonicalVariablespWithSynodicAngle}) (cfr.\ Equation \eqref{eq:FullHamiltonianN=3Dissip3PlanetsInx}), and since the synodic terms have been removed by averaging, not only $\ANGMOM$ but also $\Delta\LAMBDA_{1,2}$ is a constant of motion, so that only the ``barred'' variables $\bar {\mathbf x}=(\bar {\mathbf p},\bar {\mathbf q})=({\PSIONA_1^{(1)}},{\PSIONA_1^{(2)}},{\Delta\GAMMA_{1,2}},\Delta\GAMMA_{2,3},{\PSI_1^{(1)}},{\PSI_1^{(2)}},\delta\GAMMA_{1,2},\delta\GAMMA_{2,3})$ evolve.\footnote{
From a technical point of view, these variables are only an approximation (up to order 0 in $\mpl$) to the actual canonical variables that eliminate the non-resonant contributions. These would be the primed variables introduced later on in \eqref{eq:pqTop'q'}.} These barred variables $\bar {\mathbf x}=(\bar {\mathbf p},\bar {\mathbf q})$ are simply a subset of the variables ${\mathbf x}=({\mathbf p},{\mathbf q})$ introduced above, and represent the purely resonant degrees of freedom (this notation is only introduced to separate these variables from the synodic canonical pair $(\Delta\LAMBDA_{1,2},\delta\LAMBDA_{1,2})$; this will be a useful distinction later on).

\begin{figure*}[!t]
\centering
\begin{subfigure}[b]{0.3 \textwidth}
\centering
\includegraphics[width=1.1 \textwidth]{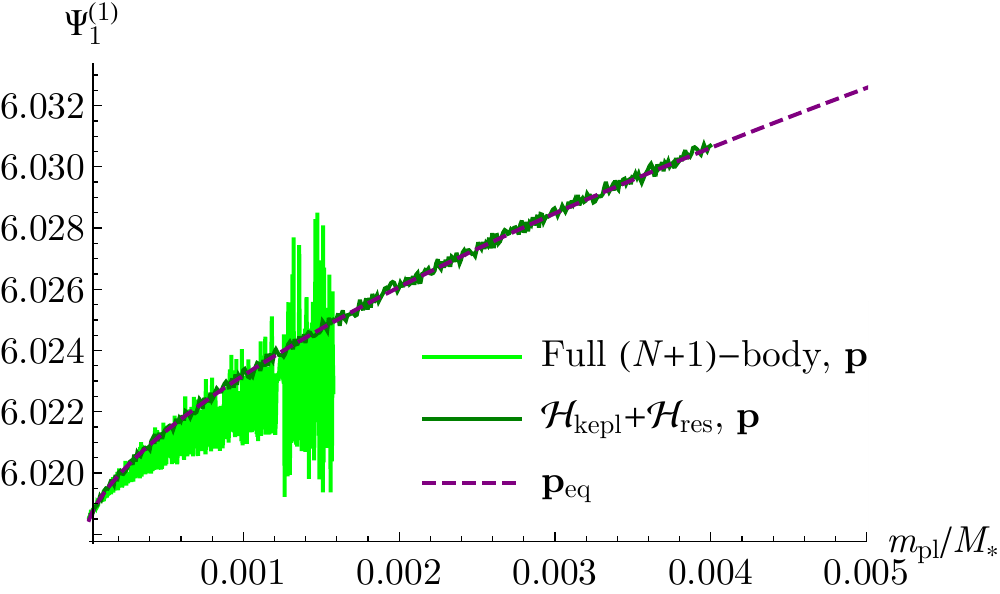}
\caption{$\PSIONA_1^{(1)}$.}
\end{subfigure}
\qquad
\centering
\begin{subfigure}[b]{0.3 \textwidth}
\centering
\includegraphics[width=1.1 \textwidth]{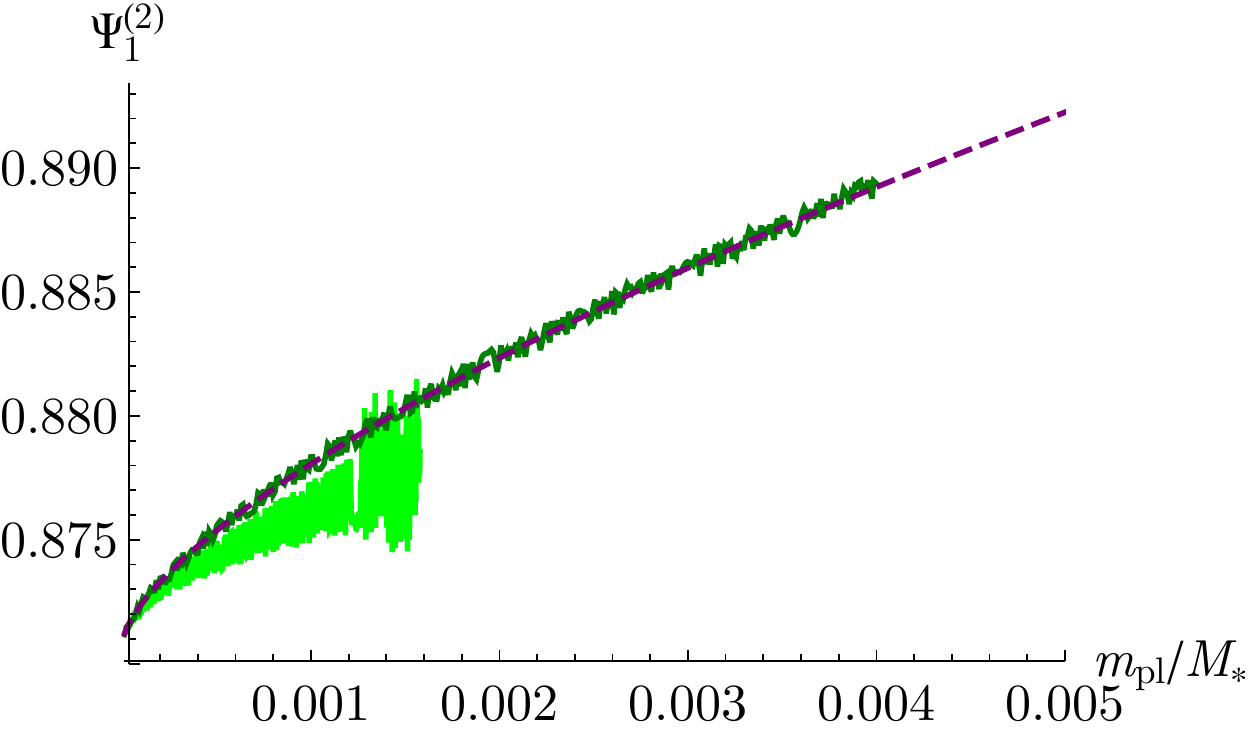}
\caption{$\PSIONA_1^{(2)}$.}
\end{subfigure}
\qquad
\centering
\begin{subfigure}[b]{0.3 \textwidth}
\centering
\includegraphics[width=1.1 \textwidth]{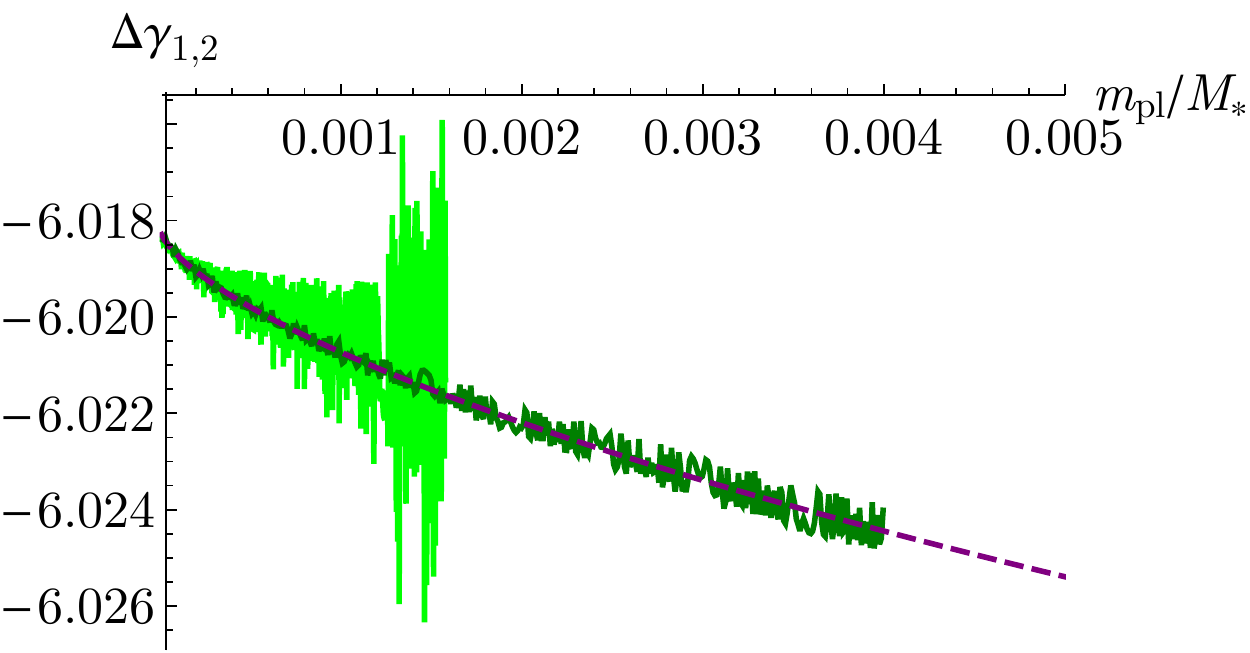}
\caption{$\Delta\GAMMA_{1,2}$.}
\end{subfigure}

\begin{subfigure}[b]{0.3 \textwidth}
\centering
\includegraphics[width=1.1 \textwidth]{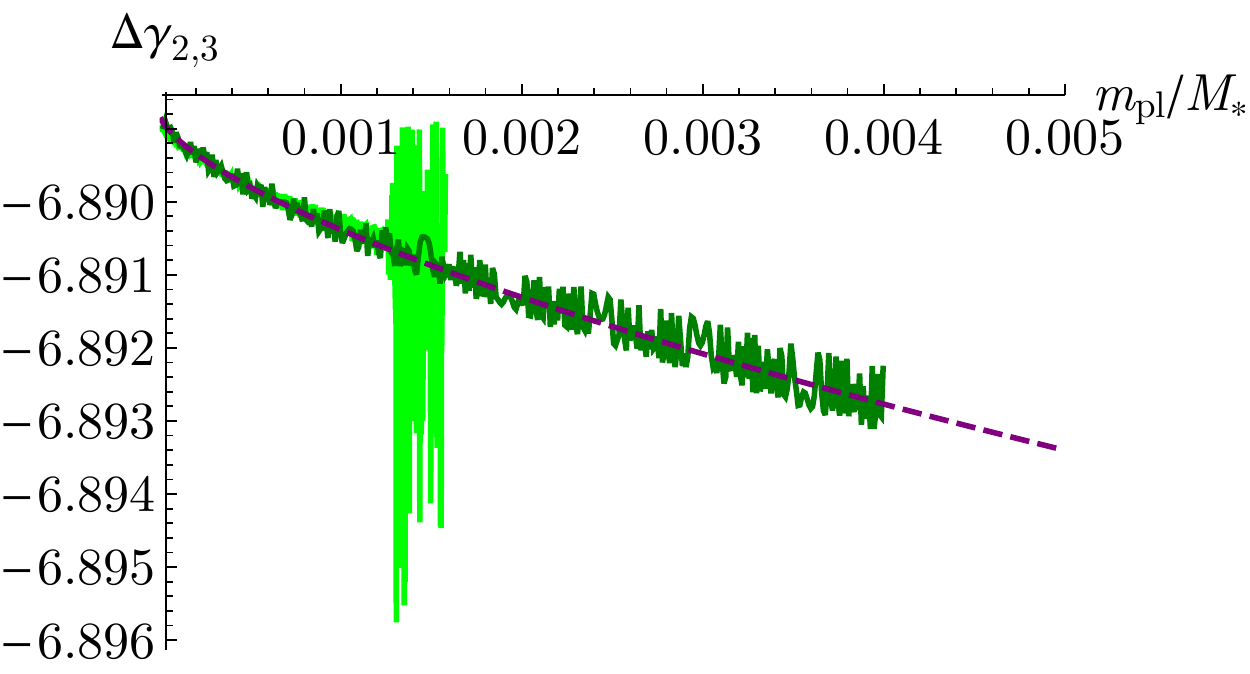}
\caption{$\Delta\GAMMA_{2,3}$.}
\end{subfigure}
\qquad
\centering
\begin{subfigure}[b]{0.3 \textwidth}
\centering
\includegraphics[width=1.1 \textwidth]{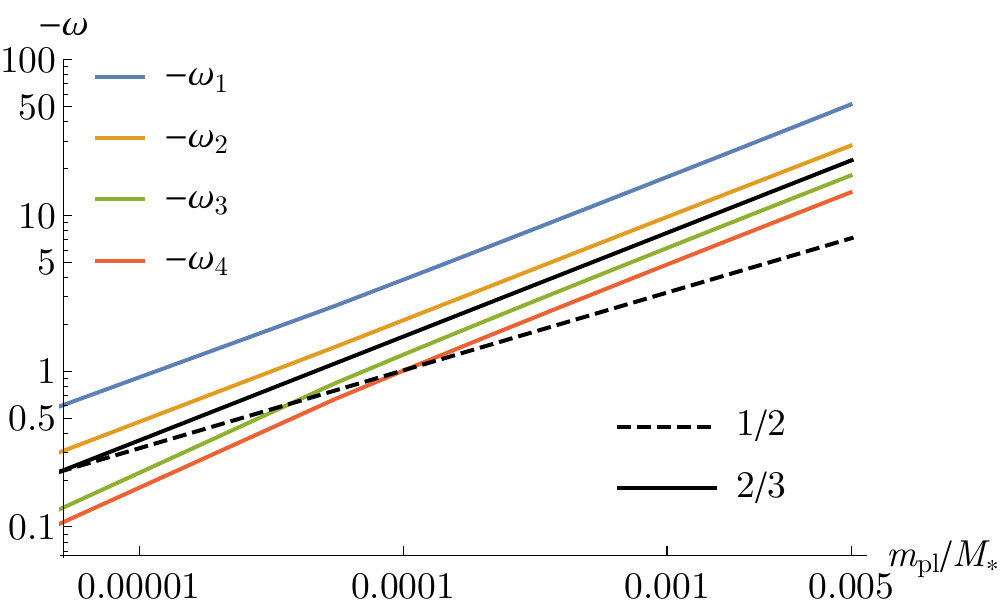}
\caption{$\boldsymbol\omega$, low $e$.}
\label{fig:3-2_3-2PurelyResonantModel.subfig:3-2_3-2_plot4FreqLowe}
\end{subfigure}
\qquad
\centering
\begin{subfigure}[b]{0.3 \textwidth}
\centering
\includegraphics[width=1.1 \textwidth]{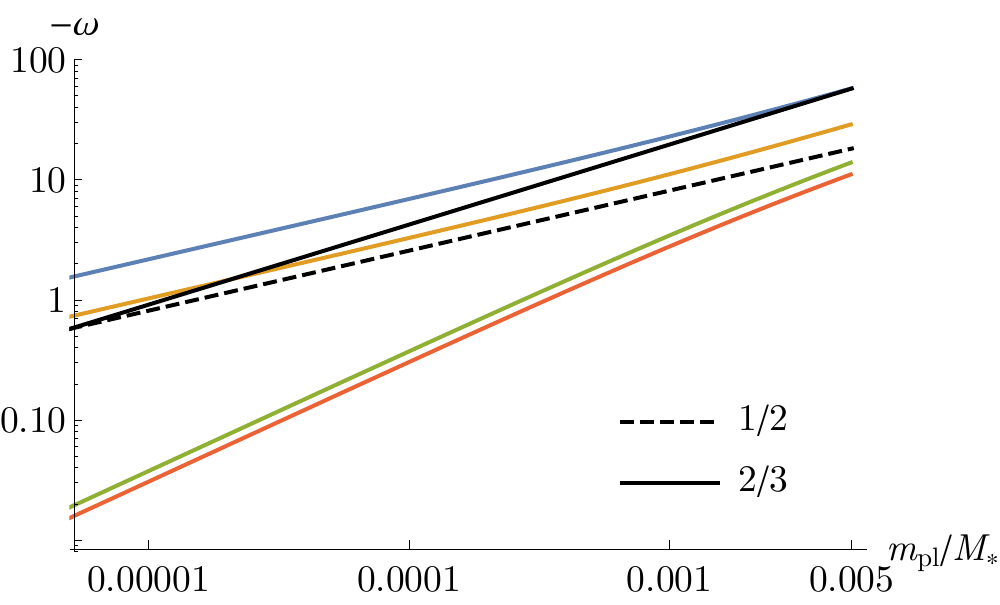}
\caption{$\boldsymbol\omega$, high $e$.}
\label{fig:3-2_3-2PurelyResonantModel.subfig:3-2_3-2_plot4FreqHighe}
\end{subfigure}
\caption{
The purely resonant evolution governed by $\bar\Ha$ in the case of three planets in a 3:2 -- 3:2 mean motion resonance chain, with the same initial conditions as in Figure \ref{fig:3-2_3-2Stability.subfig:3-2_3-2_InstabOnsetOne2Plot}. We show in dark green in panels (a) to (d) the evolution of the actions $\bar{\mathbf p}=(\PSIONA_{1}^{(1)},\PSIONA_{2}^{(1)},\Delta\GAMMA_{1,2},\Delta\GAMMA_{2,3})$ as the planetary mass $\mpl$ is slowly increasing, and we match it to the calculated equilibria $\bar{\mathbf p}_{\eq}=(\PSIONA_{1,\eq}^{(1)},\PSIONA_{2,\eq}^{(1)},\Delta\GAMMA_{1,2,\eq},\Delta\GAMMA_{2,3,\eq})$ (purple dashed line); we also add the corresponding $(N+1)$-body integration with the same initial condition (light green). A legend for panels (a) to (d) is given in panel (a). We see that the system remains stable well after the value of $\mpl/M_*\simeq\protect\input{MassAtPhenom3-2_3-2.dat}$ corresponding to the onset of excitation in Figure \ref{fig:3-2_3-2Stability.subfig:3-2_3-2_InstabOnsetOne2Plot}. Panels (e) and (f) contain the analytical explanation of the observed stability: we plot with coloured lines all the frequencies of the four degrees of freedom and we notice that they have the same sign, therefore the Hamiltonian has a maximum at the equilibrium point and for low amplitude of librations the system remains Lyapunov-stable even if the frequencies grow in absolute value. In panel (e) we used the same eccentricities that correspond to the initial conditions of panels (a) to (d), $e\simeq0.01$; in panel (f) we used higher initial eccentricities, $e\simeq0.1$. We note that the scaling law for $\omega_l(\mpl)$ changes depending on the eccentricity (see black solid and dashed lines). 
}
\label{fig:3-2_3-2PurelyResonantModel}
\end{figure*}

We integrate this Hamiltonian with a numerical integrator while slowly increasing the planetary mass at each time step as detailed above. We use again as an example the case of the 3:2 -- 3:2 chain starting with an initial planetary mass $\mpl/M_*=10^{-5}$ and we choose as initial condition that of Figure \ref{fig:3-2_3-2Stability.subfig:3-2_3-2_InstabOnsetOne2Plot}. The resulting evolution of the canonical actions $\bar{\mathbf p}$ is shown in dark green in Figure \ref{fig:3-2_3-2PurelyResonantModel}, panels (a) to (d) (the evolution of the eccentricity has been already presented in Figure \ref{fig:3-2_3-2Stability.subfig:3-2_3-2_InstabOnsetOne2Plot}).
We observe that the four resonant degrees of freedom are never unstable even up to masses significantly higher than the critical mass $(\mpl/M_*)_\crit\simeq 1.28\times 10^{-3}$ which is found in the numerical $(N+1)$-body simulations with the same initial conditions (Figure \ref{fig:3-2_3-2Stability.subfig:3-2_3-2_InstabOnsetOne2Plot}, light green evolution in Fig.\ \ref{fig:3-2_3-2PurelyResonantModel}).

We can present an analytical explanation for this. As in \cite{2018CeMDA.130...54P}, we find the stable resonant equilibrium points for $\bar\Ha(\bar{\mathbf x}; \ANGMOM, \Delta\LAMBDA_{1,2},\mpl)$ in the variables $\bar{\mathbf x}$, while keeping $\ANGMOM$ and $\Delta\LAMBDA_{1,2}$ constants and for different values of $\mpl$, yielding $\bar{\mathbf x}_{\eq}(\mpl)=\bar{\mathbf x}_{\eq}(\mpl;\ANGMOM, \Delta\LAMBDA_{1,2})$. Notice that at these low eccentricities we are interested in symmetric linearly stable equilibria only, so the equilibrium values $\bar{\mathbf q}_{\eq}$ of the angles are simply
\begin{equation}\label{eq:EquilibriumAngles3PlanetsStability}
\begin{split}
\PSI_{1,\eq}^{(1)}		&=0,\\
\PSI_{1,\eq}^{(2)}		&=0,\\
\delta\GAMMA_{1,2,\eq}	&=\pi,\\
\delta\GAMMA_{2,3,\eq}	&=\pi,
\end{split}
\end{equation} 
and we only need to solve for the equilibrium actions $\bar{\mathbf p}_{\eq}=(\PSIONA_{1,\eq}^{(1)},\PSIONA_{2,\eq}^{(1)},\Delta\GAMMA_{1,2,\eq},\Delta\GAMMA_{2,3,\eq})$. In Figure \ref{fig:3-2_3-2PurelyResonantModel}, panels (a) to (d), we superimposed the analytically calculated equilibrium points (dashed purple lines) and the numerically-obtained evolution, showing excellent agreement, which implies that the numerical solution stays on the stable equilibrium at all times.  
Then, we diagonalise the system around the equilibrium point $\bar{\mathbf x}_\eq$; since it is a stable equilibrium point, all eigenvalues are purely imaginary and the diagonalisation procedure yields a Hamiltonian of the form
\begin{equation}
\bar\Ha(\boldsymbol\xi,\boldsymbol\eta) = \sum_{l=1}^4 \frac{\omega_l}{2}(\xi_l^2+\eta_l^2) + \mathcal O(\|(\boldsymbol\xi,\boldsymbol\eta)^3\|)
\end{equation}
in cartesian coordinates $\bar{\mathbf x}=T(\boldsymbol\xi,\boldsymbol\eta)$, with $T$ a transformation matrix. Using canonical polar coordinates $(I_l,\phi_l)_{l=1,\dots,4}$ with $\left(\xi_l=\sqrt{2I_l}\cos\phi_l,~\eta_l=\sqrt{2I_l}\sin\phi_l\right)$ we get
\begin{equation}\label{eq:HbarInIphicoordinates}
\bar\Ha = \sum_{l=1}^4 \omega_l I_l + \mathcal O(\|\boldsymbol I^{3/2}\|),
\end{equation}
which appears as the sum of four decoupled harmonic oscillators plus higher order terms.
The resulting four frequencies $\omega_l$, $l=1,\dots,4$ are shown in Figure \ref{fig:3-2_3-2PurelyResonantModel.subfig:3-2_3-2_plot4FreqLowe} as a function of the planetary mass $\mpl$, and we notice right away that they all have the same sign.
This means that at vanishing amplitudes of libration the Hamiltonian has an extremum at the equilibrium point (a maximum) so that we can use the Hamiltonian itself as a Lyapunov function to deduce that the equilibrium point is Lyapunov stable for all planetary masses. This means also that if the initial amplitude of libration around the equilibrium point is small, it has to remain small at all times. \\

Since it will be useful later on, we also consider here how the libration frequencies grow with $\mpl$. This is shown in Figure \ref{fig:3-2_3-2PurelyResonantModel} panels (e) and (f). We find numerically that $\omega_{1,2}\propto\mpl^{2/3}$ at low eccentricities ($e\simeq 0.01$, panel (e)) while $\omega_{1,2}\propto\mpl^{1/2}$ at higher eccentricities ($e\simeq0.1$, panel (f)). Notice that for a pendulum-type Hamiltonian like 
\begin{equation}\label{eq:PendulumForLibrationFrequencies}
\Ha_{\mathrm{pend}}(\Sigma,\sigma) = a\Sigma^2 - \mpl b \cos\sigma
\end{equation}
the libration frequency would be $\propto\mpl^{1/2}$, so it might be interesting to ponder analytically why at low eccentricities we get a different scaling. The reason is that with changing mass we also change the corresponding equilibrium point, which means that the parameters $a$ and $b$ in the pendulum-like Hamiltonian above also depend on $\mpl$, and the real scaling would therefore be $\sqrt{ab\mpl}$. The way the equilibrium points adjust to changes in $\mpl$ here is by following lines of constant specific angular momentum (see above, and \citealt{2018CeMDA.130...54P}). Thus, with changing mass we also change the eccentricity of the corresponding equilibrium point, i.e. $b$ in \eqref{eq:PendulumForLibrationFrequencies}, as $\mpl^{1/3}$.
We finally remark that \cite{2015MNRAS.451.2589B} estimates for two planets the (highest) libration frequency, at small amplitude of librations around the resonant equilibrium point and for a value of the angular momentum at which the separatrix first appears. He finds that this frequency scales with $\left((m_1+m_2)/M_*\right)^{2/3}$: since the appearance of the separatrix happens at small eccentricities, this is consistent with our findings.

\subsection{The synodic contribution}\label{sec:StabilityN=3.subsec:SynodicContribution}
In the previous subsection we have shown that the purely resonant system is Lyapunov-stable for all planetary masses. The next natural step is therefore to introduce non-resonant contribution of the disturbing function. To lowest order in $e$, we introduce the two synodic terms \eqref{eq:RescaledSynodicTermForPlanetPair} for the inner and outer pairs that had been averaged out before, resulting in
\begin{equation}
\Hasyn=\mpl\left[C_1 \cos(\LAMBDA_1-\LAMBDA_2) + C_2 \cos(\LAMBDA_2-\LAMBDA_3)\right] = \mpl \Hasyn',
\end{equation}
with coefficients given by \eqref{eq:RescaledSynodicCCoefficients}. The full rescaled Hamiltonian written in the new variables (\ref{eq:CanonicalVariablesqWithSynodicAngle}, \ref{eq:CanonicalVariablespWithSynodicAngle}) is now
\begin{equation}
\Ha^*(\mathbf x;\ANGMOM,\mpl)=\Hakepl+\Hares+\Hasyn;
\end{equation}
we have stressed that it depends parametrically on the constant of motion $\ANGMOM$ and on the mass $\mpl$ through $\Hares=\mpl \Hares'$ and $\Hasyn=\mpl \Hasyn'$.

\begin{figure*}[!t]
\centering
\begin{subfigure}[b]{0.3 \textwidth}
\centering
\includegraphics[width=1.1 \textwidth]{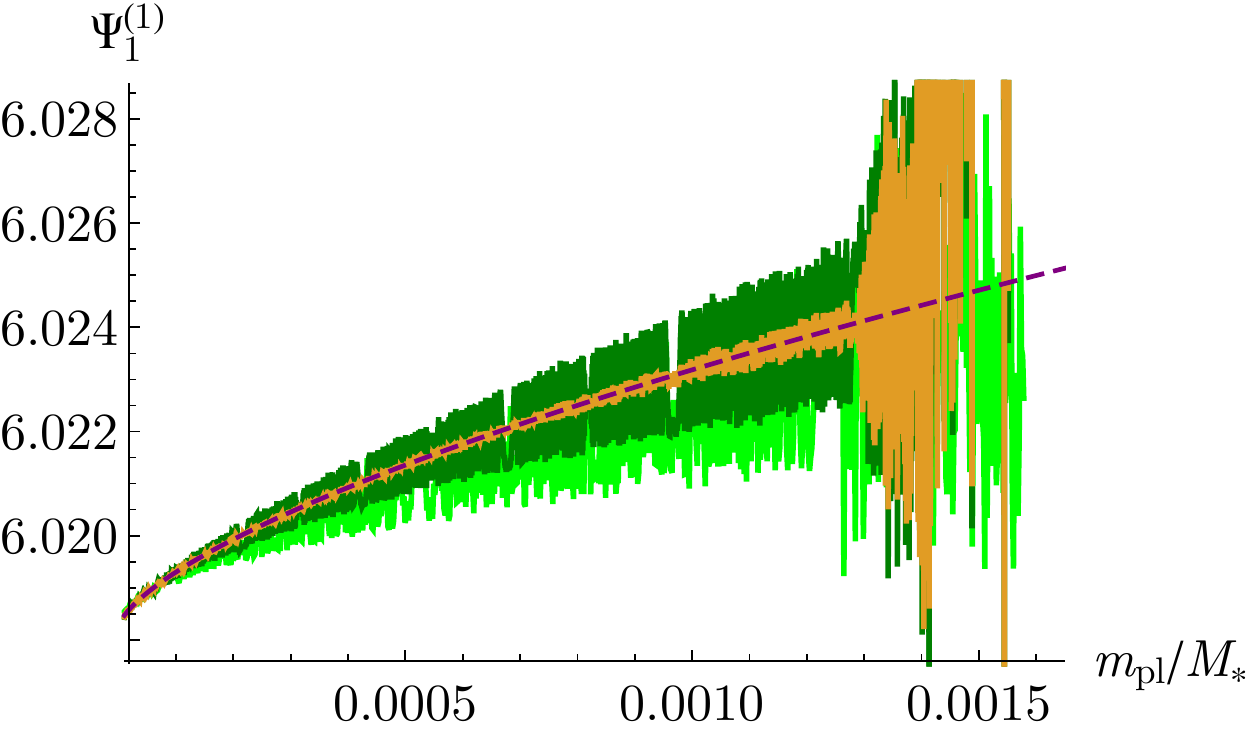}
\caption{$\PSIONA_1^{(1)}$.}
\end{subfigure}
\qquad
\centering
\begin{subfigure}[b]{0.3 \textwidth}
\centering
\includegraphics[width=1.1 \textwidth]{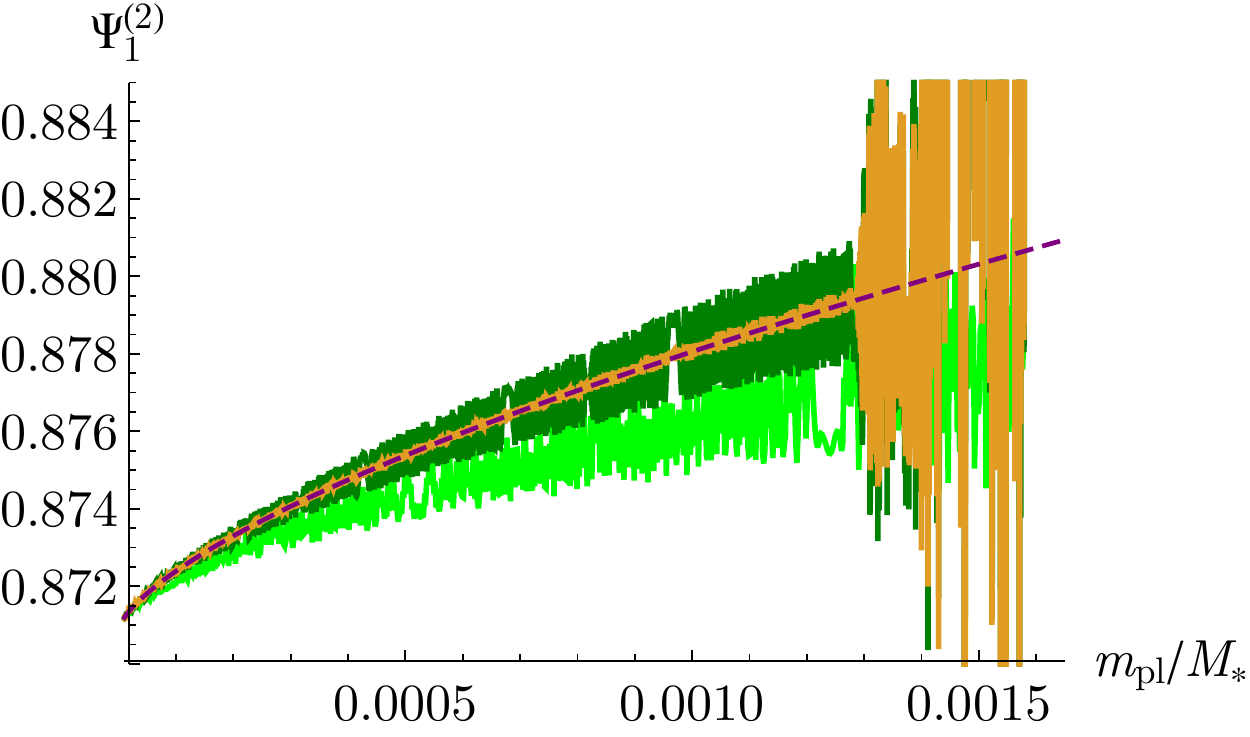}
\caption{$\PSIONA_1^{(2)}$.}
\end{subfigure}
\qquad
\centering
\begin{subfigure}[b]{0.3 \textwidth}
\centering
\includegraphics[width=1.1 \textwidth]{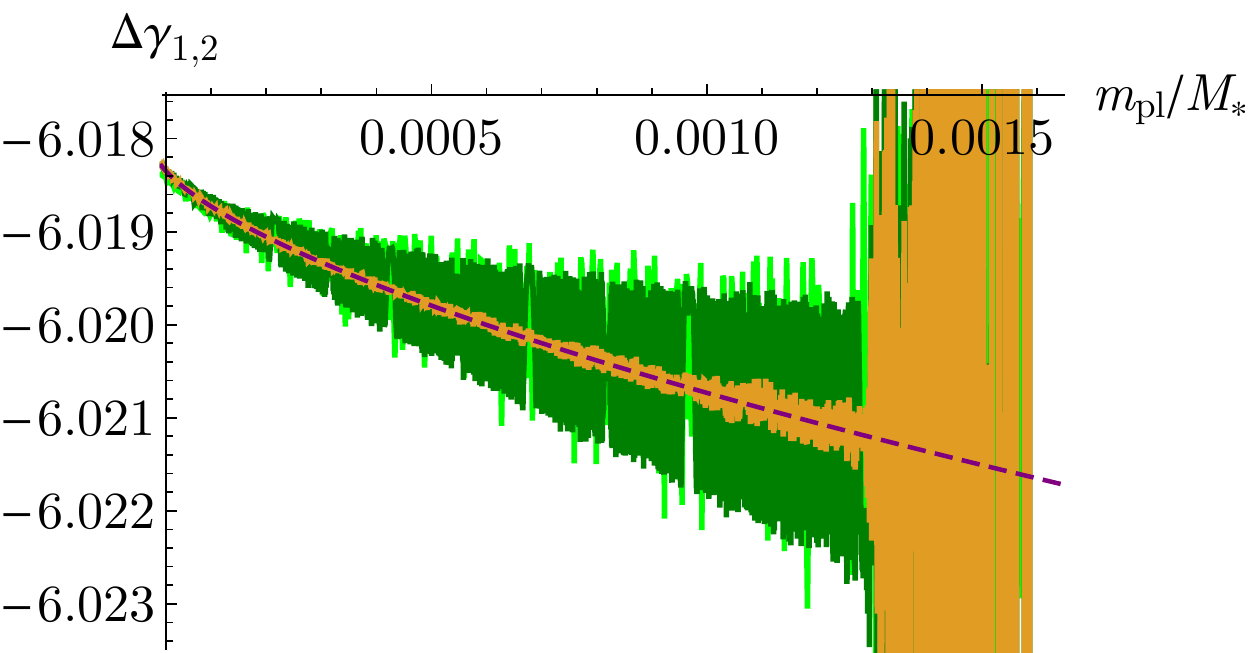}
\caption{$\Delta\GAMMA_{1,2}$.}
\end{subfigure}

\begin{subfigure}[b]{0.3 \textwidth}
\centering
\includegraphics[width=1.1 \textwidth]{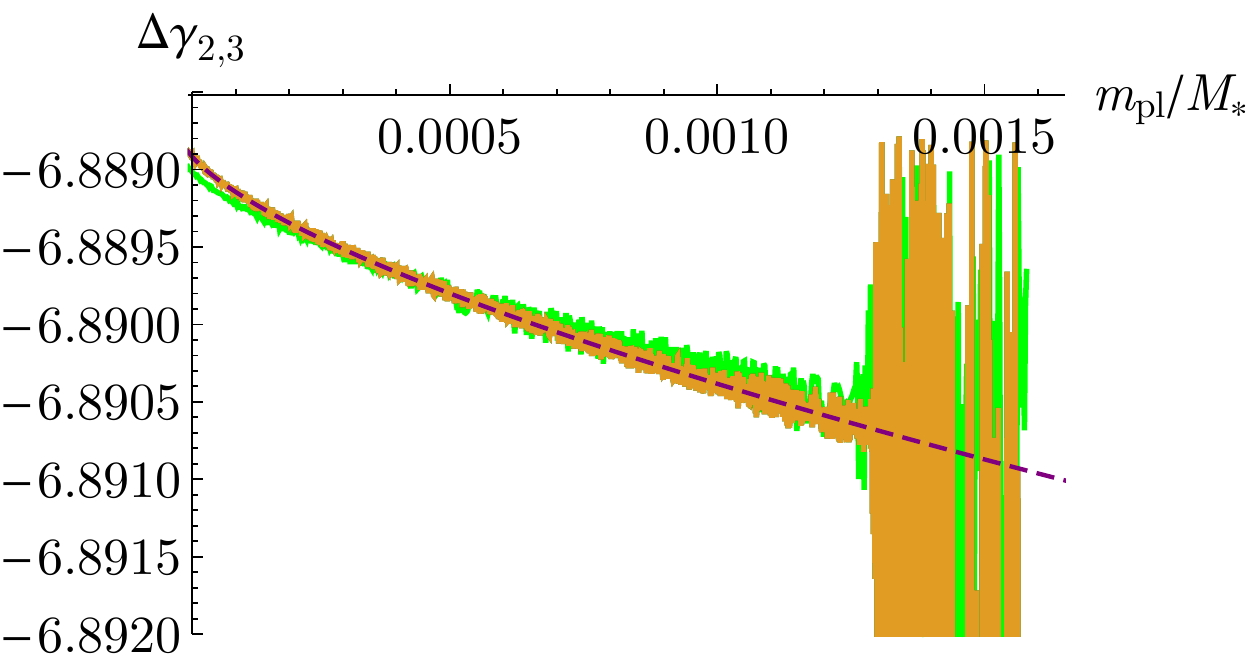}
\caption{$\Delta\GAMMA_{2,3}$.}
\end{subfigure}
\qquad
\centering
\begin{subfigure}[b]{0.3 \textwidth}
\centering
\includegraphics[width=1.1 \textwidth]{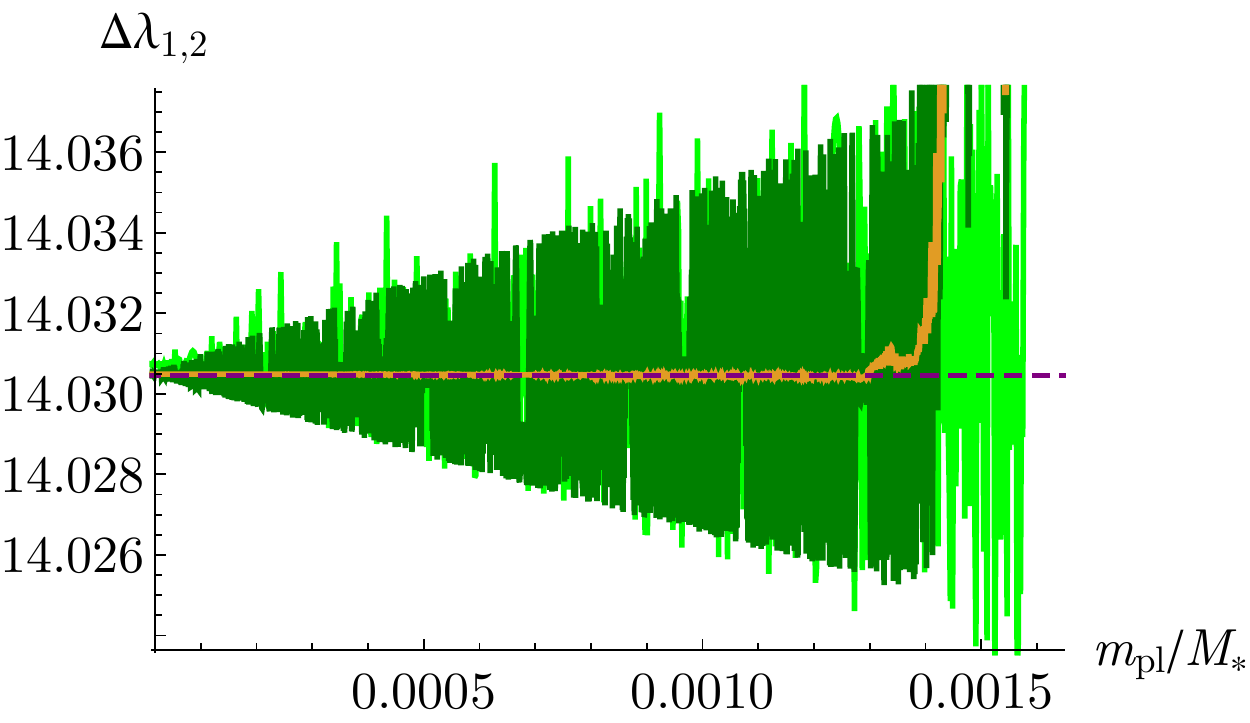}
\caption{$\Delta\LAMBDA_{1,2}$.}
\label{fig:3-2_3-2FullApproximatedModel.subfig:DeltaLambda12}
\end{subfigure}
\qquad
\centering
\begin{subfigure}[b]{0.3 \textwidth}
\centering
\includegraphics[width=1.1 \textwidth]{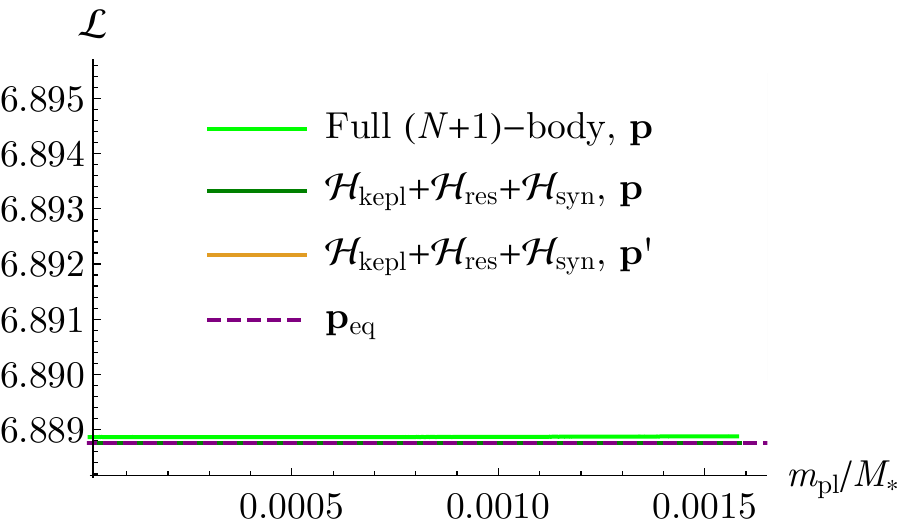}
\caption{$\ANGMOM$.}
\end{subfigure}
\caption{
Panels (a) to (e) show in dark green the evolution of the actions $\mathbf p=(\PSIONA_1^{(1)},\PSIONA_1^{(2)},\Delta\GAMMA_{1,2},\Delta\GAMMA_{2,3},\Delta\LAMBDA_{1,2})$ for the system $\Ha^*=\Hakepl+\Hares+\Hasyn$ with the same initial condition as the $(N+1)$-body integration of Figure \ref{fig:3-2_3-2Stability.subfig:3-2_3-2_InstabOnsetOne2Plot} (the evolution of these variables in the $(N+1)$-body integration is also shown here in light green for reference). 
The system follows on average the purple dashed lines, which correspond to the equilibria $\mathbf p_\eq$ for the system $\bar\Ha=\Hakepl+\Hares$. In orange we show the evolution of the averaged variables $\mathbf p'$ calculated through analytical averaging of the fast synodic frequencies, Equation \eqref{eq:3-2_3-2AnalyticallyAveragedActions}. 
Note that, for $\mpl/M_*<\protect\input{MassAtPhenom3-2_3-2.dat}$, $\mathbf p'$ has very little oscillation around the $\mathbf p_\eq$ curve, compared to the $\mathbf p$ evolution. Instead, for $\mpl/M_*>\protect\input{MassAtPhenom3-2_3-2.dat}$, the amplitude of oscillations of $\mathbf p'$ and $\mathbf p$ around $\mathbf p_\eq$ are almost the same. This reveals that, while the initial oscillation of $\mathbf p$ is entirely due to the synodic terms and is effectively removed by passing to the $\mathbf p'$ variables, it is then dominated by an increased amplitude of libration in the resonance.
The evolution of the angular momentum $\ANGMOM$ is also shown in panel (f), and it is of course a conserved quantity; panel (f) also contains the legend for all panels in this figure.}
\label{fig:3-2_3-2FullApproximatedModel}
\end{figure*}

We integrate this Hamiltonian for the 3:2 -- 3:2 chain with the same numerical scheme described before and the same initial conditions as in the previous section. This gives the evolution of the actions displayed in dark green in Figure \ref{fig:3-2_3-2FullApproximatedModel}, which is matched against the $(N+1)$-body integration with the same initial datum (lighter green) and the locations of the equilibria for $\bar\Ha$ calculated in the previous section for different planetary masses (purple dashed lines). The comparison for the eccentricity evolutions, instead of the canonical variables, has been already presented in Figure \ref{fig:3-2_3-2Stability.subfig:3-2_3-2_InstabOnsetOne2Plot}. We notice two important aspects of these plots. The first is that, initially, for all variables the evolution described by $\Ha^*$ follows on average that described by $\bar\Ha$ (compare Fig.\ \ref{fig:3-2_3-2PurelyResonantModel} with Fig.\ \ref{fig:3-2_3-2FullApproximatedModel}, and the dashed purple lines).
This can be easily understood realising that $\Ha^*$ contains fast, non-resonant angles, which, up to first order in the small parameter $\mpl$, have simply been averaged out in $\bar\Ha$; therefore, as long as the $\mathcal O(\mpl^2)$ contributions are unimportant, the only difference between the two evolutions are the short-periodic, $\mathcal O(\mpl)$ oscillations due to the $\delta\LAMBDA_{1,2}$ synodic angle. We will actually study this effect analytically below. However, as soon as the $\mathcal O(\mpl^2)$ remainder introduces important contributions to the dynamics, as in the case of the emergence of a secondary resonance, the dynamics described by the averaged $\bar\Ha$ approximation is not valid anymore. This is indeed what we see in Figure \ref{fig:3-2_3-2FullApproximatedModel}, where a phenomenon similar to the one observed in the $(N+1)$-body integrations appears, and at roughly the same value of $\mpl/M_*\simeq 1.28\times 10^{-3}$, which was not found in $\bar\Ha$. Notice that such a secondary resonance cannot be caused by an interaction of the resonant degrees of freedom $\bar{\mathbf x}$ only, as we have shown that these are stable for all values of $\mpl$. Therefore, these secondary resonances must come from an interaction between some (combination) of the four resonant degrees of freedom and the synodic degree of freedom $(\Delta\LAMBDA_{1,2},\delta\LAMBDA_{1,2})$.
In the following, we use the analytical tools of the Lie series perturbation theory in order to pinpoint the relevant secondary resonances that arise at order 2 in the planetary mass $\mpl$. We carry out the calculation for the case of three resonant planets in any resonant chain order to get the general picture, but we will focus on the case of $k^{(1)}=k^{(2)}=k$, and $k=3$ when needed.

\subsubsection{Eliminating the $\mathcal O(\mpl)$ synodic term}\label{subsubsec:EliminatingOmplSynodicTerms}
In the previous section we dropped $\Hasyn$ out by averaging the Hamiltonian. But simple averaging or dropping of harmonics is not a rigorous procedure and, as we have seen, can alter the real dynamics. Averaging is just the first step of more complex, rigorous, perturbation approach, as we describe here.

The first step is to find a canonical transformation that, to first order in $\mpl$, eliminates the synodic contribution $\mpl\Hasyn'$ from $\Ha^*$. This will introduce $\mathcal O(\mpl^2)$ terms that we want to calculate explicitly, since they contain harmonics mixing $\bar{\mathbf q}$ and $\delta\LAMBDA_{1,2}$, potentially associated to secondary resonances.

In order to eliminate $\mpl\Hasyn'$ at $\mathcal O(\mpl)$, we need to find a generating function $\chisyn$ that solves the homological equation
\begin{equation}\label{eq:HomologicalEquationForchisyn}
\{\chisyn,\Hakepl\}+\Hasyn = 0.
\end{equation}
In the above equation we used the Poisson bracket $\{\bullet,\bullet\}$ which operates on two dynamical variables $f_1$ and $f_2$ yielding a new dynamical variable $\{f_1,f_2\}$ defined by
\begin{equation}
\{f_1,f_2\}:=(\grad f_1)^\intercal J (\grad f_2),
\end{equation}
where $\grad$ is the gradient with respect to the canonical variables and $J$ is the standard symplectic matrix
\begin{equation}
J = \left[ \begin{matrix} \boldsymbol 0 & -\mathbb I_n \\ \mathbb I_n & \boldsymbol 0 \end{matrix} \right].
\end{equation}
Clearly $\chisyn$ will be of order $\mpl$ so we can write $\chisyn=\mpl \chisyn'$. From the expression for $\Hasyn$, Equation \eqref{eq:FullHamiltonianN=3Dissip3PlanetsInx.subeq:Hasyn}, we see that $\chisyn$ will have the form
\begin{align}
\chisyn	&=\mpl\Big[\frac{C_1}{\eta_1}\sin(\delta\LAMBDA_{12})\\ 
		&\quad +\frac{C_2}{\eta_2}\sin\left(\frac{1}{k^{(2)}}\big((k^{(1)}-1)\delta\LAMBDA_{12}+\PSI_1^{(1)}-\PSI_1^{(2)}-\delta\GAMMA_{1,2}\big)\right)\Big],\nonumber
\end{align}
where the divisors $\eta_1$ and $\eta_2$ are immediately found in terms of the frequencies  \eqref{eq:EtaFrequenciesFromHkepl} of the unperturbed Keplerian Hamiltonian and the combination of angles appearing in the harmonics in $\Hasyn$, yielding
\begin{equation}\label{eq:EtaDivisorsForChisyn}
\begin{split}
\eta_1	&=\eta_{\Delta\LAMBDA_{1,2}},\\
\eta_2	&=\frac{1}{k^{(2)}}\left((k^{(1)}-1)\eta_{\Delta\LAMBDA_{1,2}} + \eta_{\PSIONA_1^{(1)}} - \eta_{\PSIONA_1^{(2)}}\right).
\end{split}
\end{equation}
These divisors are not vanishing nor small, since clearly $\eta_1=n_1-n_2$, $\eta_2=n_2-n_3$ (remember that $n_i$ is the mean motion frequency of planet $i$ and that the harmonics in $\Hasyn$ in the modified Delaunay variables were simply $\LAMBDA_1-\LAMBDA_2$ and $\LAMBDA_2-\LAMBDA_3$) and the planets are evidently far from the 1:1 resonance. Therefore equation \eqref{eq:HomologicalEquationForchisyn} can indeed be solved.

Having calculated $\chisyn$, we can then write out how the Hamiltonian $\Ha^*$ transforms under the Lie series transformation $\exp(L_\chisyn)$ generated by $\chisyn$. 
Here, $\exp(L_\chisyn)$ is given by
\begin{equation}
\begin{split}
\exp(L_\chisyn)f	&= f + \mpl\{f,\chisyn'\} + \frac{\mpl^2}{2}\{\{f,\chisyn\},\chisyn\} + \dots\\
					&=\sum_{i=0}^\infty \frac{\mpl^i}{i!}L_\chisyn^if,~L_\chisyn f:=\mpl\{f,\chisyn'\}.
\end{split}
\end{equation}
The new Hamiltonian $\Ha'$ is given by $\exp(L_\chisyn)\Ha^*$, and reads, up to $\mathcal O(\mpl^2)$,
\begin{subequations}\label{eq:Hprimed}
\begin{alignat}{3}
\label{eq:Hprimed.subeq:1}
\Ha'	&=\boxed{\Hakepl}	&&+\mpl\{\Hakepl,\chisyn'\}			&&+\frac{\mpl^2}{2}\{\{\Hakepl,\chisyn'\},\chisyn'\} + \dots\\
\label{eq:Hprimed.subeq:2}
	&						&&\boxed{+\mpl \Hares'}		&&+\mpl^2\{\Hares',\chisyn'\}+\dots\\
\label{eq:Hprimed.subeq:3}
	&						&&+\mpl \Hasyn'				&&+\mpl^2\{\Hasyn',\chisyn'\}+\dots ;
\end{alignat}
\end{subequations}
as it is typical in perturbation theory via Lie transform, this transformed Hamiltonian is written in terms of the new variables $\mathbf x '$ by direct substitution of the old variables $\mathbf x$ to the new, $\mathbf x\to \mathbf x'$. The change of variable is given by $\mathbf x=\exp(L_\chisyn) \mathbf x'$ (see below, Equations \eqref{eq:pqTop'q'} and \eqref{eq:3-2_3-2AnalyticallyAveragedActions}). 
We note that the boxed terms are simply
\begin{equation}
\left.\left[\Ha_0+ \mpl \Hares'\right]\right|_{\mathbf x'} =:\left.\bar\Ha\right|_{\mathbf x'},
\end{equation}
that is $\bar\Ha$ written in the new variables $\mathbf x'$ via direct substitution. Recall that $\bar\Ha$ does not depend on $\delta\LAMBDA_{1,2}$ and so $\Delta\LAMBDA_{1,2}$ was a first integral; hence only the ``averaged variables'' $\bar{\mathbf x}'=({\PSIONA_1^{(1)}}',{\PSIONA_1^{(2)}}',{\Delta\GAMMA_{1,2}}',\Delta\GAMMA_{2,3}',{\PSI_1^{(1)}}',{\PSI_1^{(2)}}',\delta\GAMMA_{1,2}',\delta\GAMMA_{2,3}')$ remain as evolving variables (as in Subsect.\ \ref{sec:StabilityN=3.subsec:PurelyResonantDynamics}, we use a barred notation $\bar{\mathbf x}'=(\bar{\mathbf p}',\bar{\mathbf q}')$ for the purely resonant variables, the subset of $\bar{\mathbf x}'$ not including $(\Delta\LAMBDA_{1,2}',\delta\LAMBDA_{1,2}')$).
Concerning the remaining two $\mathcal O(\mpl)$ terms in \eqref{eq:Hprimed}, these actually cancel out by construction, since $\chisyn$ was chosen to satisfy \eqref{eq:HomologicalEquationForchisyn}. We can, therefore, write the transformed Hamiltonian as
\begin{equation}\label{eq:HprimedAlmostSimplified}
\Ha'=\underbrace{\boxed{\left.\Hakepl\right|_{\mathbf x'} + \mpl \left.\Hares'\right|_{\mathbf x'}}}_{\left.\bar\Ha\right|_{{\mathbf x}'},~\mathcal O(\mpl)} + \mathcal O(\mpl^2):
\end{equation}
this equation shows that $\bar\Ha$, which we have studied in the previous section, approximates $\Ha'$ only to first order in $\mpl$. 
As long as the new term $\mathcal O(\mpl^2)$ does not contain resonant terms, the variables $\bar{\mathbf p}'=({\PSIONA_1^{(1)}}',{\PSIONA_1^{(2)}}',{\Delta\GAMMA_{1,2}}',\Delta\GAMMA_{2,3}')$ closely follow the equilibrium points $\bar{\mathbf p}_\eq$ calculated from $\bar\Ha$ in Subsection \ref{sec:StabilityN=3.subsec:PurelyResonantDynamics} (i.e.\ they have oscillations around $\bar{\mathbf p}_\eq$ of order $\mathcal O(\mpl^2)$, while the oscillations of $\bar{\mathbf p}$ are $\mathcal O(\mpl)$) while $\Delta\LAMBDA_{1,2}'$ undergoes oscillations of $\mathcal O(\mpl^2)$ around the initial value $\overline{\Delta\LAMBDA_{1,2}}$ (again a conserved quantity in the purely averaged model $\bar\Ha$). This is what we observe in the numerical simulations, Figure \ref{fig:3-2_3-2FullApproximatedModel}. We can therefore simplify the calculation by writing 
\begin{align}\label{eq:DefinitionOfBarchisyn'}
\chisyn'	&= \underbrace{\chisyn'(\bar{\mathbf p}'=\bar{\mathbf p}_\eq;\ANGMOM,\Delta\LAMBDA_{1,2}'=\overline{\Delta\LAMBDA_{1,2}})}_{\barchisyn'} \\
			&\quad+ \mathcal O(|(\bar{\mathbf p}'-\bar{\mathbf p}_\eq,{\Delta\LAMBDA_{1,2}}'-\overline{\Delta\LAMBDA_{1,2}})|),\nonumber
\end{align} 
(where we called $\barchisyn'$ the first term of the last equation), and dropping the higher order terms, which correspond to small deviations from $\bar{\mathbf p}'=\bar{\mathbf p}_\eq$ and from the initial value $\overline{\Delta\LAMBDA_{1,2}}$ of $\Delta\LAMBDA_{1,2}'$.
With this approximation we can eliminate the term $\mpl^2\{\Hasyn',\barchisyn'\}$ in \eqref{eq:Hprimed.subeq:3} because now $\frac{\partial\Hasyn'}{\partial p}=\frac{\partial\barchisyn'}{\partial p}=0$ so $\{\Hasyn',\barchisyn'\}=\frac{\partial\Hasyn'}{\partial q} \frac{\partial\barchisyn'}{\partial p}-\frac{\partial\Hasyn'}{\partial p} \frac{\partial\barchisyn'}{\partial q}=0$ (of course, by the same reasoning, also the higher order terms of the Lie series for $\exp(L_\barchisyn)\Hasyn$ cancel out). The resulting Hamiltonian becomes
\begin{align}\label{eq:HprimedSimplified}
\Ha'	&=\underbrace{\boxed{\left.\Hakepl\right|_{\mathbf x'} + \mpl \left.\Hares'\right|_{\mathbf x'}}}_{\left.\bar\Ha\right|_{{\mathbf x}'},~\mathcal O(\mpl)} \\
		&\quad+ \underbrace{\left[\frac{\mpl^2}{2}\left.\{\{\Hakepl,\barchisyn'\},\barchisyn'\}\right|_{\mathbf x'} + \mpl^2\left.\{\Hares',\barchisyn'\}\right|_{\mathbf x'}\right]}_{\mathcal O(\mpl^2)}+\dots.\nonumber
\end{align}

We now explicit the transformation that to $\mathcal O(\mpl)$ eliminates the fast synodic evolution in the numerical integrations. This is given by 
\begin{equation}\label{eq:pqTop'q'}
\begin{split}
\mathbf p	&= \exp(L_\barchisyn)\mathbf p' = \mathbf p' + \mpl\{\mathbf p',\barchisyn'\} +\mathcal O(\mpl^2)\\
			&= \mathbf p' - \mpl\frac{\partial\barchisyn'}{\partial \mathbf q'}+\mathcal O(\mpl^2),\\
\mathbf q	&= \exp(L_\barchisyn)\mathbf q' = \mathbf q'.
\end{split}
\end{equation}
Notice that the angles remain unchanged since $\barchisyn$ is independent of the actions, so $\frac{\partial\barchisyn'}{\partial p'}=0$. The transformation for the actions reads, to first order in $\mpl$:
\begin{align}\label{eq:3-2_3-2AnalyticallyAveragedActions}
\PSIONA_1^{(1)}	&={\PSIONA_1^{(1)}}' - \mpl\frac{1}{k^{(2)}} \frac{C_2}{\bar{\eta}_2}\cos(\delta\LAMBDA_{2,3}),\nonumber\\
\PSIONA_1^{(2)}	&={\PSIONA_1^{(2)}}' + \mpl\frac{1}{k^{(2)}} \frac{C_2}{\bar{\eta}_2}\cos(\delta\LAMBDA_{2,3}),\nonumber\\
\Delta\GAMMA_{1,2}	&=\Delta\GAMMA_{1,2}' + \mpl\frac{1}{k^{(2)}} \frac{C_2}{\bar{\eta}_2}\cos(\delta\LAMBDA_{2,3}),\\
\Delta\GAMMA_{2,3}	&=\Delta\GAMMA_{2,3}',\nonumber\\
\Delta\LAMBDA_{1,2}	&=\Delta\LAMBDA_{1,2}' -\mpl\left[\frac{C_1}{\bar{\eta}_1}\cos(\delta\LAMBDA_{1,2}) +\frac{k^{(1)}-1}{k^{(2)}}\frac{C_2}{\bar{\eta}_2}\cos(\delta\LAMBDA_{2,3})\right],\nonumber
\end{align}
where one has to replace $\delta\LAMBDA_{2,3}$ with its expression in terms of the variables \eqref{eq:CanonicalVariablesqWithSynodicAngle}, $\delta\LAMBDA_{2,3}=\frac{1}{k^{(2)}}\big((k^{(1)}-1)\delta\LAMBDA_{1,2}+\PSI_1^{(1)}-\PSI_1^{(2)}-\delta\GAMMA_{1,2}\big)$; moreover $\bar{\eta}_1$ and $\bar{\eta}_2$ are the frequencies \eqref{eq:EtaDivisorsForChisyn} evaluated at the reference values for the actions at each $\mpl$.
We can invert these expressions to obtain $\mathbf p'$ from $(\mathbf p,\mathbf q)$, and the evolution of $\mathbf p'$ represents that of $\mathbf p$ where to first order in $\mpl$ the short periodic have been averaged out. The evolution of $\mathbf p'$ is shown in orange in Figure \ref{fig:3-2_3-2FullApproximatedModel} in our reference $N=3$, $k=3$ example, where we see that initially the averaged evolution follows closely the analytical calculation of the equilibrium points of $\bar\Ha$ for different planetary masses.

This is however only valid until a point in which the $\mathcal O(\mpl^2)$ contribution, which is still present in \eqref{eq:HprimedSimplified}, has resonant effects (which happens at $\mpl/M_*\simeq 1.28\times 10^{-3}$ in Fig.\ \ref{fig:3-2_3-2FullApproximatedModel}). Indeed, as it is typical in perturbation theory, these terms are expected to contain higher-order harmonics which were not present in the original Hamiltonian $\Ha^*=\Hakepl+\Hares+\Hasyn$: then, if these newly introduced $\mathcal O(\mpl^2)$ Hamiltonian terms contains angles which, for certain values of $\mpl$, have a vanishing or small enough frequency, they could not be eliminated by a further perturbative step because of the problem of small divisors, and may thus change the dynamics considerably. We therefore proceed to analyse these terms below.

\subsubsection{The $\mathcal O(\mpl^2)$ contribution}\label{subsubsec:TheOmpl^2contribution}
In this subsection, we look closely at the $\mathcal O(\mpl^2)$ terms in \eqref{eq:HprimedSimplified}. We are specifically interested in the harmonics that they contain, to find explicitly which combinations of angles $\bar{\mathbf q}'=({\PSI_1^{(1)}}',{\PSI_1^{(2)}}',{\delta\GAMMA_{1,2}}',{\delta\GAMMA_{2,3}}')$ and $\delta\LAMBDA_{1,2}'$ can give rise to secondary resonances at values of the planetary masses close to those where the increase in amplitude of libration is observed in the numerical integrations. Since the synodic frequency of $\delta\LAMBDA_{1,2}'$ is much higher than the libration frequencies characteristic of the angles $\bar{\mathbf q}'$, the most interesting harmonics are the ones where the lowest fraction of $\delta\LAMBDA_{1,2}'$ appears next to a combinations of $\bar{\mathbf q}'$. This is because these are the harmonic terms that will be linked to the secondary resonances that appear at lowest resonant libration frequencies, that is, by Figure \ref{fig:3-2_3-2PurelyResonantModel} panels (e) and (f), at lowest planetary mass. The following calculation is clearly general, but to simplify matters we will quickly specialise to the case of a chain of three planets with both pairs in the same resonance, $k^{(1)}=k^{(2)}=k$, as well as to the reference case $k=3$ for which the numerical integrations in Figure \ref{fig:3-2_3-2Stability} were performed.\\

We start with the main term $\{\{\Hakepl,\barchisyn'\},\barchisyn'\}$ of order $\mpl^2$ in \eqref{eq:HprimedSimplified}. Since $\Hakepl$ does not contain any angles, all secondary resonance contributions must come from combinations of the harmonics contained in $\barchisyn'$. Recall that we defined $\barchisyn'$ containing both synodic terms with harmonics $\LAMBDA_1-\LAMBDA_2$ and $\LAMBDA_2-\LAMBDA_3$, which we wrote in Equation \eqref{eq:ThreePlanetsSynodicHarmonicsInx} in terms of the new variables $\mathbf q$. Therefore, the harmonics that are included in $\{\{\Hakepl,\barchisyn'\},\barchisyn'\}$ are combinations of these synodic harmonics; more specifically, they come from the products of their cosines\footnote{\label{footnote:FormulaOnFvsChiTwice1}
This can be easily understood noting that if $\chi=\sin(q_1)+\sin(q_2)$ and $f$ depends only on the actions $p_i$ then
$$\{\{f,\chi\},\chi\}=\frac{\partial^2 f}{\partial p_1^2}\cos^2 q_1+2\frac{\partial^2 f}{\partial p_1\partial p_2}\cos q_1 \cos q_2 + \frac{\partial^2 f}{\partial p_2^2}\cos^2 q_2.$$
}.
Using the standard trigonometric identity $\cos(a)\cos(b) = \frac{1}{2}\left(\cos(a-b)+\cos(a+b)\right)$, the resulting harmonics are
\begin{equation}\label{eq:PossibleSecondaryResonanceHarmonicsInq}
\begin{split}
&2 \delta\LAMBDA_{1,2}',\\
&\big((k^{(2)}+k^{(1)}-1)\delta\LAMBDA_{1,2}' + {\PSI_1^{(1)}}' - {\PSI_1^{(2)}}' - \delta\GAMMA_{1,2}' \big)/k^{(2)},\\
&\big((k^{(2)}-k^{(1)}+1)\delta\LAMBDA_{1,2}' - {\PSI_1^{(1)}}' + {\PSI_1^{(2)}}' + \delta\GAMMA_{1,2}'\big)/k^{(2)},\\
&2\big((k^{(1)}-1)\delta\LAMBDA_{1,2}' + {\PSI_1^{(1)}}' - {\PSI_1^{(2)}}' - \delta\GAMMA_{1,2}'\big)/k^{(2)},
\end{split}
\end{equation}
so the harmonic with the lowest fraction of $\delta\LAMBDA_{1,2}'$ is $\big((k^{(2)}-k^{(1)}+1)\delta\LAMBDA_{1,2}'-{\PSI_1^{(1)}}'+{\PSI_1^{(2)}}'+\delta\GAMMA_{1,2}'\big)/k^{(2)}$. Specialising now to the case of a chain with the same resonance index $k^{(1)}=k^{(2)}=k$, this simply gives
\begin{equation}\label{eq:NeededSecondaryResonanceHarmonicsInqk1=k2=k}
\frac{1}{k}\left(\delta\LAMBDA_{1,2}'-{\PSI_1^{(1)}}'+{\PSI_1^{(2)}}'+\delta\GAMMA_{1,2}'\right).
\end{equation}
With the aid of an algebraic manipulator one can compute the full expression of $\{\{\Hakepl,\barchisyn'\},\barchisyn'\}$ and select the desired harmonic term (we used the software package Wolfram Mathematica), thus obtaining its coefficient (actually, one can see that this term emerges solely from the term $\propto\{\{1/{\LAMBDONA_2^2},\barchisyn'\},\barchisyn'\}$). We avoid writing here the full expression, which is rather cumbersome, moreover as in \eqref{eq:HprimedSimplified} we evaluate it at the reference values of the actions so the term multiplying the cosine becomes a numerical coefficient, and we write this term as 
\begin{equation}\label{eq:HaSecResKepl}
\Ha_{\secres,\kepl}=\const\times \mpl^2 \cos\big((\delta\LAMBDA_{1,2}'-{\PSI_1^{(1)}}'+{\PSI_1^{(2)}}'+\delta\GAMMA_{1,2}')/k\big).
\end{equation}

Since we want to compare the frequency of $\delta\LAMBDA_{1,2}'/k$ to that of $(-{\PSI_1^{(1)}}'+{\PSI_1^{(2)}}'+\delta\GAMMA_{1,2}')/k$, we need to consider the resonant Hamiltonian $\bar\Ha$ in the $\mathbf x'$ variables, expand the ``barred'' variables $\bar{\mathbf x}'$ around the equilibrium point characterised by the equilibrium actions $\bar{\mathbf p}_\eq$ and the equilibrium angles $\bar{\mathbf q}_\eq$ (Equation \eqref{eq:EquilibriumAngles3PlanetsStability}) as in Subsect.\ \ref{sec:StabilityN=3.subsec:PurelyResonantDynamics}, and then introduce the transformation $\bar{\mathbf x}' \to (I_l,\phi_l)_{l=1,\dots,4}$ to the action-angle variables $(\mathbf I,\boldsymbol\phi)$, which transforms $\bar\Ha$ into the sum of decoupled harmonic oscillators plus higher order terms, Equation \eqref{eq:HbarInIphicoordinates}. 
 It is also useful to translate the value of $\Delta\LAMBDA_{1,2}'$ around its initial reference value $\overline{\Delta\LAMBDA_{1,2}}$ introducing $\Delta\LAMBDA_{1,2}'=\overline{\Delta\LAMBDA_{1,2}}+\delta\Delta\LAMBDA_{1,2}'$ which is clearly a canonical transformation. Therefore, we write $\Ha_{\secres,\kepl}$ in terms of the variables $(\mathbf I,\delta\Delta\LAMBDA_{1,2}',\boldsymbol\phi,\delta\LAMBDA_{1,2}')$.
The Hamiltonian $\Ha_{\secres,\kepl}$ will now contain harmonic terms of type
\begin{equation}\label{eq:NeededSecondaryResonanceHarmonicsInqk1=k2=kw.r.tResonantVariablesphi}
\sincos (\delta\LAMBDA_{1,2}'/k+\bfh\cdotp\boldsymbol\phi),
\end{equation} 
where $\bfh\cdotp\boldsymbol\phi$ is an integer combination with coefficients $h_1,\dots,h_4\in\Z$ of the angles $\phi_1,\dots,\phi_4$, which we can calculate explicitly. Therefore, whenever $\frac{\D{}}{\D t}\left(\delta\LAMBDA_{1,2}'/k\right)=-\frac{\D{}}{\D t}\left(\bfh\cdotp\boldsymbol\phi\right)$ a secondary resonance is crossed. We can rewrite this expression as $\dot{\delta\LAMBDA_{1,2}'}/k+\bfh\cdotp\boldsymbol\omega=0$. Since the Hamiltonian has d'Alembert characteristics in each pair $(I_l,\phi_l)$, and the values of the actions $\mathbf I$ are initially (that is, before their excitation) small, the strongest secondary resonances will come from lowest integer combinations $\bfh\cdotp\boldsymbol\phi$, that is, where most $h_l$ are zero. We also note that since $\dot{\delta\LAMBDA_{1,2}'}>0$ and the frequencies $\omega_l$ are all negative, a secondary resonance term can only appear when $\bfh\cdotp\boldsymbol\omega<0$, which together with the requirement that $|\bfh|$ be small is tantamount to requiring that all non-zero integers $h_l$ are positive. Since we calculated $\omega(\mpl)$ in Subsect.\ \ref{sec:StabilityN=3.subsec:PurelyResonantDynamics}, we can calculate for each $\bfh$ the relative frequency $\left(\dot{\delta\LAMBDA_{1,2}'}/k+\bfh\cdotp\boldsymbol\omega\right)(\mpl)$ as a function of $\mpl$, and check if any of these vanish for some value of $\mpl$, which corresponds to crossing a secondary resonance.\\

\begin{figure}[!t]
\centering
\includegraphics[scale=.55]{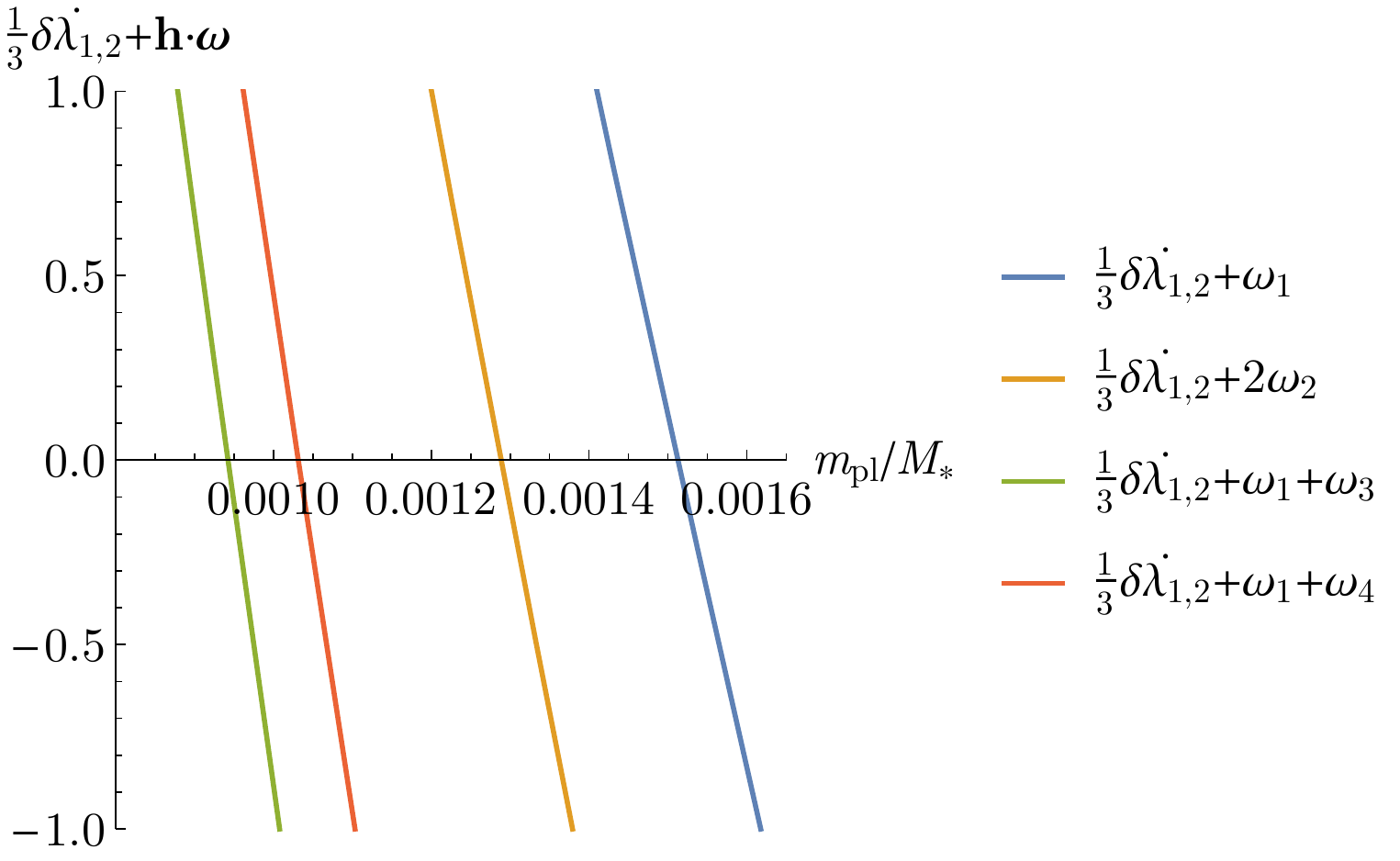}
\caption{
Frequencies of the angles $\delta\LAMBDA_{1,2}'/3+\bfh\cdotp\boldsymbol\phi$ as a function of the planetary mass in the case of the 3:2 -- 3:2 mean motion resonance chain with $a_1\simeq0.1$ (the situation depicted in Figure \ref{fig:3-2_3-2Stability.subfig:3-2_3-2_InstabOnsetOne2Plot}). Notice that the synodic frequency $\dot{\delta\LAMBDA_{1,2}'}$ varies only slightly due to the change in the equilibrium point $\bar{\mathbf x}_\eq$ for the averaged Hamiltonian $\bar\Ha$, which is followed by the full system $\Ha'$ until the second order effects become significant (cfr.\ Equation \eqref{eq:HprimedSimplified}). The main change comes from the resonant frequencies $\boldsymbol\omega$, whose dependence on the planetary mass is depicted in Figure \ref{fig:3-2_3-2PurelyResonantModel.subfig:3-2_3-2_plot4FreqLowe}. The result is that the frequencies $\left(\dot{\delta\LAMBDA_{1,2}'}/k+\bfh\cdotp\boldsymbol\omega\right)(\mpl)$ vanish within a small range of values of the planetary mass $\mpl$, meaning that a capture into a secondary resonance becomes possible. By comparing with Figure \ref{fig:3-2_3-2Stability.subfig:3-2_3-2_InstabOnsetOne2Plot}, we see that $\delta\LAMBDA_{1,2}'/3+2\phi_2$ has vanishing frequency at the same value of $\mpl/M_*\simeq\protect\input{MassAtPhenom3-2_3-2.dat}$ at which the excitation of the system occurs.
}
\label{fig:deltalambda12/3+<phi>}
\end{figure}

We carried out the calculation with the aid of the Mathematica software in the reference case $k=3$ and $a_1\simeq 0.1$, which corresponds to the evolution shown in Figure \ref{fig:3-2_3-2Stability.subfig:3-2_3-2_InstabOnsetOne2Plot} (and also Figures \ref{fig:3-2_3-2PurelyResonantModel} and \ref{fig:3-2_3-2FullApproximatedModel}). We found that $\Ha_{\secres,\kepl}$ contains, among many others, the following terms
\begin{subequations}\label{eq:PossibleSecondaryResonanceHarmonicsInphi}
\begin{align}
&1.24 \times \mpl^2 \sqrt{2 I_1} \sincos\big(\delta\LAMBDA_{1,2}'/3+\phi_1 +\phase\big),\\
\label{eq:PossibleSecondaryResonanceHarmonicsInphi.subeq:2}
&0.27 \times \mpl^2 (2 I_2)  \sincos\big(\delta\LAMBDA_{1,2}'/3+2\phi_2+\phase\big),\\
&2.39\times10^{-3}\times \mpl^2 \sqrt{2I_1}\sqrt{2I_3} \sincos\big(\delta\LAMBDA_{1,2}'/3+\phi_1+\phi_3+\phase\big),\\
&1.6 \times \mpl^2 \sqrt{2I_1}\sqrt{2I_4} \sincos\big(\delta\LAMBDA_{1,2}'/3+\phi_1+\phi_4+\phase\big).
\end{align}
\end{subequations}
The nature of these harmonics is clearly general, while the numerical coefficients are specific to the reference case $k=3$ and $a_1\simeq 0.1$ mentioned above. We then calculated for each of the harmonics in \eqref{eq:PossibleSecondaryResonanceHarmonicsInphi} their frequency $\left(\dot{\delta\LAMBDA_{1,2}'}/k+\bfh\cdotp\boldsymbol\omega\right)(\mpl)$ as a function of the mass. The results are presented in Figure \ref{fig:deltalambda12/3+<phi>}.

We immediately remark that in the case of the harmonic $\delta\LAMBDA_{1,2}'/3+2\phi_2$, the crossing of the secondary resonance happens precisely at the value of planetary mass $\mpl/M_*\simeq 1.28\times 10^{-3}$ where the numerical integrations showed the increase in amplitude of libration (see Figures \ref{fig:3-2_3-2Stability}, \ref{fig:3-2_3-2PurelyResonantModel}, \ref{fig:3-2_3-2FullApproximatedModel}). This is evidence that this phenomenon was indeed caused by the crossing of this secondary resonance.\\

Before we continue with an analytical description of the dynamics caused by this resonance, we should however go back and discuss a few technical details.

\begin{figure}[!t]
\centering
\includegraphics[scale=.71]{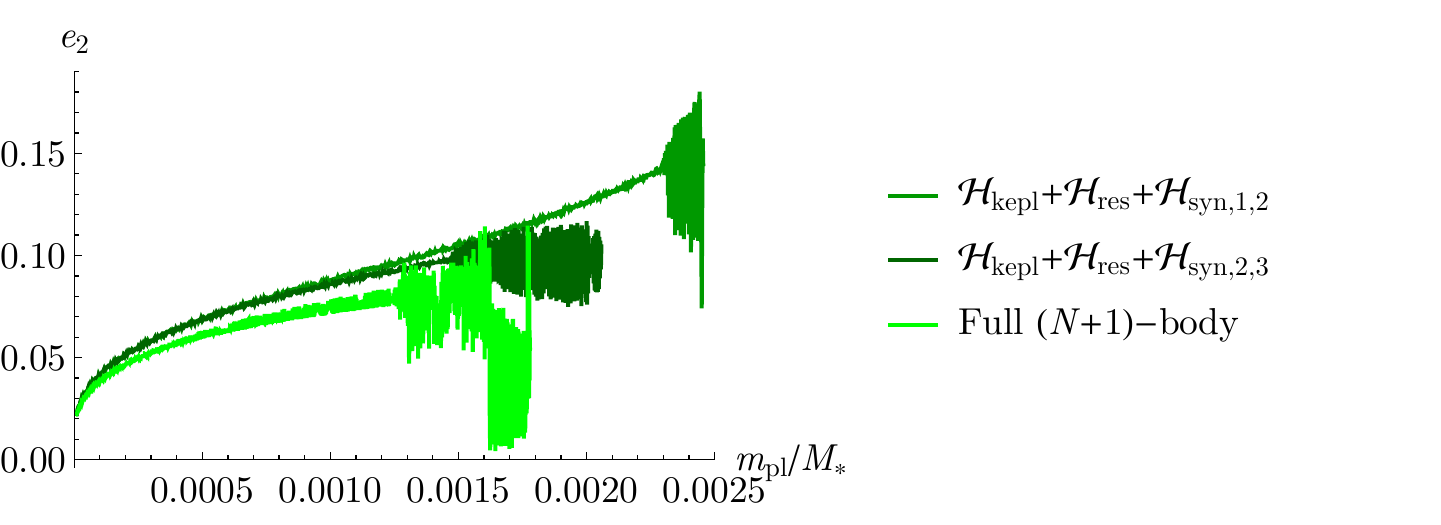}
\caption{
Comparison between the full $(N+1)$-body simulation from Figure \ref{fig:3-2_3-2Stability.subfig:3-2_3-2_InstabOnsetOne2Plot} (lightest green) and two numerical simulations with the same initial conditions where only one of the two synodic terms $\LAMBDA_1-\LAMBDA_2$ and $\LAMBDA_2-\LAMBDA_3$ appears (two darker shades of green). These two semi-synodic simulations initially appear identical, which is easily understood from the fact that they follow in average the evolution of $\bar\Ha=\Hakepl+\Hares$ which is the same for the two (cfr.\ Equation \eqref{eq:HprimedAlmostSimplified}). The important point is that in both cases, when only one synodic angle is considered, the system is not excited at value of $\mpl/M_*\simeq\protect\input{MassAtPhenom3-2_3-2.dat}$, where it is excited in the $(N+1)$-body simulation as well as in the numerical simulation which includes both synodic terms, see Figure \ref{fig:3-2_3-2Stability.subfig:3-2_3-2_InstabOnsetOne2Plot}. This shows that both synodic terms must be included in order to have a good quantitative agreement with the $(N+1)$-body simulations.}
\label{fig:3-2_3-2_NoInstabOnsetOne2SpecificSynPlot}
\end{figure}

Firstly, if we had used in $\Hasyn$ only one synodic term, not all of the the harmonics in \eqref{eq:PossibleSecondaryResonanceHarmonicsInq} would appear\footnote{
To see this, as in footnote \ref{footnote:FormulaOnFvsChiTwice1} we calculate for $\chi=\sin(q_1)$ and $f$ which depends on the actions only,
$$\{\{f,\chi\},\chi\}=\frac{\partial^2 f}{\partial p_1^2}\cos^2 q_1 = \frac{1}{2}\frac{\partial^2 f}{\partial p_1^2} (1+\cos(2q_1)).$$
Clearly we do not obtain the needed $\big((k^{(2)}-k^{(1)}+1)\delta\LAMBDA_{1,2}'-{\PSI_1^{(1)}}'+{\PSI_1^{(2)}}'+\delta\GAMMA_{1,2}'\big)/k^{(2)}$ in \eqref{eq:PossibleSecondaryResonanceHarmonicsInq} neither when $q_1=\LAMBDA_1-\LAMBDA_2=\delta\LAMBDA_{1,2}$ nor when $q_1=\LAMBDA_2-\LAMBDA_3=\frac{1}{k^{(2)}}\big((k^{(1)}-1)\delta\LAMBDA_{1,2}+\PSI_1^{(1)}-\PSI_1^{(2)}-\delta\GAMMA_{1,2}\big)$.}.
In particular, the harmonic $\delta\LAMBDA_{1,2}'/3+2\phi_2$ would not appear, so that the observed dynamical effects linked to the crossing of secondary resonances at $\mpl/M_*\simeq 1.28\times 10^{-3}$ are not expected. Indeed, we performed similar numerical integrations with only one of the synodic terms, $\LAMBDA_1-\LAMBDA_2$ and separately $\LAMBDA_2-\LAMBDA_3$, which are shown in Figure \ref{fig:3-2_3-2_NoInstabOnsetOne2SpecificSynPlot}, and there is no effect at the right value of $\mpl$. Secondary resonances do occur, but at larger values of $\mpl$, given that the generated harmonics have a larger coefficient for $\delta\LAMBDA_{1,2}'$. 

Secondly, so far we have not considered the $\mathcal O(\mpl^2)$ term $\{\Hares',\barchisyn'\}$, which is also present in \eqref{eq:HprimedSimplified}. However, with the same technique as above one can see that this term only yields harmonics of type
\begin{equation}
\begin{split}
&\delta\LAMBDA_{1,2}' \pm {\PSI_1^{(1)}}',\\
&\delta\LAMBDA_{1,2}' \pm {\PSI_1^{(2)}}',\\
&\pm\delta\LAMBDA_{1,2}' - {\PSI_1^{(1)}}' + \delta\GAMMA_{1,2}',\\
&\left((k^{(1)}-1)\delta\LAMBDA_{1,2}' - (\pm k^{(2)}-1){\PSI_1^{(1)}}' - {\PSI_1^{(2)}}' + (\pm k^{(2)}-1)\delta\GAMMA_{1,2}'\right)/k^{(2)},\\
&\left((k^{(1)}-1)\delta\LAMBDA_{1,2}' + {\PSI_1^{(1)}}' - (\pm k^{(2)}+1){\PSI_1^{(2)}}' -\delta\GAMMA_{1,2}'\right)/k^{(2)},\\
&\left((k^{(1)}-1)\delta\LAMBDA_{1,2}' + {\PSI_1^{(1)}}' - (\pm k^{(2)}+1){\PSI_1^{(2)}}' -\delta\GAMMA_{1,2}'\pm k^{(2)}\delta\GAMMA_{2,3}'\right)/k^{(2)}.\\
\end{split}
\end{equation}
Whenever $k^{(1)}\geq 3$, as in our reference case $k^{(1)}=k^{(2)}=k=3$, this does not contribute the needed harmonic \eqref{eq:NeededSecondaryResonanceHarmonicsInqk1=k2=k} with $\delta\LAMBDA_{1,2}$ appearing as a single $\delta\LAMBDA_{1,2}/k$; it will only include multiples of $\delta\LAMBDA_{1,2}/k$ and therefore to lowest order does not contribute to the secondary resonance harmonics in \eqref{eq:PossibleSecondaryResonanceHarmonicsInphi}.

Finally, in \eqref{eq:HprimedSimplified} we used the simplification $\bar{\mathbf p}'=\bar{\mathbf p}_\eq$, ${\Delta\LAMBDA_{1,2}}'-\overline{\Delta\LAMBDA_{1,2}}$ to define $\barchisyn'$ (cfr.\ Equation \eqref{eq:DefinitionOfBarchisyn'}). However, the remaining terms of $\mathcal O(|(\bar{\mathbf p}'-\bar{\mathbf p}_\eq,{\Delta\LAMBDA_{1,2}}'-\overline{\Delta\LAMBDA_{1,2}})|)$ do not contribute to the dynamics to lowest order. Indeed, concerning $\mpl\left.\{\Ha_0,\chi\}\right|_{\mathbf x'}$, this term only contains the two separate synodic harmonics already contained in $\chi$ and therefore does not yield terms linked to secondary resonances. Finally, the remaining terms in $\frac{\mpl^2}{2}\left.\{\{\Hakepl,\chisyn'\},\chisyn'\}\right|_{\mathbf x'}$ will only yield higher order terms in the actions $\mathbf I$, so we can neglect them (recall that initially the values of the actions are small since we are close to the equilibrium point). \\

\begin{figure*}[!t]
\centering
\includegraphics[scale=.7]{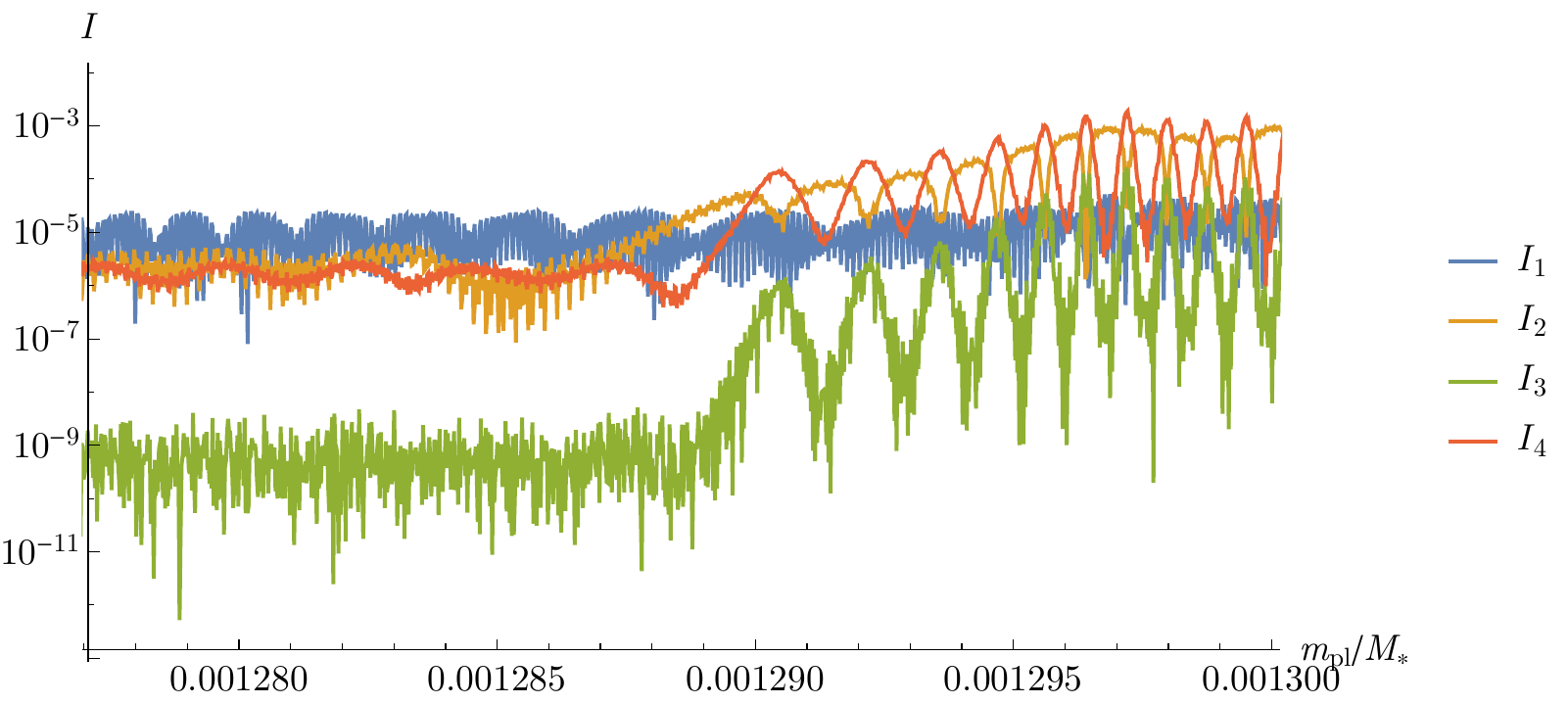}
\caption{
Evolution around $\mpl/M_*\simeq\protect\input{MassAtPhenom3-2_3-2.dat}$ of the actions $I_l$, $l=1,\dots,4$ along the reference numerical simulation shown in Figure \ref{fig:3-2_3-2Stability.subfig:3-2_3-2_InstabOnsetOne2Plot}. We see that the actions are initially relatively constant, and the system is well approximated by a Hamiltonian of the form $\sum_{l=1}^4\omega_l I_l$ (cfr.\ Equation \eqref{eq:HbarInIphicoordinates}). Then, $I_2$ increases steadily, symptom of an interaction with a secondary resonance that involves $\Theta=I_2$ as a resonant action; this is confirmed by the canonical change of coordinates \eqref{eq:TransformationForSecondaryResonancedeltalambda12+2phi2}. Soon after $I_2$ is large enough, the degrees of freedom start interacting and exchanging energy, due to the non-linear effects. 
}
\label{fig:3-2_3-2_IevoArrayFSR}
\end{figure*}

With these clarifications, we can proceed with the model of the secondary resonance linked to the angle $\delta\LAMBDA_{1,2}'/3+2\phi_2$, which, as we discussed above, has vanishing frequency exactly at the value of $\mpl$ when the increase in the amplitude of libration is observed in Figure \ref{fig:3-2_3-2Stability.subfig:3-2_3-2_InstabOnsetOne2Plot}. This realisation is further supported by Figure \ref{fig:3-2_3-2_IevoArrayFSR}. There, we plot the evolution of the actions $I_l$, $l=1,\dots,4$ along the simulation, with the planetary mass $\mpl$ on the horizontal axis. We see that initially only one action is excited, namely $I_2$, and after that the nonlinearities inherent in the system cause an exchange of energy between the degrees of freedom. This also suggests that the model that we are about to construct, which is valid only for small $\mathbf I$'s, breaks down whenever one of the actions is excited. This however presents no impediment in the description of the first phase, when the secondary resonance is encountered. One question that we wish to answer for example is whether or not there is or can be a capture in this secondary resonance or rather a jump across resonance. The integrable, low order model that we construct below can indeed answer this question.

\subsubsection{Model of the secondary resonance for $\delta\LAMBDA_{1,2}'/3+2\phi_2$}\label{subsubsec:ModelForSecondaryResonancedeltalambda12+2phi2}
In the following we detail how we can construct a model for the resonance associated with the angle $\delta\LAMBDA_{1,2}'/3+2\phi_2$ since, as we saw before, it is the one that causes the observed increase in amplitude of libration. A similar approach can be implemented for the other resonances in \eqref{eq:PossibleSecondaryResonanceHarmonicsInphi}.

We start by performing a canonical transformation which selects $\delta\LAMBDA_{1,2}'/3+2\phi_2$ as an angle. Notice that, since $\phi_2$ appears with a coefficient 2 and so $\sqrt{2I_2}$ appears as a power two in \eqref{eq:PossibleSecondaryResonanceHarmonicsInphi.subeq:2}, we have a secondary resonance of order 2; hence it is useful to define the resonant angle $\theta$ as $2\theta=\delta\LAMBDA_{1,2}'/3+2\phi_2$ in order to maintain the d'Alembert characteristics so that the Hamiltonian will not be singular at the origin. The resulting transformation is
\begin{alignat}{2}\label{eq:TransformationForSecondaryResonancedeltalambda12+2phi2}
&\Theta=I_2,						&&\quad \theta=\delta\LAMBDA_{1,2}'/6+\phi_2, \nonumber\\
&I^*_r=I_r,~r=1,3,4,					&&\quad \phi^*_r=\phi_r,~r=1,3,4,\\
&\delta\Delta\LAMBDA_{1,2}^*=\delta\Delta\LAMBDA_{1,2}' - I_2/6
								&&\quad \delta\LAMBDA_{1,2}^*=\delta\LAMBDA_{1,2}',\nonumber
\end{alignat}
whose canonicity follows immediately from the preservation of the Poisson brackets. 
We can already notice that $\Theta=I_2$ appears as the conjugated action to the angle $\theta$ associated to the secondary resonance: this explains why in Figure \ref{fig:3-2_3-2_IevoArrayFSR} it is $I_2$ which is initially excited. The other variables do not feel the resonance, except $\delta\Delta\LAMBDA_{1,2}'$ which must change according to the change in $I_2$ in order to maintain $\delta\Delta\LAMBDA_{1,2}^*$ constant; however since $I_2$ gets divided by 6 this change is minute, but nevertheless clearly visible in Figure \ref{fig:3-2_3-2FullApproximatedModel.subfig:DeltaLambda12}.
The pair $(\Theta,\theta)$ is the pair of resonant variables for this specific secondary resonance, while the others will have a faster evolution, which can be ``averaged'' away, in order to yield a 1-d.o.f.\ system that we write $\langle\Ha'\rangle_{(\boldsymbol\phi^*,\delta\LAMBDA_{1,2}^*)} (\Theta,\theta;\mathbf I^*,\delta\Delta\LAMBDA_{1,2}^*)$. The notation $\langle\bullet\rangle_{(\boldsymbol\phi^*,\delta\LAMBDA_{1,2}^*)}$ means that we eliminated perturbatively to lowest order the non-secondary-resonant contributions from the angles $(\boldsymbol\phi^*,\delta\LAMBDA_{1,2}^*)$, and we stressed that the variables $(\mathbf I^*,\delta\Delta\LAMBDA_{1,2}^*)$ will only play the role of parameters for $\langle\Ha'\rangle_{(\boldsymbol\phi^*,\delta\LAMBDA_{1,2}^*)}$.
Ultimately, the functional form of $\langle\Ha'\rangle_{(\boldsymbol\phi^*,\delta\LAMBDA_{1,2}^*)}(\Theta,\theta)$ will be that of a Andoyer Hamiltonian, that is
\begin{equation}\label{eq:AndoyerHamiltoniankForStabilityOfChains-deltalambda12+2phi2}
\langle\Ha'\rangle_{(\boldsymbol\phi^*,\delta\LAMBDA_{1,2}^*)}(\Theta,\theta)=\delta \Theta + \frac{\beta}{2} \Theta^2 + \mathcal O(\Theta^3) +c{(\sqrt{2\Theta})}^2 \cos (2\theta);
\end{equation} 
the coefficient $c$ will be of order $\mpl^2$, while $\delta$ and $\beta$ will be of order $\mpl$. Since the system is initially close to the resonant equilibrium point, $\Theta$ is small and we can drop the $\mathcal O(\Theta^3)$ terms. However, as we will see below, the parameter $\beta$ (the second derivative at $\Theta=0$) plays a crucial role in determining if there can be capture into the secondary resonance or not, so we must keep track of all $\mathcal O(\Theta^2)$ terms of the $\theta$-independent part, that is, the first two terms in \eqref{eq:AndoyerHamiltoniankForStabilityOfChains-deltalambda12+2phi2}. The main contribution to the $\theta$-independent part comes from the $\bar\Ha$ term (the $\mathcal O(\mpl)$ term in \eqref{eq:HprimedSimplified}), while $c\sqrt{2\Theta}^2 \cos (2\theta)$ comes from \eqref{eq:PossibleSecondaryResonanceHarmonicsInphi.subeq:2} and is $\mathcal O(\mpl^2)$.
Concerning the first part deriving from $\bar\Ha$, we should stress that even if $\delta\Delta\LAMBDA_{1,2}'$ appeared as a constant of motion when this Hamiltonian was treated alone, when the $\mathcal O(\mpl^2)$ is taken into account the transformation \eqref{eq:TransformationForSecondaryResonancedeltalambda12+2phi2} transforms $\delta\Delta\LAMBDA_{1,2}'$ into $\delta\Delta\LAMBDA^*_{1,2}+\Theta/6$, where $\delta\Delta\LAMBDA^*_{1,2}$ is the new constant of motion. Therefore we must keep $\delta\Delta\LAMBDA_{1,2}'$ as a variable in $\bar\Ha$ and apply \eqref{eq:TransformationForSecondaryResonancedeltalambda12+2phi2} to it. \\

With these considerations in mind we can obtain analytical insights on the dynamics, at least as long as the actions remain small (recall that before any secondary resonance is encountered, the system is very close to the equilibrium point at vanishing amplitude of libration). It is interesting for example to explore analytically if there can be a capture in this secondary resonance or not. 
Capture into resonance is possible (but not guaranteed) only if $\dot\delta\beta<0$. Intuitively, this is because near the origin one has $\dot\theta\simeq\delta+\beta\Theta$ and the resonance condition imposes that (on average) this quantity vanishes; therefore, at the centre of the resonance $\Theta=-\delta/\beta$, which only makes sense when $\beta$ and $\delta$ have opposite signs. Thus, since $\beta$ remains relatively constant (see below), only when $\dot\delta\beta$ is negative does the resonance centre appear from the origin and move at higher values of $\Theta$, while if $\dot\delta\beta$ is positive the resonance centre approaches the origin from far away, the orbit is invested by a separatrix and then the resonance disappears leaving behind an excited orbit. We already know from Figure \ref{fig:deltalambda12/3+<phi>} that, as the planetary mass increases, $\dot\theta$ goes from positive values to negative values, that is, that $\dot\delta <0$: this means a capture into this secondary resonance is possible only if $\beta>0$.

To obtain the sign of $\beta$ in \eqref{eq:AndoyerHamiltoniankForStabilityOfChains-deltalambda12+2phi2} 
we need to compute its value explicitly. We do this in steps as follows. First, we fix a value of $\mpl$ right before the observed increase of amplitude of libration, $\mpl/M_*\simeq 1.28\times 10^{-3}$, we calculate the equilibrium point $\bar{\mathbf x}_\eq=\bar{\mathbf x}_\eq (\mpl)$ and we apply the canonical diagonalisation procedure as explained in Subsect.\ \ref{sec:StabilityN=3.subsec:PurelyResonantDynamics}. This yields four pairs of cartesian canonical variables $(\boldsymbol\xi,\boldsymbol\eta)$ which replace the $\bar{\mathbf x}$: $\bar{\mathbf x}=T(\boldsymbol\xi,\boldsymbol\eta)$, with $T$ the diagonalasing matrix. Second, as in Subsect.\ \ref{sec:StabilityN=3.subsec:PurelyResonantDynamics}, we introduce canonical polar coordinates $(I_l,\phi_l)_{l=1,\dots,4}$ by $\left(\xi_l=\sqrt{2I_l}\cos\phi_l,~\eta_l=\sqrt{2I_l}\sin\phi_l\right)$. The Hamiltonian $\bar\Ha$ will then depend on the variables $(I_1,\dots,I_4,\delta\Delta\LAMBDA_{1,2}',\phi_1,\dots,\phi_4,\delta\LAMBDA_{1,2}')$. 
Third, we write $\bar\Ha$ in the variables \eqref{eq:TransformationForSecondaryResonancedeltalambda12+2phi2}; 
$\bar\Ha$ contains a term in $\Theta^2$ independent of the angles, but its coefficient is not $\beta$. To obtain the value of $\beta$ we need to perform a fourth step, and calculate $\langle\bar\Ha\rangle_{(\boldsymbol\phi^*,\delta\LAMBDA_{1,2}^*)}$, that is the perturbative elimination in $\bar\Ha$ of all the non-secondary-resonant contributions from the angles $(\boldsymbol\phi^*,\delta\LAMBDA_{1,2}^*)$, up to order 2 in $\Theta$. 
This is because, as detailed below, the elimination of these harmonics can generate terms in $\Theta^2$ independent of the angles, that need to be added to the original term to obtain $\beta$.   
To this end, we take $\bar\Ha(I_1,\dots,I_4,\delta\Delta\LAMBDA_{1,2}',\phi_1,\dots,\phi_4,\delta\LAMBDA_{1,2}')$ and expand it to order 2 with respect to the actions.
Since these terms satisfy the d'Alembert characteristics in $(\mathbf I,\boldsymbol\phi)$, we only obtain terms like
\begin{align}\label{eq:TypicalTermInHbarInIphideltaDeltalambda12}
&c_\bfalpha(\delta\Delta\LAMBDA_{1,2}') \times \sqrt{2\mathbf I}^\bfalpha\cos(\bfm\phi),\\
&\quad m_j=-\alpha_j,-\alpha_j+2,\dots,\alpha_j-2,\alpha_j, ~|\bfalpha|=1,2,3,4,\nonumber
\end{align}
where $\bfalpha\in\Nat_0^{4}$, $|\bfalpha|=\alpha_1+\dots+\alpha_4$ is restricted to $|\bfalpha|/2\le2$, 
and $c_\bfalpha(\delta\Delta\LAMBDA_{1,2}')$ is a coefficient which depends on $\delta\Delta\LAMBDA_{1,2}'$ only. These coefficients are expanded around $\delta\Delta\LAMBDA_{1,2}'=0$ to an optimal order which can be obtained in the following manner. Note that, from $\delta\Delta\LAMBDA_{1,2}'=\delta\Delta\LAMBDA_{1,2}^*+\Theta/6$ (Equation \eqref{eq:TransformationForSecondaryResonancedeltalambda12+2phi2}) each term of order $d$ in $\delta\Delta\LAMBDA_{1,2}'$ contributes a term of order $d$ in $\Theta$, so for each term of order $|\bfalpha|/2$ in $\mathbf I$ we must obtain $c_\bfalpha(\delta\Delta\LAMBDA_{1,2}')$ only up to order $\floor{2-|\bfalpha|/2}$ 
in $\delta\Delta\LAMBDA_{1,2}'$ (where $\floor{\bullet}$ is the floor function) to achieve the desired second order
with respect to all the actions.
We can then organise all terms with respect to the order of expansion in $\mathbf I$ and $\delta\Delta\LAMBDA_{1,2}'$, and write for each addend $\bar\Ha_{s/2}^j = \mathcal O(\mathbf I^{s/2}) \times \mathcal O({\delta\Delta\LAMBDA_{1,2}'}^j)$.
To get a sense of what these terms look like, we write the terms only up to order $s=2$ in $\sqrt{\mathbf I}$:
\begin{equation}
\begin{matrix}
\bar\Ha_{1/2}^0\propto\sqrt{2I_l}\cos(\phi_l), & \bar\Ha_{1/2}^1\propto\delta\Delta\LAMBDA_{1,2}' \sqrt{2 I_l}\cos(\phi_l),\\ 
\bar\Ha_{1}^0\propto I_l, & \bar\Ha_{1}^1\propto\delta\Delta\LAMBDA_{1,2}' \sqrt{2I_{l_1}}\sqrt{2I_{l_2}}\cos(\phi_{l_1}\pm\phi_{l_2});
\end{matrix}
\end{equation}
the subsequent terms of higher order in $\sqrt{\mathbf I}$ follow this structure but the possible combinations of the angles get substantially more numerous and we avoid writing them all here in the interest of brevity. Among them, there are of course also the terms $\propto  I_l^2$ appearing without angles, as well as the term $\propto {\delta\Delta\LAMBDA_{1,2}'}^2$, which contribute directly to the $\Theta^2$ term in \eqref{eq:AndoyerHamiltoniankForStabilityOfChains-deltalambda12+2phi2}.
We note that the first terms $\bar\Ha_{1/2}^0\propto\sqrt{2I_l}\cos(\phi_l)$ (corresponding to $|\bfalpha|=1$ and constant in $\delta\Delta\LAMBDA_{1,2}'$) are actually zero by definition of equilibrium point calculated at the reference value $\delta\Delta\LAMBDA_{1,2}'=0$. 
Also, the coefficients in $\bar\Ha_{1}^0$ in front of the $I_l$'s are just the frequencies $\omega_l$, since these are the ones calculated in \eqref{eq:HbarInIphicoordinates}. Therefore, in this case the role of the integrable part of the Hamiltonian for a perturbation theory step is naturally played by $\bar\Ha_{1}^0=\sum_{l=1}^4\omega_l I_l$.

To understand how the perturbative elimination of the non-resonant harmonics involving $(\boldsymbol\phi^*,\delta\LAMBDA_{1,2}^*)$ can generate terms in $\Theta^2$ independent of the angles, consider that if $f_n=\mathcal O(\mathbf I^n)$ and $\chi_m=\mathcal O(\mathbf I^m)$, then $\{f_n,\chi_m\}=\mathcal O(\mathbf I^{n+m-1})$, and so on for all the other terms of the Lie series: $\{\{f_n, \chi_m\},\chi_m\} = \mathcal O(\mathbf I^{n+2m-2})$ etc. 
When we eliminate a $\bar\Ha_{n,\pert}^j=\mathcal O(\mathbf I^{n})$ term by solving the homological equation $\{\bar\Ha_{1}^0,\chi_m\}+\bar\Ha_{n,\pert}^j=0$, we must naturally use a $\chi_n=\mathcal O(\mathbf I^{n})$. 
This introduces new terms $\{\{\bar\Ha_1^0,\chi_n\},\chi_n\}$ and $\{\bar\Ha_{n,\pert}^j,\chi_n\}$ which are $\mathcal O(\mathbf I^{2n-1})$ (given that $\bar\Ha_1^0$ is $\mathcal O(\mathbf I)$). Thus, terms of order 2 can be generated for example if $n=3/2$. We actually need to calculate explicitly only those terms that yield a $\Theta^2$ independent of the angles, as the others would be eliminated further. Such terms derive from $\bar\Ha_{3/2}^0$ (which governs the non-linear interactions between the four resonant degrees of freedom around the equilibrium point, which Figure \ref{fig:3-2_3-2_IevoArrayFSR} proves to be strong) and $\bar\Ha_{1/2}^1$ (which describes the fact that 
the equilibrium point $\bar{\mathbf p}_\eq$ of $\bar\Ha$ shifts as $\delta\Delta\LAMBDA_{1,2}'$ changes under the effects of the $O(\mpl^2)$ terms).

We implemented this procedure with the aid of the algebraic manipulator Mathematica. In our reference case $k=3$ and $a_1\simeq 0.1$ at a mass $\mpl/M_*\simeq 1.28\times 10^{-3}$ right before the development of the excitation of the resonant degrees of freedom (Figure \ref{fig:3-2_3-2_IevoArrayFSR}) this yields 
\begin{equation}\label{eq:barHaAveragedOverphi*anddeltalambda*-deltalambda12+2phi2}
\langle\bar\Ha\rangle_{(\boldsymbol\phi^*,\delta\LAMBDA_{1,2}^*)}(\Theta)= \delta \Theta+\frac{\beta}{2} \Theta^2,\quad \delta\simeq7.74\times 10^{-3},~\beta\simeq101.
\end{equation} 
The fact that $\delta$ is positive and small is consistent with the fact that we put ourselves right before the development of the excitation (cfr.\ Figure \ref{fig:deltalambda12/3+<phi>}). The fact that $\beta\sim100$ is positive yields an analytical confirmation that there can be capture into this secondary resonance. 

After we have obtained $\langle\bar\Ha\rangle_{(\boldsymbol\phi^*,\delta\LAMBDA_{1,2}^*)}$ (cfr.\ Equation \eqref{eq:barHaAveragedOverphi*anddeltalambda*-deltalambda12+2phi2}), we can easily complete the determination of the model \eqref{eq:AndoyerHamiltoniankForStabilityOfChains-deltalambda12+2phi2} for this secondary resonance. To do this, with the help of the algebraic manipulator Mathematica we use the canonical transformation \eqref{eq:TransformationForSecondaryResonancedeltalambda12+2phi2} applied to the term \eqref{eq:PossibleSecondaryResonanceHarmonicsInphi.subeq:2} which contains the resonant harmonic $2\theta$, and we obtain 
\begin{align}\label{eq:HaAveragedOverphi*anddeltalambda*-deltalambda12+2phi2}
\langle\Ha'\rangle_{(\boldsymbol\phi^*,\delta\LAMBDA_{1,2}^*)}(\Theta,\theta)	&=\delta \Theta+\frac{\beta}{2} \Theta^2+c{(\sqrt{2\Theta})}^2 \cos (2\theta+2\pi/6),\nonumber\\
																				&\quad \delta\simeq7.74\times 10^{-3},~\beta\simeq101,~c\simeq -7.8\times10^{-4}.
\end{align}
A phase is introduced which does not change the dynamics and could easily be eliminated by a simple rotation.
\begin{figure*}[!t]
\centering
\includegraphics[scale=.6]{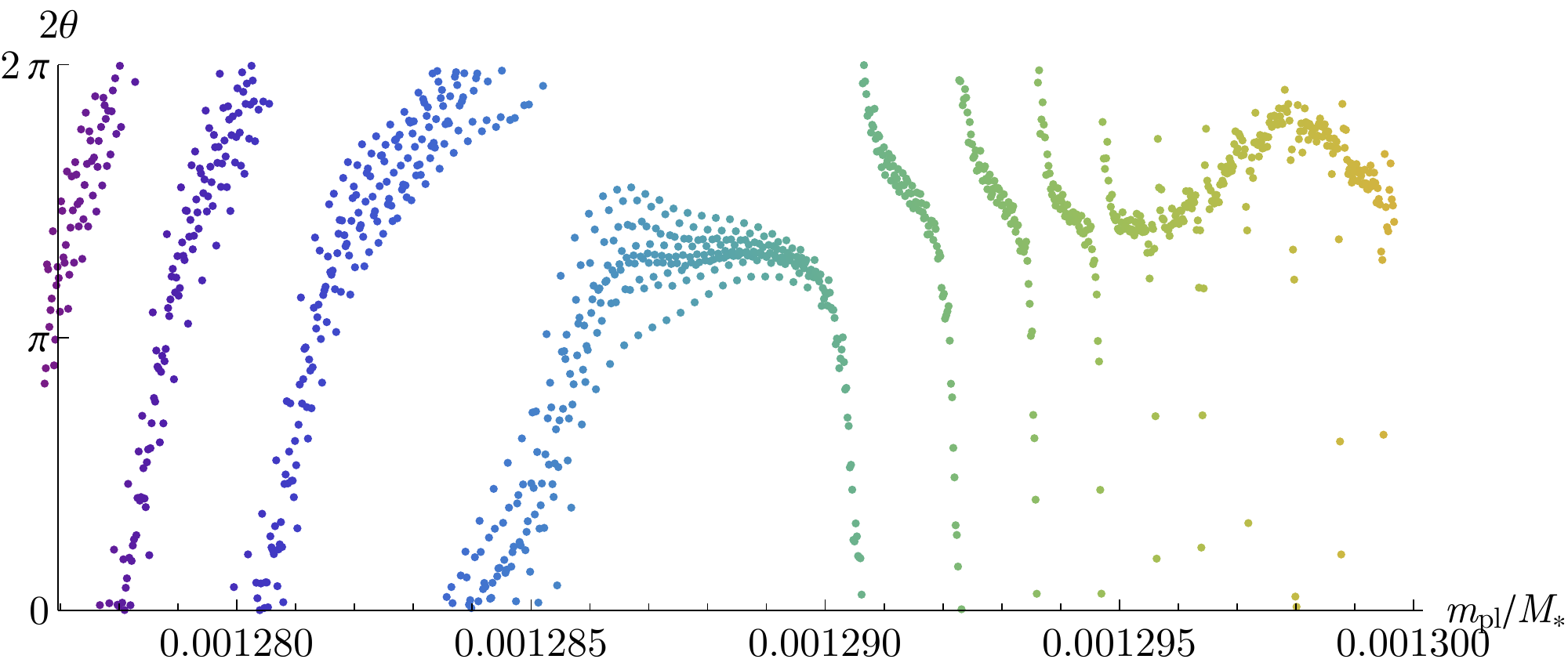}
\caption{Evolution around $\mpl/M_*\simeq\protect\input{MassAtPhenom3-2_3-2.dat}$ of the angle $\theta$, the resonant angle for the secondary resonance encountered (we plot $2\theta$ instead of $\theta$, as explained in the main text), in panel (a). This figure should be compared to Figure \ref{fig:3-2_3-2_IevoArrayFSR}: the action conjugated to $\theta$ is $\Theta=I_2$ (Equation \eqref{eq:TransformationForSecondaryResonancedeltalambda12+2phi2}), and we see that when $\theta$ start librating $I_2$ increases, indicating that the system has captured into this secondary resonance. As in Figure \ref{fig:3-2_3-2_IevoArrayFSR}, after the actions get excited the integrable approximation to the dynamics is not valid anymore.
The colour of the dots in this figure only serve as a legend for the value of the planetary mass: we use the same colour-coding in Figure \ref{fig:3-2_3-2_FigureForCaptureFSR}, where we take snapshots of the evolution of the pair $(\Theta,\theta)$ at different values of $\mpl$.
}
\label{fig:3-2_3-2_twothetaevoFSR}
\end{figure*}
We can now compare the evolution predicted by this model to the numerical integration of $\Ha^*=\Hakepl+\Hares+\Hasyn$.
The evolution of the action $\Theta=I_2$ is already shown in Figure \ref{fig:3-2_3-2_IevoArrayFSR}. We plot in Figure \ref{fig:3-2_3-2_twothetaevoFSR} the evolution of the angle $2\theta$ (which produces a numerical evolution that is graphically more legible than that of $\theta$). 
One can see that the angle starts librating at the same value of $\mpl$ where the conjugated action $\Theta=I_2$ starts increasing in Figure \ref{fig:3-2_3-2_IevoArrayFSR}: this shows that there is a passage across the resonance. Note that in such dynamics, the orbit finds itself close to the separatrix after the passage through the resonance, the adiabatic principle is not applicable and the orbit can end up in the inner circulation region (in any case, when the higher order interaction terms between the actions become too strong, a simple description of the dynamics becomes hopeless).
In order to get a better sense of the dynamical interaction with this secondary resonance, we can fix different values for $\delta$ in \eqref{eq:HaAveragedOverphi*anddeltalambda*-deltalambda12+2phi2} and look at the corresponding phase diagrams. Notice that changing $\delta$ essentially corresponds to changing $\mpl$; we also checked that for different planetary masses near $\mpl/M_*\simeq 1.28\times 10^{-3}$ the coefficients $\beta$ and $c$ do not change considerably, so we keep them fixed to obtain a qualitatively correct description of the dynamical portraits. 

\begin{figure*}[!t]
\centering
\begin{subfigure}[b]{0.275 \textwidth}
\centering
\includegraphics[width=1.1 \textwidth]{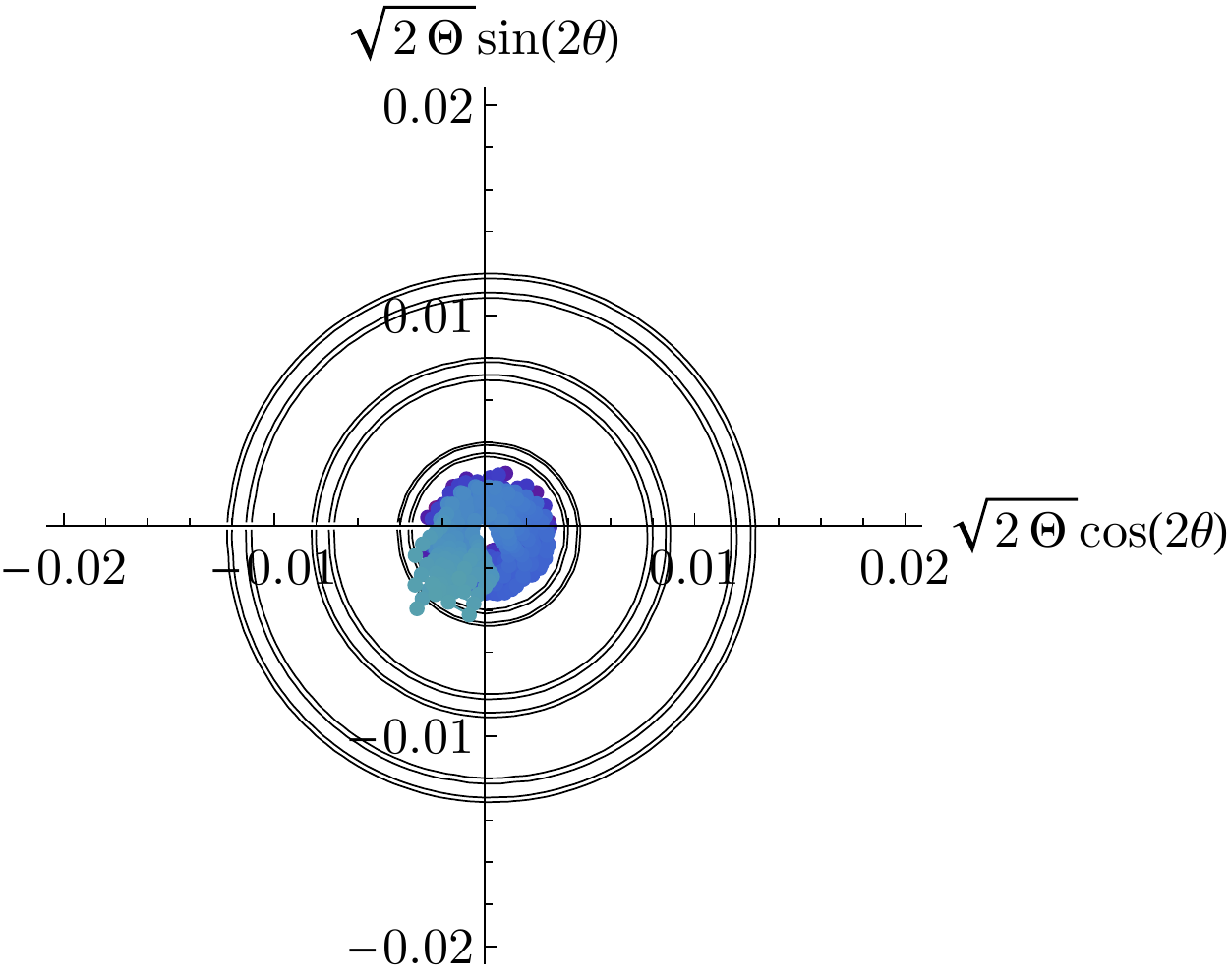}
\caption{}
\end{subfigure}
\qquad
\centering
\begin{subfigure}[b]{0.275 \textwidth}
\centering
\includegraphics[width=1.1 \textwidth]{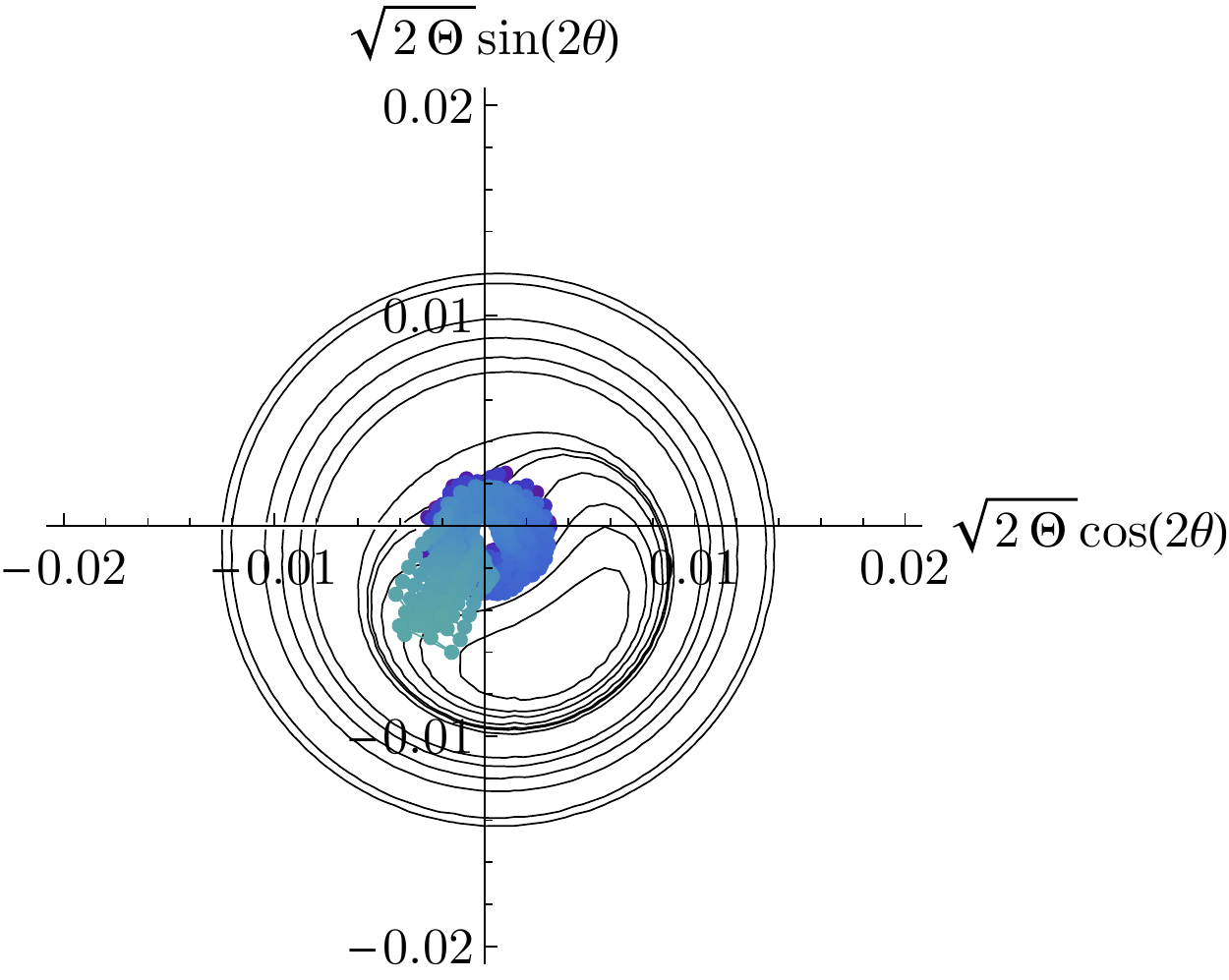}
\caption{}
\end{subfigure}
\qquad
\centering
\begin{subfigure}[b]{0.275 \textwidth}
\centering
\includegraphics[width=1.1 \textwidth]{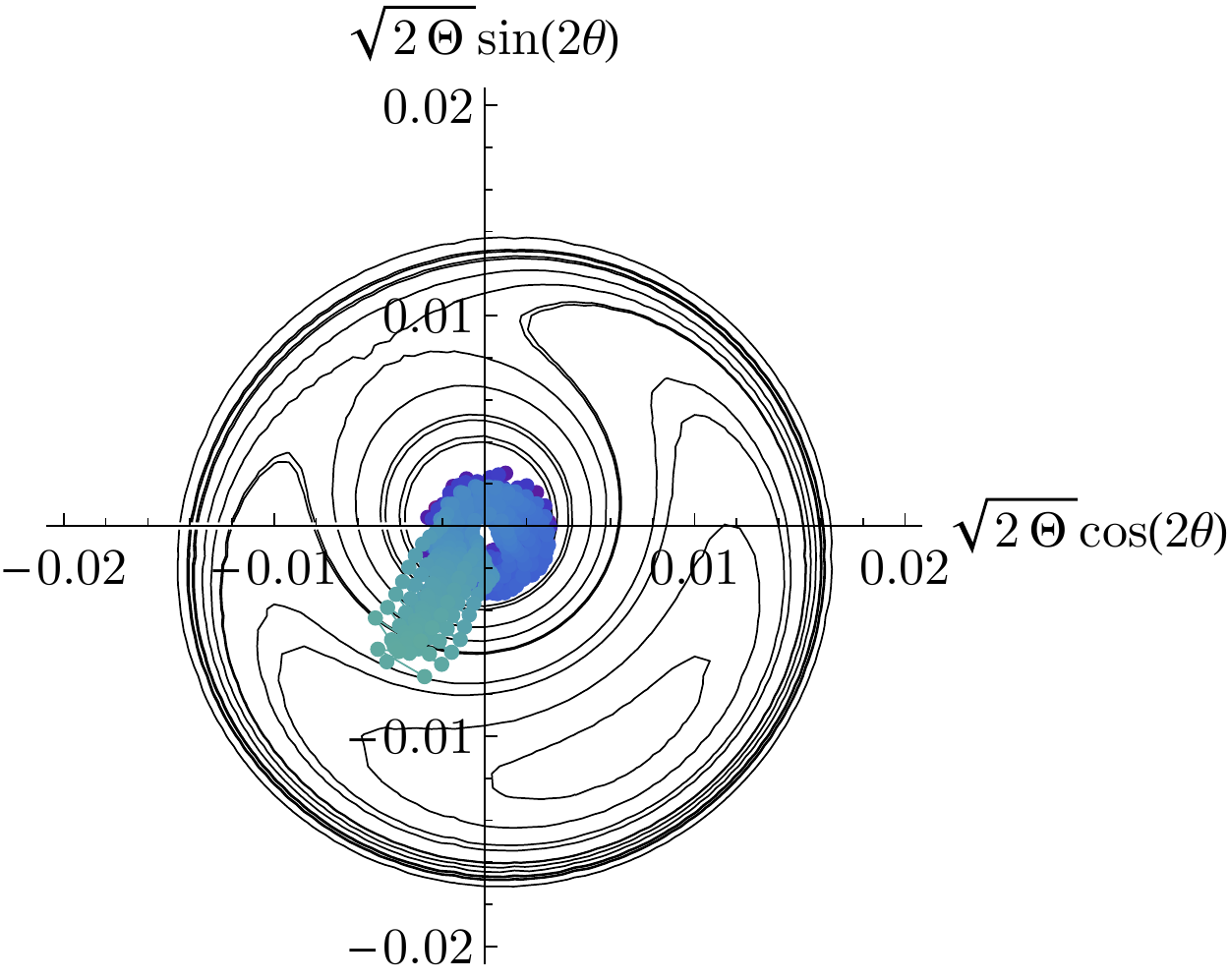}
\caption{}
\end{subfigure}

\begin{subfigure}[b]{0.275 \textwidth}
\centering
\includegraphics[width=1.1 \textwidth]{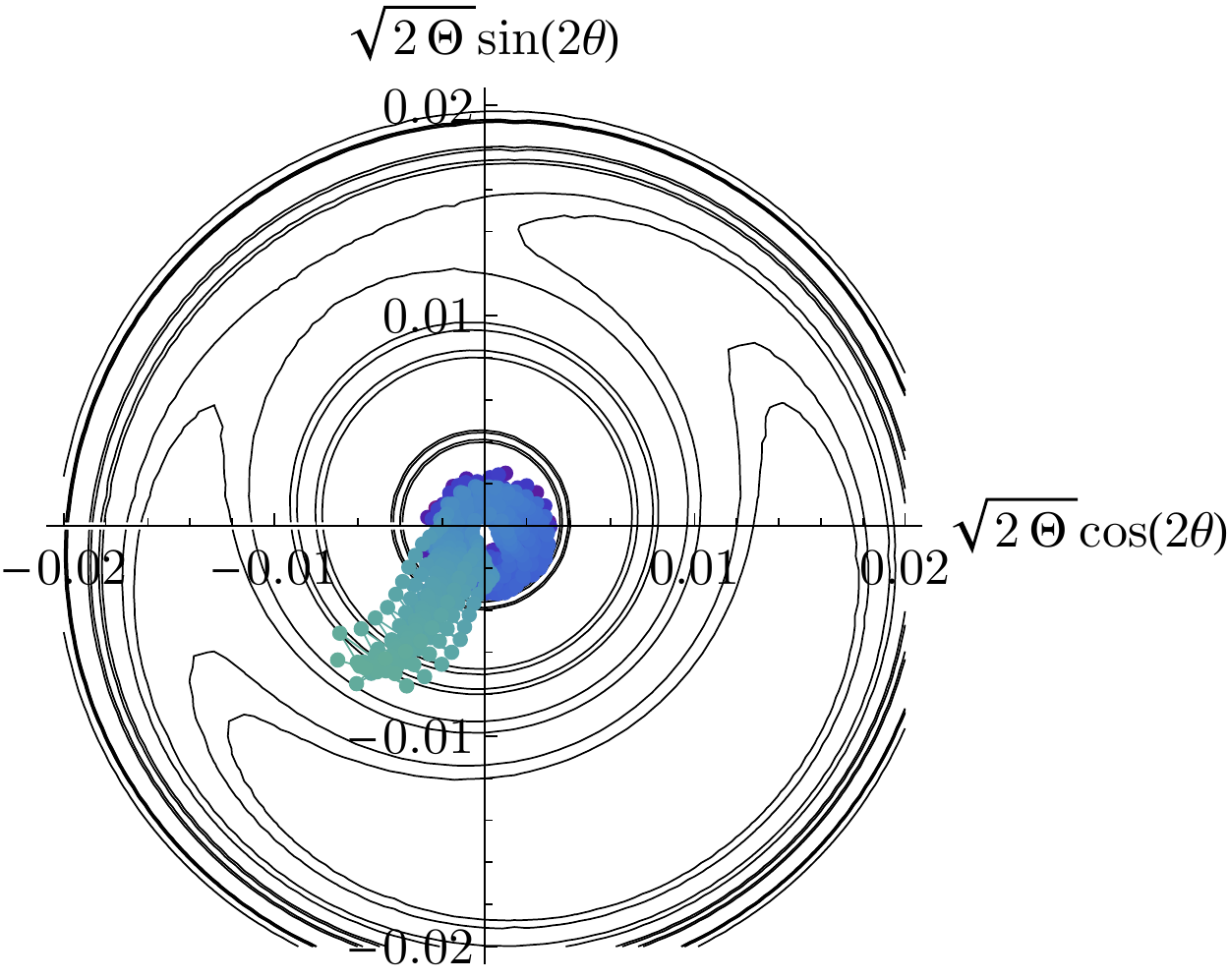}
\caption{}
\end{subfigure}
\qquad
\centering
\begin{subfigure}[b]{0.275 \textwidth}
\centering
\includegraphics[width=1.1 \textwidth]{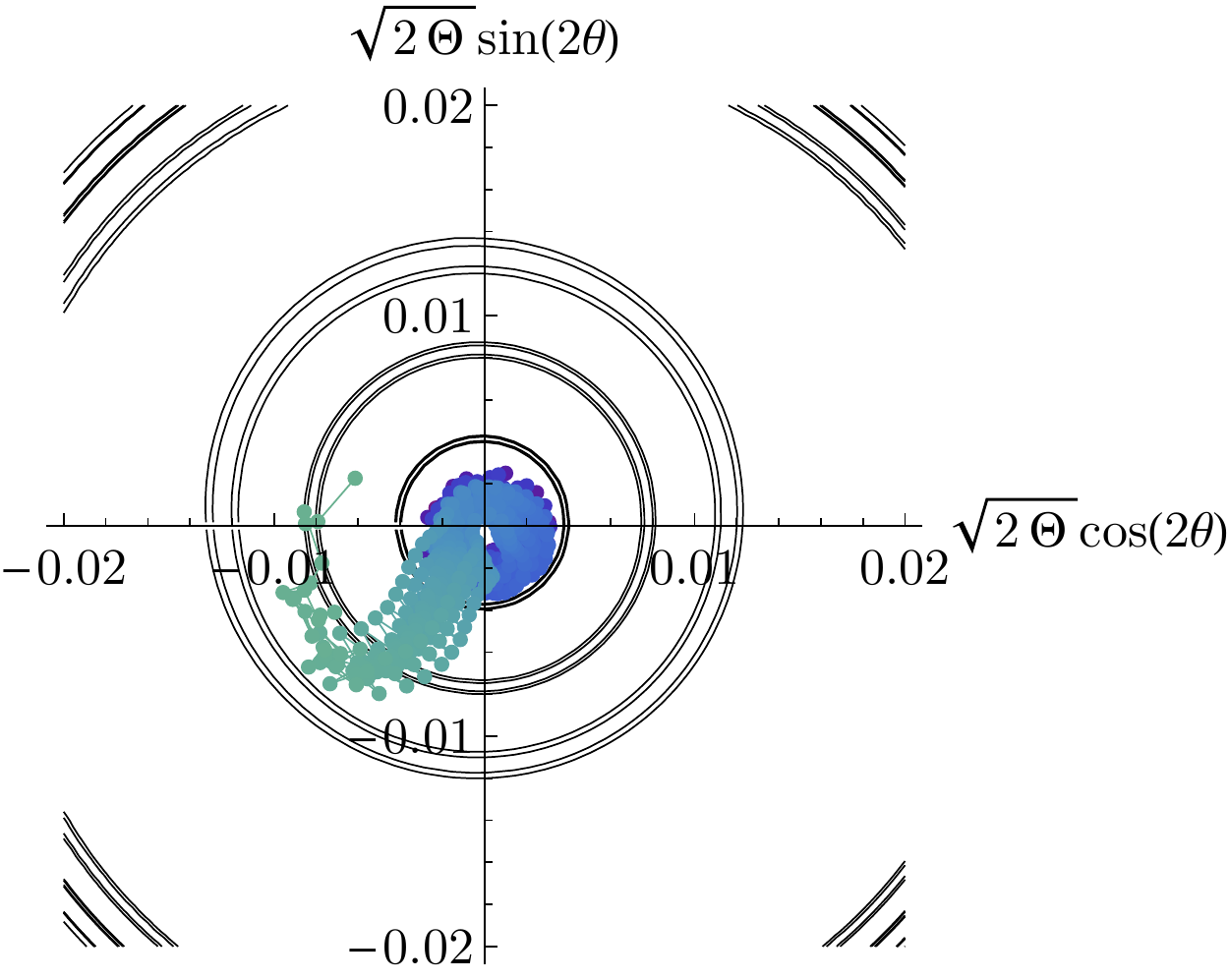}
\caption{}
\end{subfigure}
\qquad
\centering
\begin{subfigure}[b]{0.275 \textwidth}
\centering
\includegraphics[width=1.1 \textwidth]{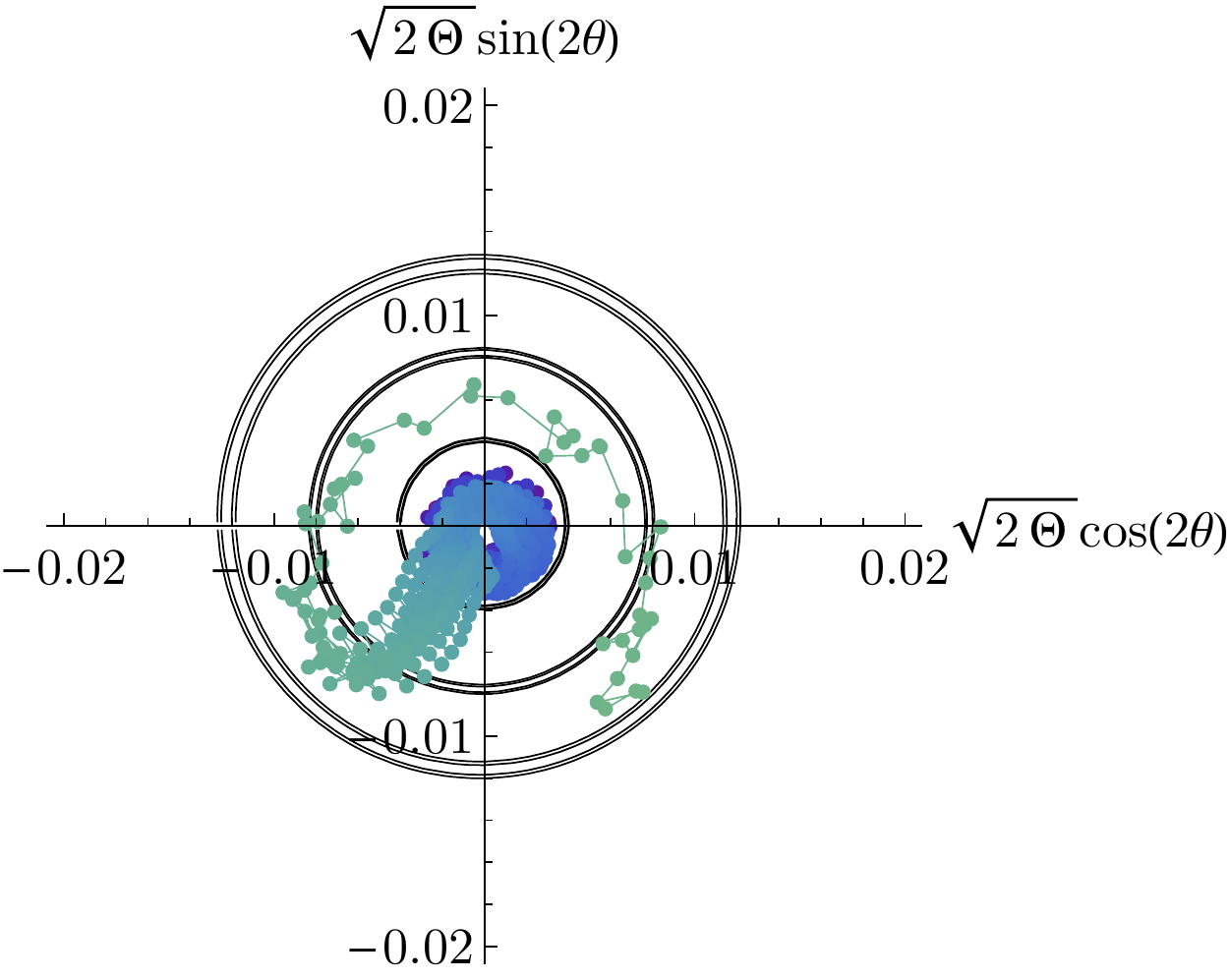}
\caption{}
\end{subfigure}
\caption{
Contour plots of the Hamiltonian of the integrable model \eqref{eq:HaAveragedOverphi*anddeltalambda*-deltalambda12+2phi2} for the secondary resonance involving $2\theta=\delta\LAMBDA_{1,2}'/3+2\phi_2$ and $\Theta=I_2$, shown in panels (a) to (f) in canonical cartesian coordinates at different values of the parameter $\delta$. The change in $\delta$ represents the change in the planetary mass $\mpl$ implemented along the integration. The dots represent the evolution of the system along the numerical integration, and their colour indicate the value of $\mpl$ using the same colour scheme as in Figure \ref{fig:3-2_3-2_twothetaevoFSR}.}
\label{fig:3-2_3-2_FigureForCaptureFSR}
\end{figure*}

\begin{figure}[!t]
\centering
\includegraphics[scale=.6]{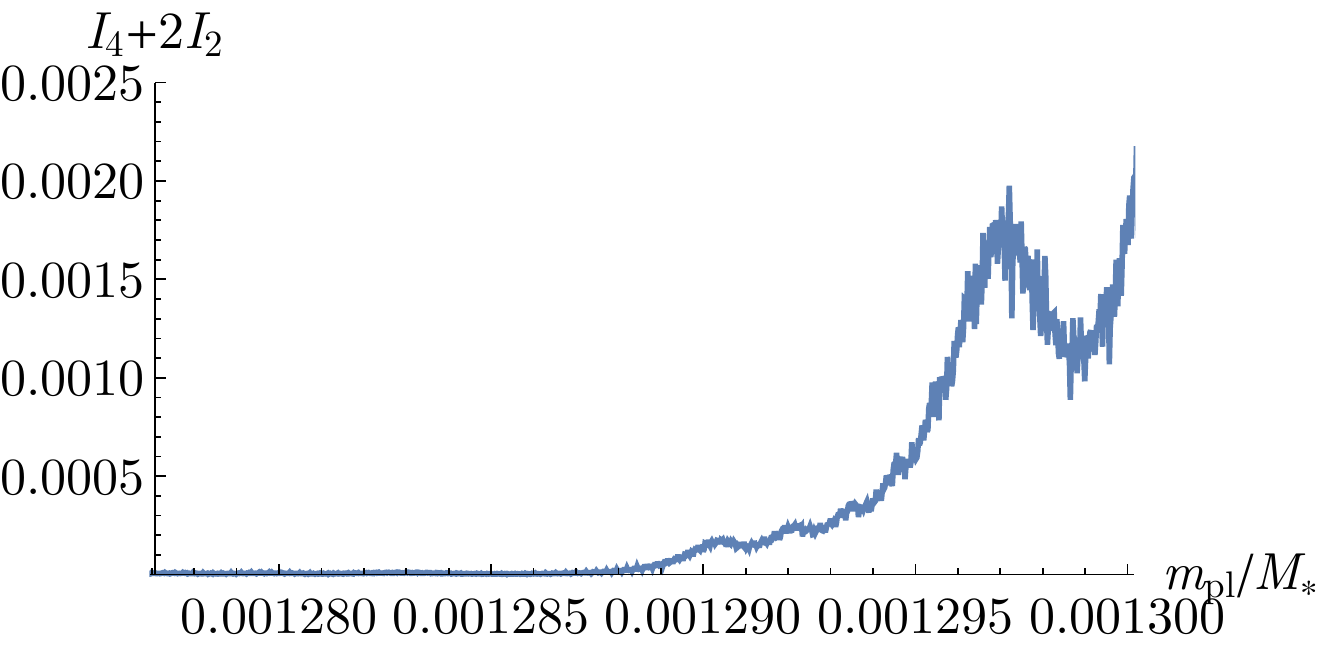}
\caption{Evolution of the action $I_4+2I_2$, which is a constant of motion relative to the harmonic term $\phi_2-2\phi_4$, as well as the resonant action conjugated to the slow angle $\delta\LAMBDA_{1,2}'/3+\phi_2+2\phi_4$, cfr.\ Eq.\ \eqref{eq:CdVForI4+2I2AsResonantAction}. 
}
\label{fig:3-2_3-2_IevoArrayFSRI4+2I2}
\end{figure}

Figure \ref{fig:3-2_3-2_FigureForCaptureFSR} shows the level plots of the Hamiltonian \eqref{eq:HaAveragedOverphi*anddeltalambda*-deltalambda12+2phi2}, for different values of $\delta$ (i.e.\ of the frequency of $\delta\LAMBDA_{1,2}'/3+2\phi_2$ at $\Theta=0$), in the variables $(X=\sqrt{2\Theta}\cos(2\theta),Y=\sqrt{2\Theta}\sin(2\theta))$; we also overplot the evolution of $(X,Y)$ obtained from the simulation (a combination of Figures \ref{fig:3-2_3-2_IevoArrayFSR} and \ref{fig:3-2_3-2_twothetaevoFSR}), truncated at the value of the planetary mass corresponding to the same $\delta$ used to plot the phase diagrams. 
Initially, there is only one stable centre at the origin (panel (a)) and the orbit circulates anti-clockwise around it with constant amplitude. Then, we see that a resonant island bifurcates from the origin in the bottom-right quadrant of the phase diagram (panel (b)), which is followed by the dynamical evolution.
Almost immediately after, a second bifurcation occurs at roughly the same $\delta$, so the inner circulation region starts to grow around the origin and catches up with the orbit (panels (c) and (d)). After crossing the inner separatrix, the dynamical evolution drops off the resonant island, falls inside the inner circulation region and and the angle $2\theta$ starts to circulate in clockwise fashion (panels (e) and (f)). This missed capture into resonance is one of the two probabilistic fates for a second order resonance when $\dot\delta\beta < 0$ and when the two bifurcations occur at close values of $\delta$. 
However, in this specific case we checked that this evolution is actually the result of more complicated interactions among the variables $(I_l,\phi_l)$ themselves, as well as another secondary resonance involving the variables $(I_l,\phi_l)$ and $(\delta\Delta\LAMBDA_{1,2}',\delta\LAMBDA_{1,2}')$. First, from Figure \ref{fig:3-2_3-2_IevoArrayFSR}, one can see that after the initial increase of $I_2$, $I_4$ starts increasing also, after which there are wide oscillations of $I_2$ and $I_4$ in opposite phase. This is symptomatic of the effect of the term $\sqrt{I_2}I_4\cos(\phi_2-2\phi_4)$, which is quasi-resonant because $\omega_2\simeq2\omega_4$ (Figure \ref{fig:3-2_3-2PurelyResonantModel.subfig:3-2_3-2_plot4FreqLowe}). To prove this, we plot in Figure \ref{fig:3-2_3-2_IevoArrayFSRI4+2I2} the action $I_4+2I_2$, which is the constant of motion relative to this harmonic term: we see that the aforementioned coupled oscillations undergone by $I_2$ and $I_4$ are completely eliminated. On the other hand, for $\mpl/M_*>1.297\times10^{-3}$ we see a much longer period large oscillation, which diverges towards the end of the integration. We interpret this as evidence of a transition of the system into the secondary resonance with argument $\delta\LAMBDA_{1,2}'/3+\phi_2+2\phi_4$ (which also has a small frequency, since $\delta\LAMBDA_{1,2}'/3+2\phi_2$ and $-\phi_2+2\phi_4$ are both slow angles). The reason is that (up to a constant) $I_4+2I_2$ can also be seen as a conjugated action of $\delta\LAMBDA_{1,2}'/3+\phi_2+2\phi_4$ through the canonical change of variables
\begin{alignat}{2}\label{eq:CdVForI4+2I2AsResonantAction}
&(I_4+2I_2)/4,						&&\quad \delta\LAMBDA_{1,2}'/3+\phi_2+2\phi_4, \nonumber\\
&(I_4-2I_2)/4,						&&\quad -\delta\LAMBDA_{1,2}'/3-\phi_2+2\phi_4,\\
&\delta\Delta\LAMBDA_{1,2}' - I_2/3,		&&\quad \delta\LAMBDA_{1,2}'.\nonumber
\end{alignat}
We see that after the angle $\theta=\delta\LAMBDA_{1,2}'/6+\phi_2$ leaves the first resonance (at mass $\mpl/M_*\simeq1.29\times 10^{-3}$, see Figure \ref{fig:3-2_3-2_twothetaevoFSR}) the action $2I_2+I_4$ keeps growing, which indicates a transition to this new resonance involving $\delta\LAMBDA_{1,2}'/3+\phi_2+2\phi_4$.

This analysis shows that the evolution presented above is very rich, and does not allow any simple description of it. In any case, Figure \ref{fig:3-2_3-2_FigureForCaptureFSR} does not leave any doubt that the initial growth of $I_2$ is due to the interaction with the secondary resonance associated to the angle $\delta\LAMBDA_{1,2}'/3+2\phi_2$, and that the simple model we have derived yields an effective understanding of the evolution, at least at a qualitative level.\\

The takeaway is the following. We showed that the numerical integration of the system $\Ha^*=\Hakepl+\Hares+\Hasyn$ presents an evolution that is similar to that obtained in the full $(N+1)$-body simulations where the resonant degrees of freedom get excited; we checked that the purely resonant system instead does not undergo the same evolution, and gave an analytical explanation to this fact. We then showed analytically that a set of secondary resonances are present in the $\Ha^*$ system, which involve a fraction of the synodic frequency and combination of the resonant frequencies, and which appear at order two in the planetary mass. Then, we found the specific secondary resonance that is encountered in the numerical integration of $\Ha^*$; we built an integrable model for this resonance valid as long as the actions remain small, and confirmed analytically that there can be capture into this resonance. Finally, we verified that the numerical evolution we obtained in the numerical integration corresponds to a temporary capture into the considered secondary resonance, followed by a rich and fascinating series of interactions with additional secondary resonances.

\section{Mass-limit for stability as a function of number of planets and resonance index}\label{sec:StabilityN>3}
We now have all the information needed to derive the general result anticipated at the end of Section \ref{sec:OriginOfInstability}, namely the dependence of the maximal planetary mass ensuring stability as a function of $N$ and $k$ (i.e.\ the planet number and resonant index).
We sketch below how the results found in the previous section can be generalised to the case of $N\geq3$ equal-mass planets in a given $k$:$k-1$ mean motion resonance chain. 

Following the development presented in Section \ref{sec:HamiltonianModel}, but for an arbitrary case of $N$ planets, we start introducing the Hamiltonian $\Ha^*=\Hakepl+\Hares+\Hasyn$, where $\Hares$ contains the resonant interactions between all $N-1$ pairs of neighbouring planets, and $\Hasyn$ contains terms of type $\cos(\LAMBDA_i-\LAMBDA_{i+1})$, $i=1,\dots,N-1$; for both $\Hares$ and $\Hasyn$ we will consider interaction terms up to order one in the eccentricities, as we did in the previous section.
We then make use of a canonical transformation analogous to (\ref{eq:CanonicalVariablesqWithSynodicAngle}, \ref{eq:CanonicalVariablespWithSynodicAngle}): we introduce the resonant angles $\PSI_1^{(i)}=k\LAMBDA_{i+1}-(k-1)\LAMBDA_i+\GAMMA_i$ and the angles $\delta\GAMMA_{i,i+1}=\GAMMA_i-\GAMMA_{i+1}$ for each planet pair, $i=1,\dots,N-1$ (this gives $2(N-1)$ resonant degrees of freedom), then we have the synodic angle $\delta\LAMBDA_{1,2}=\LAMBDA_1-\LAMBDA_2$, and finally a $\GAMMA_N'=-\GAMMA_N$ which will not appear in the Hamiltonian (its conjugated action will again be the total orbital angular momentum). 

We can immediately generalise the result of Subsection \ref{sec:StabilityN=3.subsec:PurelyResonantDynamics} and state that the $2(N-1)$ purely resonant degrees of freedom are Lyapunov stable at low amplitude of libration around the resonant equilibrium point for any number of planets $N$. This is because, when adding an outer resonant pair to the system with the same resonance index $k$, the Hamiltonian simply repeats itself, since to first order in the eccentricities we are only considering the mutual planetary perturbations due to immediately neighbouring planets and the structure of each term is the same, namely \eqref{eq:RescaledResonantInteractionHamiltonian}. So, each planet is either the inner or outer planet, or a middle planet as in the case already considered of a three-planets system. Therefore, all resonant libration frequencies will always have the same sign, and the reasoning of Subsection \ref{sec:StabilityN=3.subsec:PurelyResonantDynamics} stands.

As in the case of three resonant planets, we thus conclude that the instability must be due to an interaction between the synodic degree of freedom and the purely resonant degrees of freedom. Then, it is natural to investigate when a regime of secondary resonances analogous to \eqref{eq:NeededSecondaryResonanceHarmonicsInqk1=k2=k} and \eqref{eq:PossibleSecondaryResonanceHarmonicsInphi} can be encountered. To answer this question, we proceeded analytically following the steps of Subsect.\ \ref{subsubsec:TheOmpl^2contribution}. We introduce a generating Hamiltonian $\chisyn$ which eliminates the synodic contribution $\Hasyn$, so $\chisyn$ in Delaunay variables will have harmonic terms $\sin(\LAMBDA_i-\LAMBDA_{i+1})$, $i=1,\dots,N-1$. 
Transforming $\Ha^*=\Hakepl+\Hares+\Hasyn$ with the Lie series generated by $\chisyn$ eliminates $\Hasyn$ to first order in $\mpl$ (the planetary mass for all planets), but introduces new terms to order 2 in $\mpl$ (among which the most important is $\{\{\Hakepl,\chisyn\},\chisyn\}$, like in the case $N=3$); these newly introduced terms will contain a fraction of the synodic angle ${\delta\LAMBDA_{1,2}}$, and we are interested in the smallest fraction of $\delta\LAMBDA_{1,2}$ that appears. 
Like in the case of three planets, the term $(\mpl^2/2)\{\{\Hakepl,\chisyn\},\chisyn\}$ combines together all synodic angles $\LAMBDA_i-\LAMBDA_{i+1}$. Notice that in the new coordinates $\PSI_1^{(i)}$, $\delta\GAMMA_{i,i+1}$ and $\delta\LAMBDA_{1,2}$, each $\LAMBDA_i-\LAMBDA_{i+1}$ can be written as $\LAMBDA_i-\LAMBDA_{i+1}=\left(\frac{k-1}{k}\right)^{i-1}(\LAMBDA_1-\LAMBDA_2)=\left(\frac{k-1}{k}\right)^{i-1}\delta\LAMBDA_{1,2}$ plus terms including $\PSI_1^{(j)}$ and $\delta\GAMMA_{j,j+1}$. However we do not need to keep track of the $\PSI_1^{(j)}$'s and $\delta\GAMMA_{j,j+1}$'s since we are only interested in the way the angle $\delta\LAMBDA_{1,2}$ appears in the $\mathcal O(\mpl^2)$ terms. The smallest fraction of $\delta\LAMBDA_{1,2}$ will be generated by combining the synodic angles relative to the two outermost pairs $\LAMBDA_{N-2}-\LAMBDA_{N-1}$ and $\LAMBDA_{N-1}-\LAMBDA_N$, since already they contain the smallest fraction of $\delta\LAMBDA_{1,2}$. Multiplying them together (using $\cos(a)\cos(b) = \frac{1}{2}\left(\cos(a-b)+\cos(a+b)\right)$) yields a harmonic term of type
\begin{equation}
\cos\left(\left(\left(\frac{k-1}{k}\right)^{N-3} - \left(\frac{k-1}{k}\right)^{N-2} \right) \delta\LAMBDA_{1,2} + \dots \right),
\end{equation}
where the $+\dots$ terms represents a combination of $\PSI_1^{(j)}$'s and $\delta\GAMMA_{j,j+1}$'s, in which again we are not interested. Therefore, the lowest synodic frequency that appears in the $\mathcal O(m^2)$ term is 
\begin{equation}\label{eq:SmallestSynodicFrequencyAtOrder2InmplN>=3}
\frac{1}{k} \left(\frac{k-1}{k}\right)^{N-3}\dot{\delta\lambda_{1,2}}\simeq \frac{1}{k^2} \left(\frac{k-1}{k}\right)^{N-3}n_1.
\end{equation}
This is the fraction of the synodic frequency which can resonate with the libration frequencies $\omega_l$ of the resonant degrees of freedom. Since $\omega_l$ increase with $\mpl$ (as $\omega\sim \mpl^{1/2}$ or $\mpl^{2/3}$ according to the eccentricities), there will be a critical mass after which a regime of secondary resonances is encountered, which can excite the system and cause its instability. Since the factor $\frac{1}{k} \left(\frac{k-1}{k}\right)^{N-3}$ multiplying $\dot{\delta\lambda_{1,2}}$ decreases with increasing $N$ and $k$, the conclusion is that {\it the regime of secondary resonances between synodic and resonant degrees of freedom is encountered at lower masses for increasing $k$ and/or increasing $N$, and therefore the critical mass $(\mpl/M_*)_\crit$ allowed for stability decreases with increasing $N$ and $k$}. This gives an analytical explanation to the numerical findings of \cite{2012Icar..221..624M}.

\section{Conclusions}\label{sec:Conclusions}
In this paper, we have considered the stability of chains of mean motion resonances, in relation to the observed exoplanet population. Previous works have demonstrated that the paucity of resonances in the exoplanets sample is not in contradiction with the scenario of capture into mean motion resonance during planet migration in the disc phase, if post-disc instability rates are as high as 90\% \citep{2017MNRAS.470.1750I,2019arXiv190208772I}. This motivates a detailed study on the stability of these chains.
Previous numerical investigations pointed out that there is a critical planetary mass above which the instability time of resonant systems is comparable to that of non-resonant ones, and that this limit mass decreases with increasing number of planets and/or increasing index $k$ of the resonance \citep{2012Icar..221..624M}. The dynamical origin of these instabilities was however not discussed.
In this paper we thus investigated analytically and numerically the origin of these instabilities.

From the numerical perspective, we used numerical experiments where we first put low-mass planets deep in resonance (at low level of excitation of the resonant modes) and secondly we slowly (and fictitiously) increased the planetary mass to follow the low-amplitude regime until the onset of instability. 
We confirmed that the instability for three resonant planets occurs at smaller masses than in the two-planet case, and we identified a novel dynamical mechanism which excites the amplitude of libration of the resonant degrees of freedom. The excited systems can then become unstable by suffering close encounters and collisions. 

Therefore, we investigated analytically this phenomenon, using a simplified Hamiltonian which reproduced well the observed excitation of the system. Carrying out the calculation explicitly in the case $k=3$, we showed that the observed excitation is due to a set of secondary resonances between a combination of the resonant libration frequencies and a fraction of the synodic frequency. We identified the specific secondary resonance that caused the effect observed in the numerical integrations, and built a simple integrable model for this resonance which captures qualitatively the dynamics until the excitation of the system is too severe, showing for example that there can be a capture into this specific resonance. This technique can be generalised to the other secondary resonances. 

We therefore proposed that in the numerical simulations the systems become unstable due to a crossing of this type of secondary resonances, which excites the planets' orbits and leads to a phase of close encounters and collisions. This gives a critical mass at which a regime of secondary resonances is encountered, and after which the system can be destabilised. This scheme can then be generalised to an arbitrary number of planets $N$ and/or an arbitrary index of the first-order mean motion resonance $k$ of the chain. One can easily calculate for different $N$'s and $k$'s the lowest fraction of the synodic frequency that can resonate with the resonant frequencies, and see that it decreases with increasing values of $N$ and $k$ (Eq.\ \eqref{eq:SmallestSynodicFrequencyAtOrder2InmplN>=3}). Consequently, the regime of secondary resonances between synodic and resonant degrees of freedom is encountered when the resonant libration frequencies are slower. Because the libration frequencies grow with the planetary mass, this implies that the instability of the resonant chain occurs at lower masses for increasing $k$ and/or increasing $N$, and therefore the critical mass allowed for stability decreases with $N$ and with $k$. This gives an analytical explanation to the numerical findings of \cite{2012Icar..221..624M}.

The takeaway is that we now have a dynamical understanding of the origin of the instabilities observed in the numerical experiments of \cite{2012Icar..221..624M,2019arXiv190208772I}, which captures the trend in the dependence of the critical mass allowed for stability on the index of the resonance $k$ and the number of planets $N$. Having understood this mechanism, we will be able to perform a more focused and quantitative analysis on the threshold of stability of resonant chains with different $N$, $k$ and $\mpl$, and produce an explicit criterion for the stability against secondary resonances of the type described here. This will be the subject of future work.

\newcommand{\amp}{AMP}

\begin{appendix}
\section{Simulating capture into mean motion resonance via convergent migration}\label{app:NBodyCapture}

In this appendix we summarise the numerical recipes used to implement disc-planet interaction effects which mimic type-I migration. Formul\ae\ for the eccentricity damping and migration timescales are taken from \citep{2008A&A...482..677C}. We first briefly recall the setup for two planets which was used in \cite{2018CeMDA.130...54P}, and then describe the case of three (or more) resonant planets.\\

The effect of the disc onto a planet of mass $m$ of the Super-Earth/Mini-Neptune type (the so-called type-I migration) is split into eccentricity damping and migration, typically inward. We first define a  type-I migration factor
\begin{equation}\label{eq:twave}
\tau_\wave = \frac{M_*}{m} \frac{M_*}{\Sigma a^2} \frac{h^4}{\sqrt{\GravC M_*/a^3}},
\end{equation}
where $\Sigma=\Sigma(r) \propto r^{-\alphaSigma}$ is  
the surface density and $h = h(r) = H/r \propto r^\betah$ is the aspect ratio of the protoplanetary disc, both evaluated at the location of the planet.
The eccentricity damping is modeled as
\begin{equation}
\dot{e}_\damp = -\frac{e}{\tau_e},
\end{equation}
where $\tau_e$ is given, in the limit of vanishing eccentricities, by
\begin{equation}\label{eq:tau_e}
\tau_e 	\simeq \frac{\tau_\wave}{0.780};
\end{equation}
the change in angular momentum (a negative torque) is modeled as 
\begin{equation}\label{eq:dotAngMomMig}
\dot\ANGMOM_\mig = -\frac{\ANGMOM}{\tau_\mig},
\end{equation}
which yields 
\begin{equation}\label{eq:adotovera}
\frac{\dot a}{a} = 2\frac{\dot\ANGMOM}{\ANGMOM} + \frac{2 e \dot e}{1-e^2} 
                 = - \frac{1}{\tau_a} - p \frac{e^2}{\tau_e},
\end{equation}
where
\begin{equation}
\tau_a=\frac{\tau_\mig}{2},
\end{equation}
and, again in the limit of vanishing eccentricities, $p\simeq2$ and
\begin{equation}
\tau_\mig\simeq2 \frac{\tau_\wave}{(2.7 + 1.1 \alphaSigma)} h^{-2}.
\end{equation}
Moreover, we smoothly reverse the sign of the torque at the desired location of the inner edge of the disc in order to stop inward migration, simulating the effects of a cavity.\\

When two planets are embedded in the disc, the inner planet stops migrating at the edge of the disc and the outer planet continues to migrate inward until it approaches a mean motion resonance with the first, see \citep{2018CeMDA.130...54P} for the details. By balancing the eccentricity excitation due to the resonance and the eccentricity damping provided from the disc, one finds that the equilibrium eccentricities at the equilibrium captured state are given by
\begin{equation}\label{eq:e2EquilibriumWithTrap}
\frac{(R^{3/2}-1)}{\tau_{\mig,2}}-\frac{R e_1^2}{\tau_{e,1}} - \frac{e_2^2}{\tau_{e,2}} = 0,
\end{equation}
where $R=a_2/a_1$ and $\tau_{\mig,2}$ is migration rate of the outer planet (cfr.\ Eq.\ (A.20) in \citealt{2018CeMDA.130...54P}). Since in resonance $e_1\propto e_2$ with a factor that only depends on the masses of the planets, one has that the final equilibrium eccentricity at the capture state is of the order of $\sim H/r$, the aspect ratio of the disc.
Thus, by changing the aspect ratio of the disc (or equivalently the eccentricity damping timescale) one can reach a resonant configuration with virtually any desired value of the eccentricities. 
Modifying the disc structure is not an issue here, since the the sole role of the first phase of the numerical experiments described in Section \ref{sec:NumericalMaps} is to obtain a deeply resonant configuration with a desired eccentricity, in order to subsequently study the stability of the obtained configuration as a function of the planetary mass in the second phase.\\

The case of capture of three (or more) planets in resonance at different desired eccentricities is similar, with only one minor difference. Because the planets capture in resonance in sequence (first planet 1 and 2, then planet 3, etc.) if $\tau_e$ is large, $e_1$ and $e_2$ can grow significantly before planet 3 enters in resonance. This can force large secular eccentricity oscillations of planet 3, which may preclude its resonant capture (see e.g.\ \cite{2015MNRAS.451.2589B} on criteria for resonant capture). 
We therefore use the following numerical recipe that allows us to capture all $N$ planets at the desired resonance and at any reasonably eccentric configurations. We first capture all planets at small eccentricities, that is with small $\tau_e$. Then we slowly increase the value of $\tau_e$ while the planets are already locked in resonance: since the strength of the resonant interaction stays the same while $K=\tau_a/\tau_e$ decreases, Equation \eqref{eq:e2EquilibriumWithTrap} shows that the planets will adjust to the change in $\tau_e$ by becoming more eccentric. By doing so adiabatically we obtain, at the fixed initial planetary mass, resonant chains with the same small amplitude of libration around the resonant equilibrium point with different equilibrium eccentricities. This method does not follow the real dynamical evolution of planetary systems, however we reiterate that the role of this first phase is simply to put the planets deeply into a desired resonant chain with a desired eccentricity (i.e.\ angular momentum) in order to subsequently study the stability of the obtained configuration as a function of the planetary mass.

\end{appendix}


\begin{thebibliography}{9}

\bibitem[Batygin(2015)]{2015MNRAS.451.2589B} Batygin, K.\ 2015.\ Capture of planets into mean-motion resonances and the origins of extrasolar orbital architectures.\ Monthly Notices of the Royal Astronomical Society 451, 2589.

\bibitem[Batygin and Morbidelli(2013)a]{2013A&A...556A..28B} Batygin, K., Morbidelli, A.\ 2013.\ Analytical treatment of planetary resonances.\ Astronomy and Astrophysics 556, A28.

\bibitem[Batygin and Morbidelli(2013)b]{2013AJ....145....1B} Batygin, K., Morbidelli, A.\ 2013.\ Dissipative Divergence of Resonant Orbits.\ The Astronomical Journal 145, 1.

\bibitem[Batygin and Adams(2017)]{2017AJ....153..120B} Batygin, K., Adams, F.~C.\ 2017.\ An Analytic Criterion for Turbulent Disruption of Planetary Resonances.\ The Astronomical Journal 153, 120.

\bibitem[Chatterjee et al.(2016)]{2016IAUFM..29A..30C} Chatterjee, S., Krantzler, S.~O., Ford, E.~B.\ 2016.\ Period Ratio Distribution of Near-Resonant Planets Indicates Planetesimal Scattering.\ IAU Focus Meeting 29A, 30.

\bibitem[Cresswell and Nelson(2008)]{2008A&A...482..677C} Cresswell, P., Nelson, R.~P.\ 2008.\ Three-dimensional simulations of multiple protoplanets embedded in a protostellar disc.\ Astronomy and Astrophysics 482, 677.

\bibitem[Deck and Batygin(2015)]{2015ApJ...810..119D} Deck, K.~M., Batygin, K.\ 2015.\ Migration of Two Massive Planets into (and out of) First Order Mean Motion Resonances.\ The Astrophysical Journal 810, 119.

\bibitem[Fressin et al.(2013)]{2013ApJ...766...81F} Fressin, F., and 8 colleagues 2013.\ The False Positive Rate of Kepler and the Occurrence of Planets.\ The Astrophysical Journal 766, 81.

\bibitem[Gillon et al.(2016)]{2016Natur.533..221G} Gillon, M., and 14 colleagues 2016.\ Temperate Earth-sized planets transiting a nearby ultracool dwarf star.\ Nature 533, 221.

\bibitem[Gillon et al.(2017)]{2017Natur.542..456G} Gillon, M., and 29 colleagues 2017.\ Seven temperate terrestrial planets around the nearby ultracool dwarf star TRAPPIST-1.\ Nature 542, 456.

\bibitem[Gladman(1993)]{1993Icar..106..247G} Gladman, B.\ 1993.\ Dynamics of Systems of Two Close Planets.\ Icarus 106, 247.

\bibitem[Howard et al.(2012)]{2012ApJS..201...15H} Howard, A.~W., and 66 colleagues 2012.\ Planet Occurrence within 0.25 AU of Solar-type Stars from Kepler.\ The Astrophysical Journal Supplement Series 201, 15.

\bibitem[Izidoro et al.(2017)]{2017MNRAS.470.1750I} Izidoro, A., and 7 colleagues 2017.\ Breaking the chains: hot super-Earth systems from migration and disruption of compact resonant chains.\ Monthly Notices of the Royal Astronomical Society 470, 1750. 

\bibitem[Izidoro et al.(2019)]{2019arXiv190208772I} Izidoro, A., and 6 colleagues 2019.\ Formation of planetary systems by pebble accretion and migration: Hot super-Earth systems from breaking compact resonant chains.\ arXiv e-prints arXiv:1902.08772.

\bibitem[Laskar(1990)]{1990mmmc.conf...63L} Laskar, J.\ 1990.\ {\it Syst{\`e}mes de Variables et El{\'e}ments.\ Modern Methods in Celestial Mechanics}, Comptes Rendus de la 13{\`e}me Ecole Printemps d'Astrophysique de Goutelas (France), 24-29 Avril, 1989.~Edited by Daniel Benest and Claude Froeschle.~ Gif-sur-Yvette: Editions Frontieres, 1990., p.63 63. 

\bibitem[Laskar and Robutel(1995)]{1995CeMDA..62..193L} Laskar, J., Robutel, P.\ 1995.\ Stability of the Planetary Three-Body Problem. I. Expansion of the Planetary Hamiltonian.\ Celestial Mechanics and Dynamical Astronomy 62, 193.

\bibitem[Luger et al.(2017)]{2017NatAs...1E.129L} Luger, R., and 32 colleagues 2017.\ A seven-planet resonant chain in TRAPPIST-1.\ Nature Astronomy 1, 129.

\bibitem[Marchal and Bozis(1982)]{1982CeMec..26..311M} Marchal, C., Bozis, G.\ 1982.\ Hill Stability and Distance Curves for the General Three-Body Problem.\ Celestial Mechanics 26, 311.

\bibitem[Masset et al.(2006)]{2006ApJ...642..478M} Masset, F.~S., Morbidelli, A., Crida, A., Ferreira, J.\ 2006.\ Disk Surface Density Transitions as Protoplanet Traps.\ The Astrophysical Journal 642, 478.

\bibitem[Matsumoto et al.(2012)]{2012Icar..221..624M} Matsumoto, Y., Nagasawa, M., Ida, S.\ 2012.\ The orbital stability of planets trapped in the first-order mean-motion resonances.\ Icarus 221, 624.

\bibitem[Mayor et al.(2011)]{2011arXiv1109.2497M} Mayor, M., and 13 colleagues 2011.\ The HARPS search for southern extra-solar planets XXXIV. Occurrence, mass distribution and orbital properties of super-Earths and Neptune-mass planets.\ arXiv e-prints arXiv:1109.2497.

\bibitem[Michtchenko et al.(2008)]{2008MNRAS.387..747M} Michtchenko, T.~A., Beaug{\'e}, C., Ferraz-Mello, S.\ 2008.\ Dynamic portrait of the planetary 2/1 mean-motion resonance - I. Systems with a more massive outer planet.\ Monthly Notices of the Royal Astronomical Society 387, 747.

\bibitem[Millholland et al.(2017)]{2017ApJ...849L..33M} Millholland, S., Wang, S., Laughlin, G.\ 2017.\ Kepler Multi-planet Systems Exhibit Unexpected Intra-system Uniformity in Mass and Radius.\ The Astrophysical Journal 849, L33.

\bibitem[Mills et al.(2016)]{2016Natur.533..509M} Mills, S.~M., Fabrycky, D.~C., Migaszewski, C., Ford, E.~B., Petigura, E., Isaacson, H.\ 2016.\ A resonant chain of four transiting, sub-Neptune planets.\ Nature 533, 509.

\bibitem[Morbidelli(2002)]{2002mcma.book.....M} Morbidelli, A.\ 2002, {\it Modern celestial mechanics : aspects of solar system dynamics}, by Alessandro Morbidelli.~London: Taylor {\amp} Francis, 2002, ISBN 0415279399,

\bibitem[Morbidelli et al.(2008)]{2008A&A...478..929M} Morbidelli, A., Crida, A., Masset, F., Nelson, R.~P.\ 2008.\ Building giant-planet cores at a planet trap.\ Astronomy and Astrophysics 478, 929.

\bibitem[Murray \& Dermott(1999)]{1999ssd..book.....M} Murray, C.~D., \& Dermott, S.~F.\ 1999, {\it Solar system dynamics} by C.D.~Murray and S.F.~McDermott.~(Cambridge, UK: Cambridge University Press),  ISBN 0-521-57295-9 (hc.), ISBN 0-521-57297-4 (pbk.).,    

\bibitem[Obertas et al.(2017)]{2017Icar..293...52O} Obertas, A., Van Laerhoven, C., Tamayo, D.\ 2017.\ The stability of tightly-packed, evenly-spaced systems of Earth-mass planets orbiting a Sun-like star.\ Icarus 293, 52.

\bibitem[Ogihara et al.(2015)]{2015A&A...578A..36O} Ogihara, M., Morbidelli, A., Guillot, T.\ 2015.\ A reassessment of the in situ formation of close-in super-Earths.\ Astronomy and Astrophysics 578, A36.

\bibitem[Petigura et al.(2013)]{2013PNAS..11019273P} Petigura, E.~A., Howard, A.~W., Marcy, G.~W.\ 2013.\ Prevalence of Earth-size planets orbiting Sun-like stars.\ Proceedings of the National Academy of Science 110, 19273.

\bibitem[Petit et al.(2018)]{2018A&A...617A..93P} Petit, A.~C., Laskar, J., Bou{\'e}, G.\ 2018.\ Hill stability in the AMD framework.\ Astronomy and Astrophysics 617, A93.

\bibitem[Pichierri et al.(2018)]{2018CeMDA.130...54P} Pichierri, G., Morbidelli, A., Crida, A.\ 2018.\ Capture into first-order resonances and long-term stability of pairs of equal-mass planets.\ Celestial Mechanics and Dynamical Astronomy 130, 54.

\bibitem[Poincar{\'e}(1892)]{1892mnm..book.....P} Poincar{\'e}, H.\ 1892, {\it Les  m{\'e}thodes nouvelles de la m{\'e}canique c{\'e}leste}, Gauthier-Villars, Paris, 1892  

\bibitem[Ramos et al.(2017)]{2017A&A...602A.101R} Ramos, X.~S., Charalambous, C., Ben{\'\i}tez-Llambay, P., Beaug{\'e}, C.\ 2017.\ Planetary migration and the origin of the 2:1 and 3:2 (near)-resonant population of close-in exoplanets.\ Astronomy and Astrophysics 602, A101.

\bibitem[Rogers(2015)]{2015ApJ...801...41R} Rogers, L.~A.\ 2015.\ Most 1.6 Earth-radius Planets are Not Rocky.\ The Astrophysical Journal 801, 41.

\bibitem[Terquem and Papaloizou(2007)]{2007ApJ...654.1110T} Terquem, C., Papaloizou, J.~C.~B.\ 2007.\ Migration and the Formation of Systems of Hot Super-Earths and Neptunes.\ The Astrophysical Journal 654, 1110.

\bibitem[Weiss et al.(2018)]{2018AJ....155...48W} Weiss, L.~M., and 12 colleagues 2018.\ The California-Kepler Survey. V. Peas in a Pod: Planets in a Kepler Multi-planet System Are Similar in Size and Regularly Spaced.\ The Astronomical Journal 155, 48.

\bibitem[Winn and Fabrycky(2015)]{2015ARA&A..53..409W} Winn, J.~N., Fabrycky, D.~C.\ 2015.\ The Occurrence and Architecture of Exoplanetary Systems.\ Annual Review of Astronomy and Astrophysics 53, 409.

\bibitem[Xu et al.(2018)]{2018MNRAS.481.1538X} Xu, W., Lai, D., Morbidelli, A.\ 2018.\ Migration of planets into and out of mean motion resonances in protoplanetary discs: overstability of capture and non-linear eccentricity damping.\ Monthly Notices of the Royal Astronomical Society 481, 1538.

\bibitem[Zeng et al.(2019)]{2019PNAS..116.9723Z} Zeng, L., and 15 colleagues 2019.\ Growth model interpretation of planet size distribution.\ Proceedings of the National Academy of Science 116, 9723.

\end{thebibliography}
\end{document}